\documentclass[floats,floatfix,showpacs,amssymb,prd,twocolumn,superscriptaddress,nofootinbib,nolongbibliography,reprint]{revtex4-2}

\usepackage{amssymb,amsmath,verbatim,mathtools,needspace,enumitem,etoolbox,graphicx,physics,microtype,afterpage,xspace,tabularx,lmodern,multirow}
\usepackage{gensymb}
\usepackage[dvipsnames, usenames]{xcolor}
\definecolor{linkcolor}{rgb}{0.0,0.3,0.5}
\usepackage[unicode, colorlinks=true, linkcolor=linkcolor, citecolor=linkcolor, filecolor=linkcolor, urlcolor=linkcolor, linktocpage, breaklinks]{hyperref}
\usepackage[all]{hypcap}
\usepackage[T1]{fontenc}
\usepackage[utf8]{inputenc}
\usepackage[usenames,dvipsnames]{xcolor}
\hypersetup{colorlinks=true,citecolor=romared,linkcolor=romared,urlcolor=romared}

\definecolor{romared}{RGB}{142,0,28}

\newcommand{\be}{\begin{equation}}
\newcommand{\ee}{\end{equation}}

\def\be{\begin{equation}}
\def\ee{\end{equation}}
\newcommand{\beq}{\begin{eqnarray}}
\newcommand{\eeq}{\end{eqnarray}}

\usepackage{aas_macros}
\usepackage{makecell}
\usepackage{soul}

\newcommand{\PPE}{\mbox{\tiny ppE}}
\newcommand{\GR}{\mbox{\tiny GR}}
\newcommand{\DIP}{\mbox{\tiny dipole}}
\newcommand{\dCS}{\mbox{\tiny dCS}}
\newcommand{\EdGB}{\mbox{\tiny EdGB}}

\newcommand{\MG}{\mbox{\tiny MG}}
\newcommand{\NC}{\mbox{\tiny NC}}

\newcommand{\BHE}{\mbox{\tiny BHE}}
\newcommand{\Gd}{\mbox{\tiny $\dot{G}$}}

\newcommand{\DET}{\mbox{\tiny det}}
\newcommand{\OPT}{\mbox{\tiny opt}}
\newcommand{\THRESH}{\mbox{\tiny thr}}
\newcommand{\SPACE}{\mbox{\tiny space}}
\newcommand{\TERR}{\mbox{\tiny terr}}
\newcommand{\MERG}{\mbox{\tiny merger}}
\newcommand{\MBH}{\mbox{\tiny MBH}}
\newcommand{\SOBH}{\mbox{\tiny SOBH}}
\newcommand{\OBS}{\mbox{\tiny obs}}
\newcommand{\ML}{\mbox{\tiny ML}}
\newcommand{\software}[1]{\texttt{#1}}
\newcolumntype{Y}{>{\centering\arraybackslash}X}
\newcommand{\tabcolsepcustom}{4pt}

\begin{document}
\title{Probing Fundamental Physics with Gravitational Waves: The Next Generation}

\begin{abstract}
Gravitational wave observations of compact binary mergers are already providing stringent tests of general relativity and constraints on modified gravity.
Ground-based interferometric detectors will soon reach design sensitivity and they will be followed by third-generation upgrades, possibly operating in conjunction with space-based detectors. 
How will these improvements affect our ability to investigate fundamental physics with gravitational waves?
The answer depends on the timeline for the sensitivity upgrades of the instruments, but also on astrophysical compact binary population uncertainties, which determine the number and signal-to-noise ratio of the observed sources.
We consider several scenarios for the proposed timeline of detector upgrades and various astrophysical population models. Using a stacked Fisher matrix analysis of  binary black hole merger observations, we thoroughly investigate future theory-agnostic bounds on modifications of general relativity, as well as bounds on specific theories. 
For theory-agnostic bounds, we find that ground-based observations of stellar-mass black holes and LISA observations of massive black holes can each lead to improvements of 2--4 orders of magnitude with respect to present gravitational wave constraints, while multiband observations can yield improvements of 1--6 orders of magnitude.
We also clarify how the relation between theory-agnostic and theory-specific bounds depends on the source properties.
\end{abstract}

\author{Scott E. Perkins}
\email{scottep3@illinois.edu}
\affiliation{Illinois Center for Advanced Study of the Universe (iCASU), Department of Physics, University of Illinois at Urbana-Champaign, Champaign, Illinois 61820 USA}
\author{Nicol\'as Yunes}
\email{nyunes@illinois.edu}
\affiliation{Illinois Center for Advanced Study of the Universe (iCASU), Department of Physics, University of Illinois at Urbana-Champaign, Champaign, Illinois 61820 USA}
\author{Emanuele Berti}
\email{berti@jhu.edu}
\affiliation{Department of Physics and Astronomy, Johns Hopkins University,
3400 N. Charles Street, Baltimore, Maryland, 21218, USA}
\date{\today}
\maketitle

\tableofcontents

\clearpage

\section{Introduction}\label{sec:intro}

Einstein's general relativity (GR) has been wildly successful. 
The agreement with the observed perihelion precession of Mercury and the 1919 eclipse expedition to verify the prediction of relativistic light-bending around the Sun were the beginning of a century of thorough vetting~\cite{Will:2014kxa}. 
The theory has passed every experimental test so far, and it was recently validated in the strong-field regime, most notably through the imaging of a black-hole (BH) shadow in the electromagnetic spectrum by the Event Horizon Telescope~\cite{Akiyama:2019cqa} and through the observation of coalescing binary black holes (BBHs) by the LIGO/Virgo Collaboration~\cite{LIGOScientific:2019fpa,TheLIGOScientific:2016src}. 

One century of experimental triumphs did not deter theoretical work on observationally viable extensions of GR for mainly two sets of reasons~\cite{Berti:2015itd}.
The first is observational: some of the most outstanding open questions in physics might be explained by modifying the gravitational sector. 
For example, one could introduce an additional scalar field to the gravitational action~\cite{Charmousis:2011bf,Charmousis:2011ea} or allow the graviton to be massive~\cite{deRham:2010tw,DAmico:2011eto,deRham:2010kj} to explain the late-time acceleration of the Universe~\cite{Riess:1998cb,Perlmutter:1998np} without invoking the cosmological constant or dark energy. 
The second set of reasons is theoretical: string theory and other ultraviolet completions of the Standard Model usually add higher-order curvature corrections to the Einstein-Hilbert action, implying deviations from GR at high energies and large curvatures~\cite{Polchinski:1998rq,Polchinski:1998rr,Fujii:2003pa}.
Therefore it is important to systematically test the assumptions underlying GR, which are often summarized in terms of Lovelock's theorem~\cite{Berti:2015itd,Alexander:2017jmt}. More specifically, GR assumes that
the gravitational interaction is mediated by the metric tensor alone;
the metric tensor is massless;
spacetime is four-dimensional;
the theory of gravity is position-invariant and Lorentz-invariant; and
the gravitational action is parity-invariant.
There is no \emph{a priori} reason why these assumptions should be true, and therefore it is reasonable to explore alternatives to GR by systematically questioning each of them~\cite{Yunes:2013dva,Berti:2015itd}.
Our study is motivated by a combination of these two reasons: we will focus on theories that may address long-standing problems in physics, while questioning the validity of the main assumptions behind GR.

The LIGO-Virgo-KAGRA network of Earth-based detectors just completed their third observing run (O3). 
A fourth observing run (O4) is planned in 2022, and future observations will combine data from LIGO Hanford~\cite{TheLIGOScientific:2014jea}, LIGO Livingston~\cite{TheLIGOScientific:2014jea}, Virgo~\cite{TheVirgo:2014hva}, KAGRA~\cite{Akutsu:2020his}, LIGO India~\cite{LIGOIndia}, and third-generation (3g) detectors such as Cosmic Explorer (CE)~\cite{ 2015PhRvD..91h2001D} and the Einstein Telescope~\cite{Punturo:2010zza}. 
The space-based observatory LISA~\cite{2017arXiv170200786A}, scheduled for launch in 2034, will extend these observations to the low-frequency window. 
As existing ground-based detectors are improved, new ones are built and space-based detectors are deployed, our ability to test GR will be greatly enhanced, but to what level? 

The main goal of this study is to combine the anticipated timeline of technological development for Earth- and space-based gravitational-wave (GW) detectors with astrophysical models of binary merger populations to determine what theories will be potentially ruled out (or validated) over the next three decades.
We estimated parameters by running $\sim 10^8$ Fisher matrix calculations using waveform models including the effects of precession~\cite{Hannam:2013oca,Khan:2015jqa,Husa:2015iqa}.
Our null hypothesis is that GR correctly describes our Universe, and that all modifications must reduce to GR in some limit for the coupling constants of the modified theory~\cite{Yunes:2013dva}.
Under this assumption, we employ the parameterized post-Einsteinian (ppE) framework~\cite{Yunes:2009ke,Cornish:2011ys,Sampson:2013lpa,Chatziioannou:2012rf} to place upper limits on the magnitudes of any modification, assuming future GW observations to be consistent with GR. 
As our GW observatories are most sensitive to changes in the GW phase, we ignore modifications to the GW amplitude, an approximation that has been shown to be very good~\cite{Tahura:2019dgr}.

\subsection*{Executive Summary}

For the reader's convenience, here we provide an executive summary of the main results of this lengthy study. 

\noindent
\textbf{(i) We use public catalogs of BBH populations observable by LISA and by different combinations of terrestrial networks over the next thirty years, and extract merger rates and detection-weighted source parameter distributions.}

While this was not the main goal of this work, we did require astrophysical population models to realistically model GW science over the next three decades.
In the pursuit of constructing forecasts of constraints on GR, we developed useful statistics concerning the distribution of intrinsic parameters for detectable merging BBHs for a variety of population models and detectors.

Useful quantities calculated here and related to BBH mergers are the expected detection rates for a large selection of population models and detector networks. These rates are listed in Table~\ref{tab:rates}, and discussed in Secs.~\ref{sec:T_pdet} and~\ref{sec:S_pdet}.
Detection rates depend not only on the population model, but also on the detector network.
For LISA, we follow the method outlined in Ref.~\cite{Gerosa:2019dbe} to compute detection rates for multiband and massive black hole (MBH) sources.

We constructed synthetic catalogs by filtering the datasets coming from the full population models based on their signal-to-noise ratio (SNR). This yields a detection-weighted distribution of source parameters (discussed in Sec.~\ref{sec:catalog_creation})
which is useful to understand detection bias and to understand the typical sources accessible by different networks over the next three decades.
In Figs.~\ref{fig:source_properties} and~\ref{fig:source_properties_SPACE} we show these distributions for a large selection of detection network/population model combinations, considering both stellar-origin black holes (SOBHs) and MBHs.

The main conclusions of this analysis are summarized in Fig.~\ref{fig:sources_per_year}, which shows the typical detection rates and SNR distributions for different source models and networks.
This plot contains key information on the relative constraining performance of different population model/detector network combinations, which will be important for the following discussion of tests of GR.

\noindent
\textbf{(ii) We find that improvements over existing GW constraints on theory-agnostic modifications to GR range from 2 to 4 orders of magnitude for ground-based observations, from 2 to 4 orders of magnitude for LISA observations of MBHs, and from 1 to 6 orders of magnitude for multiband observations,  depending on what terrestrial network upgrades will be possible, on LISA's mission lifetime, and on the astrophysical distribution of merging BBHs in the Universe.}

The main issue addressed in this work is the scientific return on investment of future detector upgrades in terms of future explorations of strong gravity theories beyond GR. What future detectors and network upgrades are most efficient at constraining beyond-GR physics? 
Our models use astrophysical populations of SOBHs and MBHs and three reasonable development scenarios for ground-based detectors (ranging from optimistic to pessimistic) to try and answer this question. We first consider generic (theory-agnostic) modifications of GR, and then focus on specific classes of theories that test key assumptions underlying Einstein's theory.

Our primary conclusions for generic modifications to GR are summarized in Fig.~\ref{fig:SMBH_time} and in Sec.~\ref{sec:general_mod}, where we show bounds on generic deviations from GR at a variety of post-Newtonian (PN) orders, separated by the class of source and marginalized over the detector configurations and population models.
A term in the GW phase that is proportional to $\left( \pi \mathcal{M} f\right)^{b/3}$, where $\mathcal{M}$ is the chirp mass of the binary and $f$ is the GW frequency, is said to be of $(b+5)/2$ PN order.
While the range in constraints between the different models and scenarios is large, we have plotted constraints from current pulsar and GW tests of GR for comparison, where available and competitive.
There are several trends present in this figure, most notably:

\begin{itemize}
\item[1)] SOBH multiband sources observed by both LISA and terrestrial networks are the most effective at setting bounds on negative PN effects, outperforming all other classes of sources by at least an order of magnitude.
This observation must be tempered, however, because no multiband sources are observed at all in some of the scenarios we have analyzed.
The detection rate of multiband sources is an open question~\cite{Gerosa:2019dbe,Moore:2019pke}. 
We hope that their importance for tests of GR, outlined here and elsewhere~\cite{Barausse:2016eii,Cutler:2019krq,Gnocchi:2019jzp,Carson:2019rda,Carson:2019kkh,Toubiana:2020vtf,Liu:2020nwz}, will stimulate further work on this class of sources.

\item[2)] The MBH mergers observed by LISA outperform SOBH sources observed only in the terrestrial band for negative PN orders in the more pessimistic ground-based detector scenarios. 
For most negative PN orders, LISA MBH observations perform at least comparably to the most optimistic terrestrial network scenario, and greatly outperform the other two terrestrial scenarios analyzed in this work.

\item[3)] Terrestrially observed SOBH sources are most effective at constraining positive PN effects, outperforming MBHs and multiband sources.
Furthermore, for positive PN effects, the difference between the different terrestrial network scenarios closes dramatically. 
The constraining power between the different terrestrial networks shrinks, spanning a range of 4 orders of magnitude at negative PN orders but showing significant overlap for positive PN orders. 
This suggests that highly sensitive detectors are less important for constraining deviations that first enter at positive PN order, as opposed to negative PN order.
\end{itemize}

In terms of what detectors would have the highest return on investment, LISA's contribution to constraints on negative PN effects is quite high. 
Multiband sources are, by far, the most effective testbeds for fundamental physics in the early inspiral of GW signals, but even in the absence of multiband sources (a realistic concern), MBH sources perform as well or better than even the most optimistic terrestrial network scenario we examined. 
The difference in terrestrial network scenarios is fairly drastic for negative PN effects, and so ground-based detector upgrades would play an important role if LISA were not available.
The strongest improvement occurs in our most optimistic scenario (including CE and ET), but there is also a clear separation between the ``pessimistic'' and ``realistic'' scenarios.

Terrestrial networks perform the best for positive PN effects, but not by orders of magnitude.
Even at positive PN orders, LISA MBH sources are still as effective as the more pessimistic terrestrial network scenarios.
Furthermore, while constraining positive PN effects, no single terrestrial network scenario drastically outperforms the others: there is a clear hierarchy between the three scenarios, but with significant overlap.

These conclusions are also summarized in Table~\ref{tab:generic_summary}, where we show a concise overview of current constraints on generic ppE parameters coming from observations of pulsars~\cite{Nair:2020ggs} and GWs~\cite{LIGOScientific:2019fpa}, and we compare them against forecasts from our simulations.
\begin{table}
\begin{tabularx}{\linewidth}{ @{\hspace{\tabcolsepcustom}}c | @{\hspace{\tabcolsepcustom}}c | @{\hspace{\tabcolsepcustom}}c | @{\hspace{\tabcolsepcustom}}c  }
\hline \hline  
\makecell{PN order \\ (ppE $b$)}& \makecell{Current \\ Constraint} & \makecell{Best (Worst) \\ Constraint} &\makecell{Best (Worst) \\ Source Class} \\ \hline \hline
-4 (-13)& $-$ & $10^{-25}$ ($10^{-14}$) & MB (T) \\ \hline
-3.5 (-12)& $-$ & $10^{-23}$ ($10^{-14}$) & MB (T) \\ \hline
-3 (-11)& $-$ & $10^{-21}$ ($10^{-12}$) & MB (T) \\ \hline
-2.5 (-10)& $-$ & $10^{-19}$ ($10^{-11}$) & MB (T) \\ \hline
-2 (-9)& $-$ & $10^{-17}$ ($10^{-10}$) & MB (T) \\ \hline
-1.5 (-8)& $-$ & $10^{-15}$ ($10^{-9}$) & MB (T) \\ \hline
-1 (-7)& $2\times10^{-11}$ & $10^{-13}$ ($10^{-11}$) & MB (MBH) \\ \hline
-0.5 (-6)& $1.4\times10^{-8}$ & $10^{-11}$ ($10^{-8}$) & MB (T) \\ \hline
0 (-5)& $1.0\times10^{-5}$ & $10^{-7}$ ($10^{-5}$) & MBH (T) \\ \hline
.5 (-4)& $4.4\times10^{-3} {}^{\ast}$ & $10^{-7}$ ($10^{-5}$) & MB (T) \\ \hline
1 (-3)& $2.5\times10^{-2} {}^{\ast}$ & $10^{-6}$ ($10^{-4}$) & MB/T (T) \\ \hline
1.5 (-2)& $0.15{}^{\ast}$ & $10^{-5}$ ($10^{-3}$) & T (MB) \\ \hline
2 (-1)& $0.041{}^{\ast}$ & $10^{-4}$ ($10^{-2}$) & T (MB) \\ \hline
\hline
\end{tabularx}
\caption{Summary of the constraints we predict on the theory-agnostic ppE modification parameter $\beta$ as a function of the PN order parameter $b$, as defined in Eqs.~(\ref{eq:ppE1}) and (\ref{eq:ppE2}) below.
	We compare these constraints against current constraints from pulsar tests~\cite{Nair:2020ggs} and GW observations from the LVC~\cite{LIGOScientific:2019fpa}, denoted by $({}^{\ast})$. 
        The LVC analysis used a slightly different formalism, so we mapped their results to the ppE framework for 4 specific sources (GW150914, GW170104, GW170608, and GW170814), we computed the standard deviation of        the Markov Chain Monte Carlo (MCMC) samples, and then combined the posteriors assuming a normal distribution to obtain a rough order-of-magnitude estimate of current ppE bounds from the LVC results.        
	The columns list, from left to right: the PN order of each particular modification, the current constraint (if one exists), the best and worst constraints from our simulations, and the class of astrophysical sources those constraints come from.
	All the constraints are $1\sigma$ bounds, and we only show worst-case constraints that still improve on existing bounds.	
	The source class acronyms are as follows: MB stands for multiband observations of SOBHs, T stands for terrestrial-only observations of SOBHs, and MBH stands for space-based detection of MBHs.  }
      \label{tab:generic_summary}
\end{table}

\noindent
\textbf{(iii) LISA and future terrestrial network constraints on theory-agnostic modifications to GR  follow trends which depend on the PN order, the underlying population of sources, and the detector network.}

Using suitable approximations, we derive analytical expressions that help to elucidate the reason for the hierarchy of constraining power observed in our simulations.
We first examine single observations, and show how different source properties influence the constraints. We then attempt to quantify the importance of stacking multiple observations to develop a cumulative constraint from an entire catalog of observations.

In Sec.~\ref{sec:ind_scaling} [Eqs.~\eqref{eq:single_source_scaling} and \eqref{eq:single_source_scaling2}] we show that, to leading order, the relative constraining power of one class of sources over another depends on the binary masses and on the initial frequency of observation, raised to a power which depends on the PN order in question.
As this power changes sign going from negative to positive PN orders, this scaling explains why multiband and MBH sources are more competitive at negative PN orders, while terrestrial networks are more effective at positive PN orders.
This trend is succinctly summarized in Fig.~\ref{fig:source_class_scaling}.

Besides single-source trends, in Sec.~\ref{sec:multiple_source_scaling} we quantify the effect of stacking observations and the benefit of large catalogs.
In Fig.~\ref{fig:Neff_pn} we show that, as the PN order of the modification goes from negative to positive, the number of single observations meaningfully contributing to the cumulative bound from a catalog rises exponentially. 
This helps to further explain the improvement of terrestrial-only catalogs over LISA catalogs for higher PN orders: the very large catalogs coming from third-generation detectors are effectively leveraged to produce much stronger bounds, but only for positive PN orders.
As shown in Fig.~\ref{fig:Neff_distribution}, this depends on the relation between the three parameters of primary concern (the SNR, the chirp mass, and the constraint), and on how their relation evolves as a function of the PN order.

These considerations help us understand the behavior observed in our simulations.
The single-source scaling implies that MBHs and multiband sources should be more efficient at negative PN orders, because of the typical masses and initial frequencies of the observations. 
At positive PN orders the balance shifts in favor of terrestrial-only catalogs, further enhanced by the fact that large catalogs bear much more weight for positive PN effects.

The considerations made above also explain the significant overlap of different terrestrial detection scenarios at positive PN orders, and their separation at negative PN orders: negative PN effects are well constrained by single, loud events (favoring the most optimistic detector scenarios), while positive PN effects benefit from large catalogs. 
As detection rates are comparable for all three terrestrial scenarios, they perform comparably for positive PN effects.

\noindent
\textbf{(iv) We quantify the expected improvement over current constraints on theory-specific coupling parameters. We derive trends for theory-specific scalings and find that some conclusions following from generic modifications must be \emph{reversed}.}

The analysis of generic deviations from GR is a good theory-agnostic diagnostic tool for estimating the efficacy of future efforts to constrain fundamental physics. This is useful to perform null tests of GR, but at the end of the day, tests of GR focused on specific contending candidates provide the most meaningful physical insights~\cite{Chua:2020oxn}.
Many of the trends observed for generic modifications remain valid when considering specific theories, but the scaling relations we observe in our simulations can change significantly for some of our target theories. 

A bird's eye summary of our conclusions can be found in Table~\ref{tab:theory_summary}. 
There we identify the current bound on theory-specific parameters, our predicted bounds after thirty years, and the class of sources which is most effective at improving the bounds.
In this table we only include constraints obtained from actual data with a robust statistical analysis, in an effort to limit our comparisons to reliable experimental limits (as opposed to forecasts, simulations, etcetera).
In-depth results by source class and trend derivations are presented in Sec.~\ref{sec:specific_theories}.
We refer the reader to that section for a detailed discussion of individual theories.
In broad terms, the process of mapping generic constraints to theory-specific parameters can impose significant modifications to the trends observed in the analysis of generic constraints.
These modifications can be significant enough to completely reverse the conclusions derived from generic deviations. This should temper any interpretation of our conclusions from general modifications.
We also remark that our analysis for specific theories is far from comprehensive: there is, in principle, a very large number of GR modifications that have different mappings to ppE parameters, and therefore different trends in connection with source distributions.

\begin{table*}
\begin{tabularx}{\linewidth}{ @{\hspace{\tabcolsepcustom}}c | @{\hspace{\tabcolsepcustom}}c | @{\hspace{\tabcolsepcustom}}c | @{\hspace{\tabcolsepcustom}}c | @{\hspace{\tabcolsepcustom}}c }
\hline \hline
Theory & Parameter & Current bound & \makecell{Most (Least) Stringent \\ Forecasted Bound} & \makecell{Most (Least) \\Constraining Class} \\ \hline \hline
Generic Dipole & $\delta \dot{E}$  & $1.1\times10^{-3}$~\cite{Yunes:2016jcc,Chamberlain:2017fjl}${}^{\ast}$ & $10^{-11} $ ($10^{-6}$)  & MB (T)\\ \hline  
Einstein-dilaton-Gauss-Bonnet & $\sqrt{\alpha_{\EdGB}}$ &  \makecell{$1$ km~\cite{Yagi:2012gp}\\ $3.4$ km~\cite{Nair:2019iur}${}^{\ast}$ }& $10^{-3}$ ($1$) km& T (MBH) \\ \hline 
Black Hole Evaporation & $\dot{M}$ & -- &  $10^{-8}$ ($10^{2}$) $M_{\odot}/$yr & MB (T)\\ \hline 
Time Varying G & $\dot{G}$ & $10^{-13}-10^{-12} \text{ yr}^{-1}$~\cite{Bambi:2005fi,Copi:2003xd,Manchester:2015mda,KONOPLIV2011401,2010A.A...522L...5H} & $10^{-9}$ ($10$) yr${}^{-1}$& MB (T) \\ \hline 
Massive Graviton & $m_g$ & \makecell{$10^{-29}\text{eV}$~\cite{Hare:1973px,Goldhaber:1974wg,Talmadge:1988qz,Desai:2017dwg}\\ $10^{-23}\text{eV}$~\cite{LIGOScientific:2019fpa,Brito:2013wya}${}^{\ast}$} & $10^{-26}$ ($10^{-24}$) eV & MBH (MB) \\ \hline 
dynamic Chern Simons & $\sqrt{\alpha_{\dCS}}$ & $5.2$ km~\cite{Silva:2020acr} & $10^{-2}$ ($10$) km & T (MB) \\ \hline 
Non-commutative Gravity & $\sqrt{\Lambda}$ & $2.1$ $l_p$~\cite{Kobakhidze:2016cqh}${}^{\ast}$& $10^{-3}$ ($10^{-1}$) $l_p$& T (MB) \\ \hline \hline
\end{tabularx}
\caption{
Summary of forecasted constraints on specific modifications of GR. 
The source class acronyms are the same as in Table~\ref{tab:generic_summary}.
A (${}^{\ast}$) symbol denotes constraints coming from previous BBH observations, as opposed to other experimental evidence.
When necessary, we have mapped all existing constraints to $1\sigma$ constraints by assuming the posterior to be normally distributed.
We only show worst-case constraints that improve on existing GW bounds.
For consistency with previous work, $\dot{M}$ is given in units of $M_{\odot}/$yr, while we use geometrical units (so that $\delta \dot{E}$ is dimensionless) for the generic dipole radiation bound.
Note that the necessary factor for transforming between the two is $c^3/G =6.41\times10^{12} M_{\odot}/{\rm yr} $.
The time derivative of the gravitational constant, $\dot{G}$, is normalized to the current value of $G$, and it does indeed have units of yr${}^{-1}$ in geometrical units (where $G=c=1$). 
}\label{tab:theory_summary}
\end{table*}

Our conclusions on the best return of investment from GW detector development from the generic modification analysis \emph{generally} hold also for specific theories.
EdGB gravity (Sec.~\ref{sec:res_edgb}) and massive graviton theories (Sec.~\ref{sec:res_mg}) are two notable exceptions: in these cases, the dependence of the theory-agnostic parameters on source mass, spin and distance implies that the  generic modifications predictions (at $-1$PN and 1PN orders, respectively) must be reversed.

The remainder of the paper presents the calculations summarized above in much more detail. The plan of the paper is as follows.
In Sec.~\ref{sec:detector_networks} we give details on the detector networks implemented in this work. This section includes information about the proposed timelines of detector development, as well as the specific sensitivity curves we have implemented at each stage.
In Sec.~\ref{sec:detector_prob} we discuss the statistics with which this network is used to filter astrophysical populations, including the calculation of detection probabilities for both terrestrial and space-based detectors.
In Sec.~\ref{sec:methodology} we describe the population models, then discuss the calculation of detection rates and the creation of our synthetic catalog.
In Sec.~\ref{sec:stat}  we outline the statistics of parameter estimation procedures and waveform models, including a brief overview of Fisher analysis and the modified-GR waveforms implemented in this study.
In Sec.~\ref{sec:results} we present the results of our numerical investigation, as well as an analytical analysis to break down certain trends that have appeared in our findings.
Finally, in Sec.~\ref{sec:conclusions} we discuss limitations of this study and directions for future work.
To improve readability, some technicalities about Bayesian inference and Fisher matrix calculations, the mapping of the ppE formalism to specific theories and our waveform models are relegated to Appendices~\ref{app:Fisher}, \ref{sec:theories} and~\ref{sec:imr_vs_ins}, respectively.
Throughout this paper we will use geometrical units ($G=c=1$) and we assume a flat Universe with the cosmological parameters inferred by the Planck Collaboration~\cite{Ade:2015xua}.

\begin{table*}[t]
\begin{tabular}[b]{>{\centering}p{0.15\textwidth}>{\centering} p{0.15\textwidth}>{\centering} p{0.35\textwidth} p{0.1\textwidth}}
\hline \hline
{Year} & {Detectors} & Noise curves& {Moniker(s)}\\ \hline \hline
\multirow{4}{*}{2022-2023~\cite{Aasi:2013wya}} & \multirow{1}{*}{LIGO Hanford} & \multirow{1}{*}{Advanced LIGO design~\cite{ligo_SN_forecast}} & \multirow{4}{*}{HLVKO4} \\ 
	& \multirow{1}{*}{LIGO Livingston} & \multirow{1}{*}{Advanced LIGO design} &  \\ 
	& \multirow{1}{*}{Virgo} & \multirow{1}{*}{Advanced Virgo+ phase 1~\cite{ligo_SN_forecast}} &  \\ 
	& \multirow{1}{*}{KAGRA} & \multirow{1}{*}{KAGRA 80Mpc or 128Mpc~\cite{ligo_SN_forecast}} &  \\ \hline
\multirow{5}{*}{\makecell{2025-2030~\cite{Aasi:2013wya} \\ (one year observations \\ in alternating years) }} & \multirow{1}{*}{LIGO Hanford} & \multirow{1}{*}{Advanced LIGO A+~\cite{ligo_SN_forecast}} & \multirow{5}{*}{\makecell{HLVKIO5 \\ HLVKIO6 \\ HLVKIO7} } \\ 
	& \multirow{1}{*}{LIGO Livingston} & \multirow{1}{*}{Advanced LIGO A+} &  \\ 
	& \multirow{1}{*}{Virgo} & \multirow{1}{*}{Advanced Virgo+ phase 2 high or low~\cite{ligo_SN_forecast}} &\\ 
	& \multirow{1}{*}{KAGRA} & \multirow{1}{*}{KAGRA 80Mpc or 128Mpc} &  \\ 
	& \multirow{1}{*}{LIGO India} & \multirow{1}{*}{Advanced LIGO A+} &  \\ \hline
\multirow{5}{*}{\makecell{2032-2035 \\ (one year observations \\ in alternating years)}} & \multirow{1}{*}{LIGO Hanford} & \multirow{1}{*}{Advanced LIGO Voyager~\cite{Voyager_detector}} & \multirow{5}{*}{\makecell{HLVKIO8 \\ HLVKIO9 }} \\ 
	& \multirow{1}{*}{LIGO Livingston} & \multirow{1}{*}{Advanced LIGO Voyager} &  \\ 
	& \multirow{1}{*}{Virgo} & \multirow{1}{*}{Advanced Virgo+ phase 2 high or low} &  \\ 
	& \multirow{1}{*}{KAGRA} & \multirow{1}{*}{KAGRA 80Mpc or 128Mpc} &  \\ 
	& \multirow{1}{*}{LIGO India} & \multirow{1}{*}{Advanced LIGO Voyager} &  \\ \hline \hline
\multicolumn{4}{c}{ Scenario 1} \\ \hline\hline
\multirow{4}{*}{2035-2039~\cite{Baker:2019nia,Reitze:2019dyk}} & \multirow{1}{*}{Cosmic Explorer} & \multirow{1}{*}{CE phase 1~\cite{CE_psd}} & \multirow{4}{*}{CEKL} \\ 
	& \multirow{1}{*}{Einstein Telescope} & \multirow{1}{*}{ET-D~\cite{Hild:2010id}} &  \\ 
	& \multirow{1}{*}{KAGRA} & \multirow{1}{*}{KAGRA 128Mpc} &  \\ 
	& \multirow{1}{*}{LISA} & \multirow{1}{*}{LISA~\cite{Cornish:2018dyw,Tanay:2019knc}} &  \\ \hline
\multirow{4}{*}{2039-2045~\cite{Baker:2019nia,Reitze:2019dyk}} & \multirow{1}{*}{Cosmic Explorer} & \multirow{1}{*}{CE phase 1} & \multirow{4}{*}{CEKLext} \\ 
	& \multirow{1}{*}{Einstein Telescope} & \multirow{1}{*}{ET-D} &  \\ 
	& \multirow{1}{*}{KAGRA} & \multirow{1}{*}{KAGRA 128Mpc} &  \\ 
	& \multirow{1}{*}{LISA} & \multirow{1}{*}{LISA} &  \\ \hline
\multirow{3}{*}{2045-2050~\cite{Baker:2019nia,Reitze:2019dyk}} & \multirow{1}{*}{Cosmic Explorer} & \multirow{1}{*}{CE phase 2~\cite{CE_psd}} & \multirow{4}{*}{CEK} \\ 
	& \multirow{1}{*}{Einstein Telescope} & \multirow{1}{*}{ET-D} &  \\ 
	& \multirow{1}{*}{KAGRA} & \multirow{1}{*}{KAGRA 128Mpc} &  \\ \hline \hline
\multicolumn{4}{c}{ Scenario 2} \\ \hline\hline
\multirow{4}{*}{2035-2039} & \multirow{1}{*}{Cosmic Explorer} & \multirow{1}{*}{CE phase 1} & \multirow{4}{*}{CVKL} \\ 
	& \multirow{1}{*}{Virgo} & \multirow{1}{*}{Advanced Virgo+ phase 2 high} &  \\ 
	& \multirow{1}{*}{KAGRA} & \multirow{1}{*}{KAGRA 128Mpc} &  \\ 
	& \multirow{1}{*}{LISA} & \multirow{1}{*}{LISA} &  \\ \hline
\multirow{4}{*}{2039-2045} & \multirow{1}{*}{Cosmic Explorer} & \multirow{1}{*}{CE phase 1} & \multirow{4}{*}{CVKLext} \\ 
	& \multirow{1}{*}{Virgo} & \multirow{1}{*}{Advanced Virgo+ phase 2 high} &  \\ 
	& \multirow{1}{*}{KAGRA} & \multirow{1}{*}{KAGRA 128Mpc} &  \\ 
	& \multirow{1}{*}{LISA} & \multirow{1}{*}{LISA} &  \\ \hline
\multirow{3}{*}{2045-2050} & \multirow{1}{*}{Cosmic Explorer} & \multirow{1}{*}{CE phase 2} & \multirow{4}{*}{CVK} \\ 
	& \multirow{1}{*}{Virgo} & \multirow{1}{*}{Advanced Virgo+ phase 2 high} &  \\ 
	& \multirow{1}{*}{KAGRA} & \multirow{1}{*}{KAGRA 128Mpc} &  \\ \hline \hline
\multicolumn{4}{c}{ Scenario 3} \\ \hline\hline
\multirow{6}{*}{2035-2039} & \multirow{1}{*}{LIGO Hanford} & \multirow{1}{*}{Advanced LIGO Voyager} & \multirow{6}{*}{\makecell{HLVKIL }} \\ 
	& \multirow{1}{*}{LIGO Livingston} & \multirow{1}{*}{Advanced LIGO Voyager} &  \\ 
	& \multirow{1}{*}{Virgo} & \multirow{1}{*}{Advanced Virgo+ phase 2 high or low} &  \\ 
	& \multirow{1}{*}{KAGRA} & \multirow{1}{*}{KAGRA 80Mpc or 128Mpc} &  \\ 
	& \multirow{1}{*}{LIGO India} & \multirow{1}{*}{Advanced LIGO Voyager} &  \\ 
	& \multirow{1}{*}{LISA} & \multirow{1}{*}{LISA} &  \\ \hline
\multirow{6}{*}{2039-2045} & \multirow{1}{*}{LIGO Hanford} & \multirow{1}{*}{Advanced LIGO Voyager} & \multirow{6}{*}{\makecell{HLVKILext }} \\ 
	& \multirow{1}{*}{LIGO Livingston} & \multirow{1}{*}{Advanced LIGO Voyager} &  \\ 
	& \multirow{1}{*}{Virgo} & \multirow{1}{*}{Advanced Virgo+ phase 2 high or low} &  \\ 
	& \multirow{1}{*}{KAGRA} & \multirow{1}{*}{KAGRA 80Mpc or 128Mpc} &  \\ 
	& \multirow{1}{*}{LIGO India} & \multirow{1}{*}{Advanced LIGO Voyager} &  \\ 
	& \multirow{1}{*}{LISA} & \multirow{1}{*}{LISA} &  \\ \hline
\multirow{5}{*}{2045-2050} & \multirow{1}{*}{LIGO Hanford} & \multirow{1}{*}{Advanced LIGO Voyager} & \multirow{5}{*}{\makecell{HLVKI+ }} \\ 
	& \multirow{1}{*}{LIGO Livingston} & \multirow{1}{*}{Advanced LIGO Voyager} &  \\ 
	& \multirow{1}{*}{Virgo} & \multirow{1}{*}{Advanced Virgo+ phase 2 high or low} &  \\ 
	& \multirow{1}{*}{KAGRA} & \multirow{1}{*}{KAGRA 80Mpc or 128Mpc} &  \\ 
	& \multirow{1}{*}{LIGO India} & \multirow{1}{*}{Advanced LIGO Voyager} &  \\ \hline \hline
\end{tabular}
\caption{
The above timeline tabulates the exact terrestrial detector evolution utilized by this study. 
There is a single timeline of detectors until 2035, when we model three separate scenarios that could play out in the next three decades: Scenario 1, 2, and 3. 
A graphical representation is shown in Fig.~\ref{fig:timeline}.
The various sensitivity curves in column 3 are shown in Fig.~\ref{fig:SN}.
    }\label{tab:timeline}
\end{table*}

\begin{figure*}[htb]
\centering
\includegraphics[clip=true, width=\textwidth]{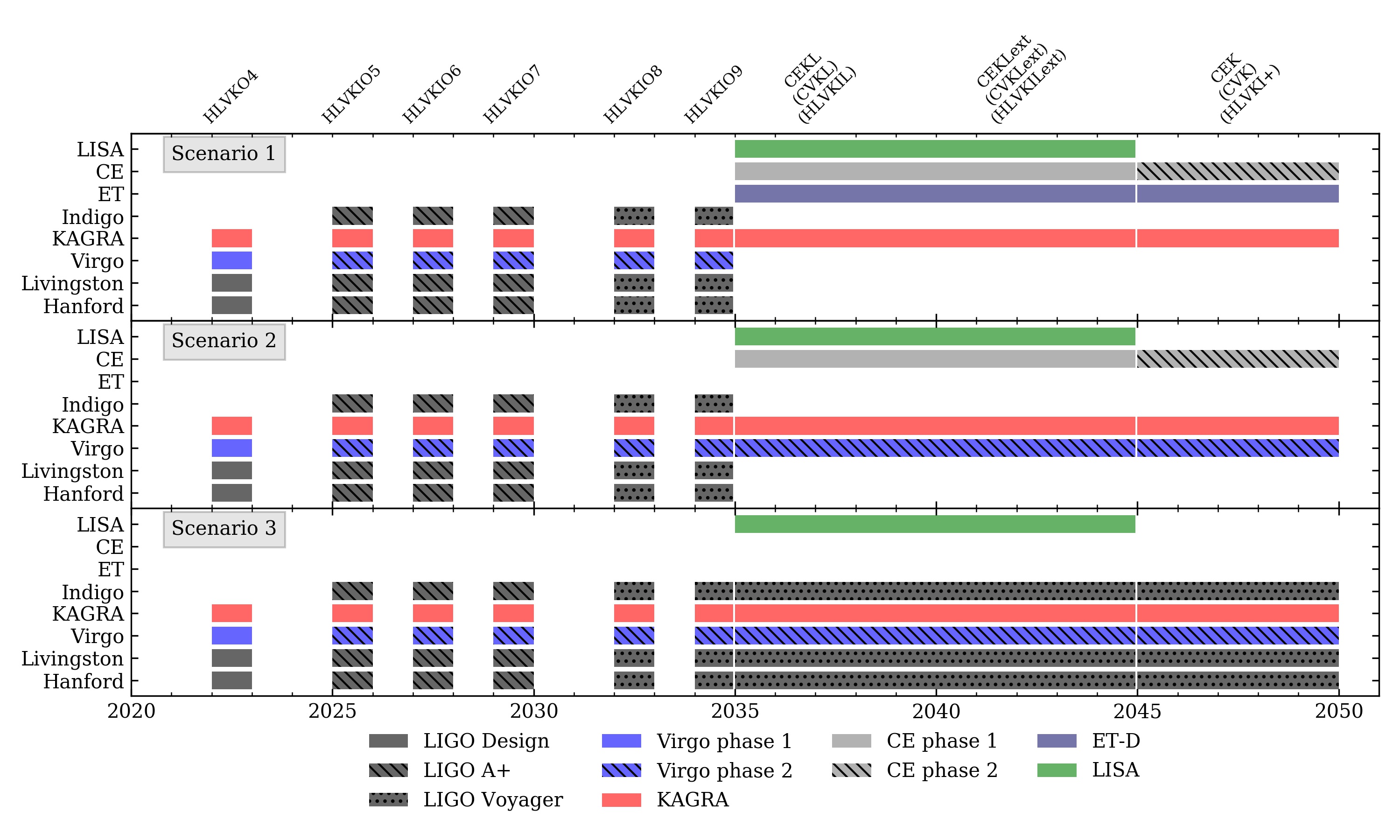}
\caption{Graphical representation of Table~\ref{tab:timeline}.
The shaded regions in the figure represent periods of active observation, and the colors/hatching corresponds to the noise curve being implemented, as shown in Fig.~\ref{fig:SN}.
}\label{fig:timeline}
\end{figure*}

\begin{figure*}[t]
\includegraphics[width=0.9\textwidth]{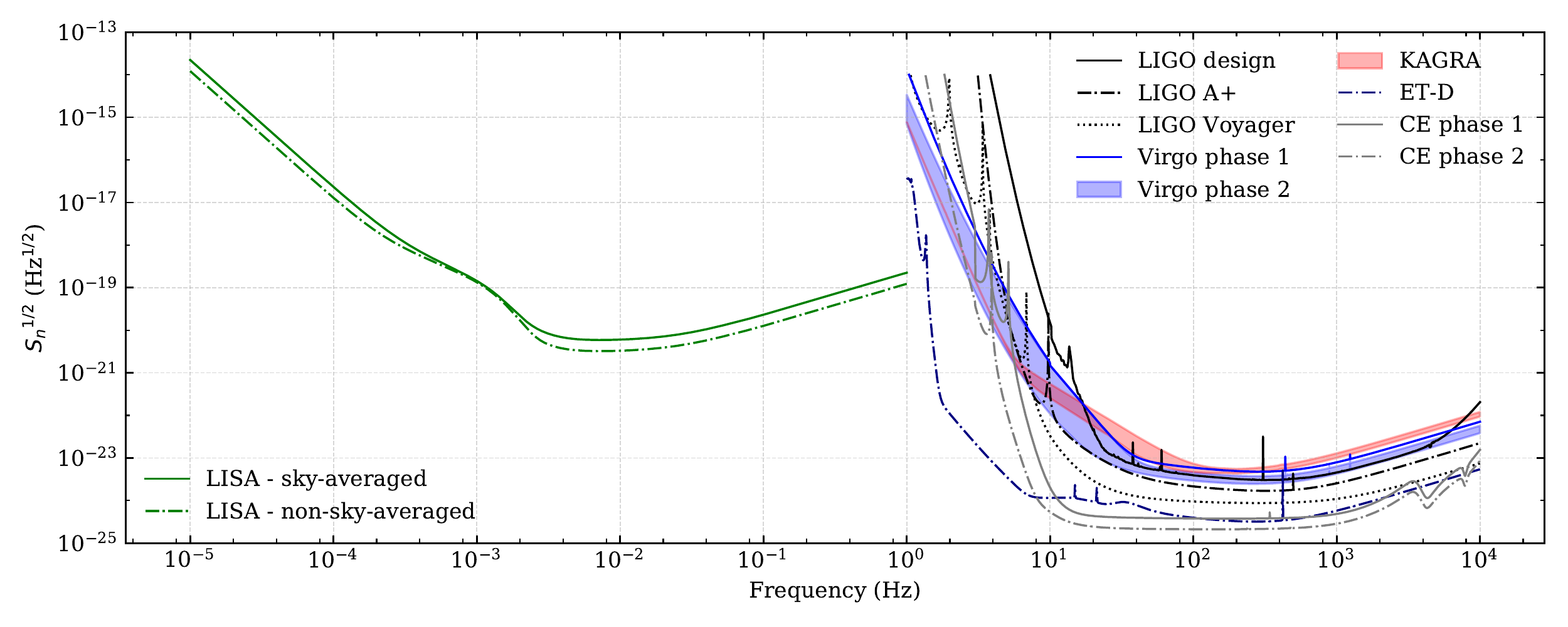}
\caption{Noise curves for the various detector configurations studied in this work. The shaded bands observed for the Virgo+ phase 2 and KAGRA sensitivities reflect uncertainties in estimates of their anticipated power spectral densities.}\label{fig:SN}
\end{figure*}

\section{Detector Networks}\label{sec:detector_networks}

The construction and enhancement of GW detectors across the world and in space is expected to proceed steadily over the next thirty years. Tests of GR using GW observations are fundamentally tied to this global timeline of detector development, so it  is important to have a realistic range of models for detector networks that spans the inevitable uncertainties intrinsic in planning experiments over such a long time.
In this section we describe potential timelines for upgrades and deployment of new detectors, our assumptions on the location of the detectors, and their expected sensitivities.

\subsection{Estimated Timeline}
\label{subsec:timeline}

Three plausible scenarios for the GW detector roadmap as of the writing of this paper are schematically presented in Fig.~\ref{fig:timeline}, with more details in Table~\ref{tab:timeline}. 
The timeline starts with the fourth observing run (O4) of the LIGO-Virgo-KAGRA detectors, which are scheduled to take data at their design sensitivities for one year starting in 2022. 
After this run, the instruments would be taken offline to be upgraded to higher sensitivity, with the next set of one-year-long observing runs starting in 2025. 
At this point, the network would also be joined by LIGO-India. 
Subsequent upgrades for the LIGO detectors to LIGO Voyager are planned for the early 2030's. 
The plans for 3g detectors are understandably more uncertain, with CE and ET potentially joining the network in 2035. 
After a 5--10 year observing run, CE is expected to be taken offline for upgrades, with a second set of runs expected in 2045. 
Meanwhile, LISA is scheduled to fly in 2034, with a minimum mission lifetime of 4 years and a possible extension by 6 additional years, for a total of 10 years of observation~\cite{Baker:2019nia}.

Given the timeline described above, one can identify several distinct periods of observations in which a different combination of detectors would be simultaneously online. 
During the O4 run, LIGO Hanford (H), LIGO Livingston (L), Virgo (V) and KAGRA (K) are expected to collect data simultaneously, creating the HLVKO4 network. 
LIGO India is expected to join the data collection effort in the late 2020's for the O5, O6 and O7 observation campaigns, creating the HLVKIO5/O6/O7 networks. 
In the early 2030's, the LIGO detectors (Hanford, Livingston, and Indigo) will be upgraded to the Voyager design, reflected in the HLVKIO8/09 networks.

The timeline beyond 2035 is quite uncertain, and we cannot model every possible scenario. 
Therefore, we chose to model three different timelines:

\begin{itemize}
\item[1)] After 2035, an optimistic detector schedule would see the Virgo and LIGO detectors replaced by the Einstein Telescope (E) and CE (C) detectors, respectively. 
Furthermore, LISA (L) is targeting around 2035 as the beginning of its data collection, with a nominal 4-year mission and an additional 6-year extension. 
These assumptions correspond to the CEKL and CEKLext networks, respectively.
We follow up the multiband observation campaigns with a final terrestrial-only observation period from 2045-2050 for the CEK network.
This timeline is shown as ``Scenario 1'' in Table~\ref{tab:timeline}. 

\item[2)] A less optimistic scenario might see one terrestrial 3g detector receive full funding and come online in the 2030's. We chose to use CE as our one 3g terrestrial detector to create the CVKL, CVKLext, and CVK networks. This is ``Scenario 2'' in Table~\ref{tab:timeline}.

\item[3)] We also consider a pessimistic scenario where no terrestrial 3g detectors will be observing before the 2050's. The network will remain at its O9 sensitivity, but it will still be joined by LISA in the 2030's.  This scenario includes the HLVKIL, HLVKILext, and HLVKI+ networks, and is denoted as ``Scenario 3'' in Table~\ref{tab:timeline}.
\end{itemize}
      
Because these last three observation periods for all three scenarios are less defined and span a wide time range, we assume an 80\% duty cycle when estimating terrestrial-only detection rates, but we use the full observation period for calculating multiband rates.

\begin{table}[t]
\begin{tabular}{c  c  c }
\hline \hline
Detector & Latitude (${}^\circ$) & Longitude (${}^\circ$) \\ \hline \hline
LIGO Hanford & 46.45 & -119.407 \\ %
LIGO Livingston & 30.56 & -90.77 \\ %
Virgo & 43.63 & 10.50 \\ %
KAGRA & 36.41 & 137.31 \\ %
LIGO India &14.23 & 76.43 \\ %
Cosmic Explorer & 40.48 & -114.52 \\ %
Einstein Telescope & 43.63 & 10.50 \\ 
\hline \hline
\end{tabular}
\caption{Detector locations used in this paper.}\label{table:locations}
\end{table}

\subsection{Estimated Sensitivity}

The detector sensitivities can be characterized in terms of their power spectral density $S_{n}$, which we present in Fig.~\ref{fig:SN}. 

We assume that the LIGO detectors will start operating at design sensitivity (``LIGO design''~\cite{ligo_SN_forecast} in Fig.~\ref{fig:SN}) in O4, but will be upgraded to the A+ configuration (``LIGO A+''~\cite{ligo_SN_forecast} in Fig.~\ref{fig:SN}) in time for the O5 observing run. 
In the early 2030's, the LIGO detectors will be upgraded to the Voyager sensitivity (``LIGO Voyager''~\cite{Voyager_detector} in Fig.~\ref{fig:SN}). 
Virgo observations begin with the Advanced Virgo+ phase 1 noise curve (``Virgo phase $1$''~\cite{ligo_SN_forecast} in Fig.~\ref{fig:SN}) in O4, and they will subsequently be upgraded to Advanced Virgo+ phase 2 (``Virgo phase 2''~\cite{ligo_SN_forecast} in Fig.~\ref{fig:SN}) beginning in O5. 
To bracket uncertainties, we consider both an optimistic (``high'') configuration and a pessimistic (``low'') configuration for Virgo+~\cite{ligo_SN_forecast}.  
We model the KAGRA detector using the ``128Mpc'' and ``80Mpc'' configurations from Ref.~\cite{ligo_SN_forecast} for optimistic and pessimistic outlooks, respectively (``KAGRA'' in Fig.~\ref{fig:SN}). 
LIGO India is planned to join the network in O5 with sensitivity well approximated by the A+ noise curve, mirroring the Hanford and Livingston detectors. 
LIGO India will follow the same development path as its American counterparts, and be upgraded to Voyager sensitivity in the early 2030's. 

The US-led 3g detector, CE, may replace the LIGO detectors in 2035 at phase 1 sensitivity (``CE phase $1$'' in Fig.~\ref{fig:SN}). 
After upgrades are completed in the early 2040's, the detector may come back online with phase 2 noise sensitivity (``CE phase 2'' in Fig.~\ref{fig:SN})~\cite{Reitze:2019dyk}. 

The European-led 3g counterpart ET could replace the Virgo detector in 2035. 
ET will be modeled with the ET-D sensitivity in this study (``ET-D'' in Fig.~\ref{fig:SN}). 
In reality, ET is comprised of 3 individual detectors arranged in an equilateral triangle, and a fully consistent treatment of ET would incorporate the three detectors separately. 
However, after testing on subsets of our populations, we concluded that modeling ET as three identical copies of one of the constituent detectors minimally impacts our estimates on constraints of modified gravity, because of the small correlations between modified gravity modifications to the phase and the extrinsic parameters of the source, like sky location and orientation.  
This approximation significantly reduces the computational resources required to perform this study, so we opted to use it when constructing the Fisher matrices themselves (as discussed in Sec.~\ref{sec:stat}). 
When calculating the detection probability, however, we do account for the three detectors separately (cf. Sec.~\ref{sec:rates}). This is because the different orientations and positions of the detectors affect the rates more than they affect parameter estimation.

For networks that include a mixture of 3g and 2g detectors, we will only model the 2g detectors with the most optimistic sensitivity curve, i.e. the ``high'' configuration for Virgo and the ``128Mpc" configuration for KAGRA. The impact of the different 2g sensitivities is small when implemented alongside a 3g detector, and the shrinking of the parameter space for our models significantly reduces the computational cost of the problem.

For LISA, we model the noise curve using the approximations in Ref.~\cite{Cornish:2018dyw}. 
At different points in this work, we required both sky-averaged and non-sky-averaged response functions to various detectors. 
For LISA this can be more complicated than terrestrial interferometers, so we plot the sky-averaged noise curve directly from Ref.~\cite{Cornish:2018dyw} (``LISA -- sky-averaged'' in Fig.~\ref{fig:SN}) and the full (non-sky-averaged) sensitivity produced in Ref.~\cite{Tanay:2019knc} (``LISA -- non-sky-averaged'' in Fig.~\ref{fig:SN}). 
However, in contrast to Ref.~\cite{Tanay:2019knc}, we do include the factor of 2 to account for the second channel, mirroring the approximation we made for ET.

\subsection{Estimated Location}
\label{sec:network_response_functions}

The relative locations of the various detectors affects the global response function, and thus it impacts the analysis performed in this paper. 
For terrestrial detectors, the various geographical locations of each site are shown in Table~\ref{table:locations}. 
The sites of detectors currently built or under construction were taken from data contained in \software{LALSuite}~\cite{lalsuite}. 
Since a site has yet to be decided upon for CE, we chose a reasonable location near the Great Basin desert, in Nevada. 
For LISA, the detector's position and orientation as a function of time must be taken into account, so we use the time-dependent response function derived in Refs.~\cite{Cutler:1997ta,Berti:2004bd}. 
Unlike those papers we use the polarization angle defined by the total angular momentum $\mathbf{J}$, instead of the orbital angular momentum $\mathbf{L}$, because the latter precesses in time, while $\mathbf{J}$ remains (approximately) constant.

\section{Statistical Methods for Population Simulations}\label{sec:detector_prob}

Both terrestrial and space-borne GW detectors have nonuniform sensitivity over the sky. This effect is important when attempting to estimate the expected detection rate and the resulting population catalog. 

Terrestrial detector networks can mitigate this selection bias by incorporating more detectors into the network, which can ``fill in'' low-sensitivity regions in the sky. 
The incorporation of the most accurate combination of detectors and their locations can be important. This is why in Sec.~\ref{sec:network_response_functions} we specified the locations used in this study. 

For space-borne detectors, some signals may be detectable for much longer than the observation period, so random sky locations map to random spacetime locations, and the effect of only seeing a portion of the signal must be accounted for.

These issues with terrestrial networks and space detectors, and their associated detection probabilities, are discussed in  Secs.~\ref{sec:T_pdet} and Sec.~\ref{sec:S_pdet}, respectively. 

We wish to calculate the probability that the GWs emitted by some source will be detected by a terrestrial network of instruments, which we will refer to as the detection probability. 
We will focus primarily on two classes of sources: SOBH binaries~\cite{Gerosa:2018wbw} and MBH binaries~\cite{Klein:2015hvg}. 
We will use publicly available SOBH population synthesis models to produce synthetic catalogs which are mainly of interest for the terrestrial network, but can also be observed as ``multiband'' events by both the terrestrial network and LISA. We will also use MBH binary simulations to create synthetic catalogs for LISA (these sources are typically well outside the frequency band accessible to terrestrial networks). Intermediate-mass BH binaries could also be of interest~\cite{Datta:2020vcj}, but we do not consider them here, mainly because their astrophysical formation models and rates have large uncertainties~\cite{Gair:2010dx,Cutler:2019krq,Jani:2019ffg}.

\begin{table*}[!htb]
\centering
\begin{tabular}{ c  c  c }
\hline \hline
Detection network & Detector locations & Detector sensitivity curve \\ \hline \hline
\multirow{3}{*}{HLVKO4} & \multirow{1}{*}{ Hanford site} & \multirow{3}{*}{Ad. LIGO design~\cite{ligo_SN_forecast}} \\ 
	& \multirow{1}{*}{ Livingston site} &  \\ 
	& \multirow{1}{*}{ Virgo site} &  \\ \hline 
\multirow{4}{*}{HLVKIO5-O7} & \multirow{1}{*}{ Hanford site} & \multirow{4}{*}{Ad. LIGO A+~\cite{ligo_SN_forecast}} \\ 
	& \multirow{1}{*}{Livingston site} &  \\ 
	& \multirow{1}{*}{Virgo site} &  \\ 
	& \multirow{1}{*}{KAGRA site} &  \\ \hline 
\multirow{4}{*}{HLVKIO8-O9} & \multirow{1}{*}{ Hanford site} & \multirow{4}{*}{Ad. LIGO Voyager~\cite{Voyager_detector}} \\ 
	& \multirow{1}{*}{Livingston site} &  \\ 
	& \multirow{1}{*}{Virgo site} &  \\ 
	& \multirow{1}{*}{KAGRA site} &  \\ \hline 
\multirow{2}{*}{CEKL(ext)} & \multirow{1}{*}{ Cosmic Explorer site} & \multirow{2}{*}{CE phase 1~\cite{CE_psd}} \\ 
	& \multirow{1}{*}{All ET sites} &  \\  \hline
\multirow{1}{*}{CVKL(ext)} & \multirow{1}{*}{ Cosmic Explorer site} & \multirow{1}{*}{CE phase 1} \\ \hline
\multirow{4}{*}{HLVKIL(ext)} & \multirow{1}{*}{ Hanford site} & \multirow{4}{*}{Ad. LIGO Voyager} \\ 
	& \multirow{1}{*}{Livingston site} &  \\ 
	& \multirow{1}{*}{Virgo site} &  \\ 
	& \multirow{1}{*}{KAGRA site} &  \\ \hline 
\multirow{2}{*}{CEK} & \multirow{1}{*}{ Cosmic Explorer site} & \multirow{2}{*}{CE phase 2~\cite{CE_psd}} \\ 
	& \multirow{1}{*}{All ET sites} &  \\  \hline
\multirow{1}{*}{CVK} & \multirow{1}{*}{ Cosmic Explorer site} & \multirow{1}{*}{CE phase 2} \\ \hline
\multirow{4}{*}{HLVKI+} & \multirow{1}{*}{ Hanford site} & \multirow{4}{*}{Ad. LIGO Voyager} \\ 
	& \multirow{1}{*}{Livingston site} &  \\ 
	& \multirow{1}{*}{Virgo site} &  \\ 
	& \multirow{1}{*}{KAGRA site} &  \\ \hline 
\end{tabular}
\caption{
Configurations used at each stage of our analysis to calculate the probability of detection for a given binary for the terrestrial detector network. 
Note that networks involving multiple detectors are labelled by the network nodes and not just their number, because the relative position of the detectors impacts the calculation of the detection probability.
Our calculation depends on the assumption that all the detectors have approximately the same sensitivity curve, and so the curve used at each stage is given in the last column. 
Because of this assumption, and the extreme disparity in sensitivity between second- and third-generation detectors, we only use the CE detector to calculate rates when CE is part of the network.}\label{table:detection_prob_table}
\end{table*}

\begin{figure}
\includegraphics[width=\linewidth]{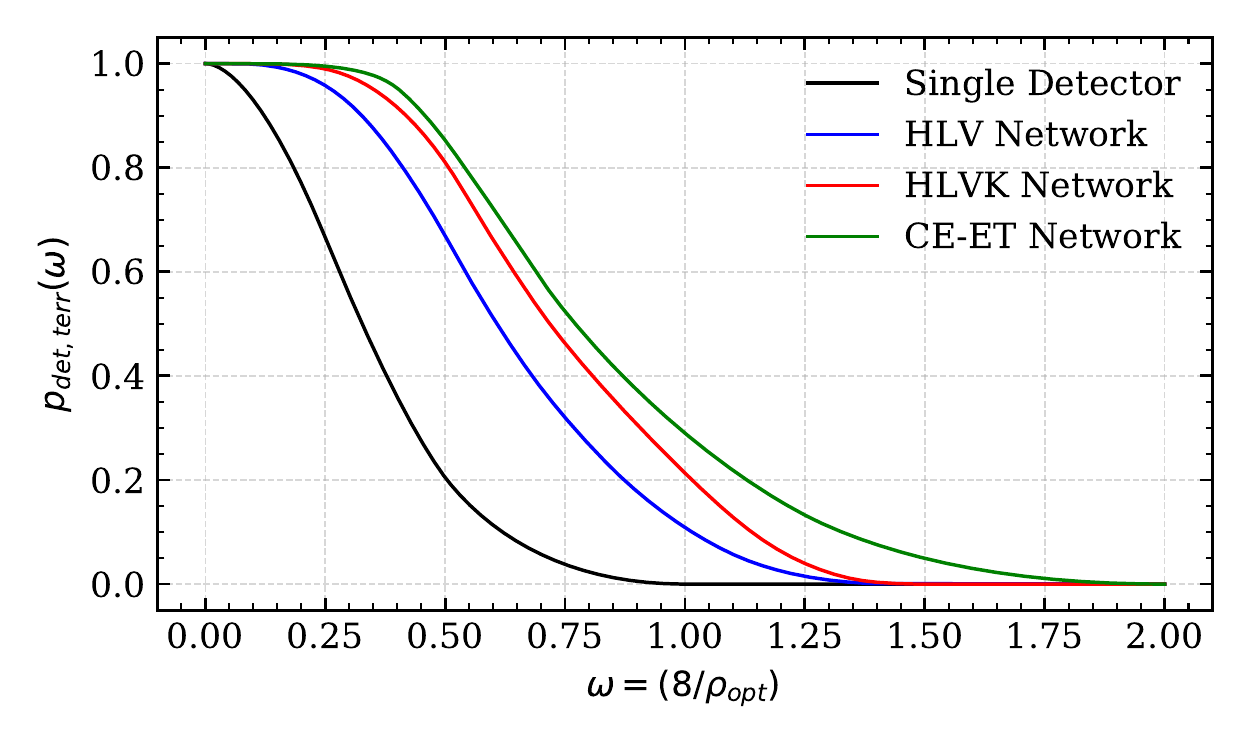}
\caption{
Detection probability $p_{\DET}$ for the four networks examined in this paper. 
The black curve is for a single detector (where global position no longer matters, so this is valid for any single right-angle Michelson interferometer). 
The blue curve is specifically for the Hanford, Livingston, and Virgo (HLV) network. 
The red curve is for the Hanford, Livingston, Virgo, and KAGRA (HLVK) network. 
Finally, the green curve represents a network comprised of CE and ET (which includes all three of the ET detectors as well as the $60^{\circ}$ angle between each set of arms).}\label{fig:pdet_curves}
\end{figure}

\subsection{Terrestrial Detection Probability}\label{sec:T_pdet}

An accurate calculation of the detection probability for each source requires injections into search pipelines. A simplifying, while still satisfactorily accurate, assumption used in most of the astrophysical literature (see e.g.~\cite{Dominik:2014yma,Finn:1992xs,Finn:1995ah}) involves computing the SNR $\rho$, defined by
\begin{equation}
  \rho^2= 4 \Re \left[ \int \frac{\tilde{h} \; \tilde{h}^{\ast}}{S_n(f)} df \right]\,,
  \label{eq:SNR}
\end{equation}
where we recall that $S_{n}(f)$ is the noise power spectral density of the detector, while $\tilde{h}=\tilde{h}(f)$ is the Fourier transform of the contraction between the GW strain and the detector response function. 
We can factor out all the detector-dependent quantities from the SNR in the form of the ``projection parameter'' $\omega$ defined as~\cite{Dominik:2014yma,Finn:1995ah}
\begin{equation}\label{eq:omega}
\omega^2 = \frac{(1 + \cos^2 \iota)^2}{4} F_{+}^2(\theta,\phi,\psi) + \cos^2{\iota} F_{\times}^2 (\theta,\phi,\psi)\,,
\end{equation}
where $\iota$ is the inclination of the binary relative to the line of sight, $\theta$ and $\phi$ are the spherical angles of the source relative to the vector perpendicular to the plane of the detector, and $\psi$ is the polarization angle. 
The single-detector antenna pattern functions $F_{+}$ and $F_{\times}$ are given by 
\begin{align}
F_{+} &= \frac{1}{2} \left( 1+\cos^2\theta \right) \cos 2\phi \cos 2\psi - \cos \theta \sin2 \phi \sin 2\psi \,, \nonumber \\
F_{\times} &= \frac{1}{2} \left( 1+\cos^2\theta \right) \cos 2\phi \sin 2\psi + \cos \theta \sin2 \phi \cos 2\psi \,.
\end{align}

With the projection-parameter approximation, we can approximate the SNR as
\begin{equation}
\rho^{2} \approx \omega^2 \rho_{\OPT}^2 \,,
\end{equation}
where $\rho_{\OPT}$ is the SNR for an optimally oriented binary with $\theta=0$, $\iota=0$, and $\psi = 0$.
This relation is approximate if the binary is precessing, so that $\iota$ is a function of time, but it is exact otherwise.

The calculation of the detection probability can then be rephrased as a search for the extrinsic source parameters that satisfy $\omega \approx \rho / \rho_{\OPT} \geq \rho_{\THRESH} / \rho_{\OPT} \equiv \omega_{\THRESH} $ for some $\rho_{\THRESH}$. The probability that $\omega$ satisfies the above criteria translates into finding the cumulative probability distribution~\cite{Dominik:2014yma}
\begin{equation}\label{eq:pdet}
p_{\DET,\TERR}(\vec{\lambda}) = \int \Theta\left(\omega'(\theta,\phi, \psi, \iota) - \omega_{\THRESH} \right)\frac{\sin \theta d\theta d\phi}{4 \pi} \frac{d \psi}{\pi} \frac{d\cos \iota}{2} \,,
\end{equation}
where $\Theta(\cdot)$ is the Heaviside function, which ultimately describes the selection effects of our terrestrial networks. This cumulative probability clearly depends on the source parameter vector $\vec{\lambda}$, inherited from $\omega_{\THRESH}=\omega_{\THRESH}(\vec{\lambda})$.

Equation~\eqref{eq:pdet} can be extended to multiple-detector networks by expanding our definition of $\omega$ to 
\begin{equation}\label{eq:omega_network}
\omega_{\text{network}}^2 = \sum_i \omega_i^2\,,
\end{equation}
where $\omega_i$ is the projection parameter for a single detector in the network, and $\omega_{\text{network}} = \rho_{\text{network-thr}} / \rho_{\OPT}$ with some threshold network SNR, $\rho_{\text{network-thr}}$, and single-detector optimal SNR, $\rho_{\OPT}$. 
In the case of a multiple-detector network, the locally defined position coordinates $\theta$ and $\phi$ are replaced with the globally defined position coordinates $\alpha$ (the right ascension angle) and $\delta$ (the declination angle). 
The polarization angle $\psi$ is changed to the globally defined polarization angle $\bar{\psi}$, which is defined with respect to an Earth-centered coordinate axis instead of the coordinate system tied to a single detector.

Evaluating Eq.~\eqref{eq:pdet} for each network, with the network projection operator defined as Eq.~\eqref{eq:omega_network}, provides a good estimation of the probability we are seeking: a weighting factor for a given binary that incorporates the sensitivity and global geometry of a given detector network, as well as the impact that the intrinsic properties of the source have on its detectability.
Importantly, the intrinsic source parameters themselves only enter into Eq.~\eqref{eq:pdet} through the calculation of $\rho_{\OPT}$ in $\omega_{\THRESH}$.
Once a threshold SNR $\rho_{\THRESH}$ is set, the detection probability function can be seen as a function of only one number $\omega_{\THRESH}$ (for a given network), through its dependence on $\rho_{\OPT}$.
As Eq.~\eqref{eq:pdet} is a four-dimensional integral and must be calculated numerically, this detail can significantly save on computational cost if we can approximate the full function $p_{\DET,\TERR}(\omega_{\THRESH})$ once for each network.
To do this, we form a grid in $\omega_{\THRESH}$ with approximately 100 grid points, and evaluate Eq.~\eqref{eq:pdet} for each grid point with $10^9$ samples uniformly distributed in $\bar{\psi}$, $\cos \iota$, $\alpha$, and $\sin \delta$.
Interpolating across the grid in $\omega_{\THRESH}$ produces an approximation for $p_{\DET,\TERR}(\omega_{\THRESH})$.
This approximation must be calculated for each specific network, as the quantity $\omega'$ in Eq.~\eqref{eq:pdet} depends on the number and relative location of the detectors, but it only needs to be evaluated once per network, rather than once per source.

The resulting probability functions for the four terrestrial networks examined in this paper are shown in Fig.~\ref{fig:pdet_curves}. 
Note that the relative location of each detector in a network impacts the form of $p_{\DET,\TERR}$, so we label the curves by the detector nodes and not just their number (i.e.~the form of $p_{\DET,\TERR}$ will be slightly different for a Hanford, Livingston, and Virgo network when compared to a Hanford, Livingson, and KAGRA network).
Furthermore, an important assumption in this calculation is that the sensitivity of each detector is identical.
This is not a good approximation when jointly considering second- and third-generation detectors, so in these cases we neglect all the 2g detectors in the network. 
The configurations used at each stage are summarized in Table~\ref{table:detection_prob_table}.

\subsection{Space Detection Probability}\label{sec:S_pdet}

For space-based detectors, which operate at much lower frequencies, the picture changes quite drastically. 
The terrestrial detection probability of Sec.~\ref{sec:T_pdet} addresses the issue of random sky location and orientation of the sources, but an important effect for detectors like LISA is the time spent in band. 
Because signals observable by LISA can be detected for much longer than the observation time $T_{\OBS}$ of the LISA mission, the time spent in the frequency range accessible to LISA will characterize the detectability of the binary.
We characterize this effect as outlined below (we refer the reader to Ref.~\cite{Gerosa:2019dbe} for a more thorough derivation and further details).

To determine the time the binary spends in the observational frequency band of LISA, we look for the roots of 
\begin{equation}\label{eq:roots_eq}
\rho(t_{\MERG}) - \rho_{\THRESH} = 0\,,
\end{equation}
where $t_{\MERG}$ is the time before merger at which the signal starts, $\rho_{\THRESH}$ is some threshold SNR, and the SNR $\rho(t_{\MERG})$ is defined as 
\begin{equation}\label{eq:rho_tmerg}
\rho(t_{\MERG})  = 4 \Re \left[
	\int^{\text{min}(f(t_{\MERG}-T_{\OBS}), 1\,\text{Hz})}_{f(t_{\MERG})} 
	\frac{\tilde{h} \; \tilde{h}^{\ast}}{S_{n}(f)} df
\right]\,.
\end{equation}
Note that, at variance with Ref.~\cite{Gerosa:2019dbe}, we use $1\,$Hz as the upper cutoff for the LISA noise curve.

Once the roots of Eq.~\eqref{eq:roots_eq} (say $T_1$ and $T_2$) have been found, we can obtain the probability of mergers for LISA via
\begin{align}\label{eq:space_merger_rate_SOBH}
p_{\DET,\SPACE}^{\SOBH} (\vec{\lambda}) &= p_{\DET,\TERR}(\vec{\lambda}) \times \text{min}\left[ \frac{T_1 - T_2}{T_{\OBS}}, \frac{T_{\text{wait}} - T_2}{T_{\OBS}}\right]
\end{align}
for SOBH binaries, and 
\begin{align}\label{eq:space_merger_rate_MBH}
p_{\DET,\SPACE}^{\MBH} (\vec{\lambda}) &= \text{min}\left[ \frac{T_1 - T_2}{T_{\OBS}}, \frac{T_{\text{wait}} - T_2}{T_{\OBS}}\right]
\end{align}
for MBH binaries. The probability $p_{\DET,\SPACE}^{\SOBH}$ is weighted by $p_{\DET,\TERR}$ because all SOBH binaries we consider for LISA are also candidate multiband events, which must be observed both by LISA and by a terrestrial network to be considered ``true'' multiband sources. In these expressions, $T_{\text{wait}}$ is some maximum waiting time for the binary to merge, which (following Ref.~\cite{Gerosa:2019dbe}) we choose to be $5 \times T_{\OBS}$ for each detector network iteration.

\subsection{Waveform Model for Population Estimates}
\label{subsec:wf-pop}

When computing the detection probability of a given source, we need a model for the Fourier transform of the time-domain response function $h = F_{+} h_{+} + F_{\times} h_{\times}$. In the terrestrial case, we implement the full precessing inspiral/merger/ringdown model \software{IMRPhenomPv2}~\cite{Hannam:2013oca,Khan:2015jqa,Husa:2015iqa} with an inclination angle of $\iota=0^\circ$ to calculate the optimal SNR, $\omega_{\OPT}$.  
For the space-based estimates in the next section, we will use the spinning (but nonprecessing) sky-averaged \software{IMRPhenomD} waveform model~\cite{Khan:2015jqa,Husa:2015iqa}, with a small modification: since we are interested in LISA rather than terrestrial, right-angle interferometers, we replace the usual factor of $2/5$ (that arises from sky-averaging) in favor of the sky-averaged LISA sensitivity curve from~\cite{Cornish:2018dyw}, which accounts for the second LISA data channel, sky-averaging, and the $60^{\circ}$ angle between the detector arms. This waveform model depends on parameters $\vec{\lambda}_{D} = [\alpha,\delta,\theta_{\rm L},\phi_{\rm L}, \phi_{\text{ref}}, t_{c,\text{ref}}, D_L, \mathcal{M}, \eta, \chi_1, \chi_2]$, where $\alpha$ is the right ascension, $\delta$ is the declination, $\theta_{\rm L}$ and $\phi_{\rm L}$ are the polar and azimuthal angles of the binary's orbital angular momentum $\mathbf{L}$ in equatorial coordinates at the reference frequency, $\phi_{\text{ref}}$ and $t_{c,\text{ref}}$ are the orbital phase and the time of coalescence at the reference frequency, $D_{L}$ is the luminosity distance, $\mathcal{M}$ and $\eta$ are the redshifted chirp mass and the symmetric mass ratio, and $\chi_i=\mathbf{\hat{L}} \cdot \mathbf{S}_i/m_i^2$ are the dimensionless spin components along $\mathbf{\hat{L}}=\mathbf{L}/|\mathbf{L}|$ with spin angular momentum $\mathbf{S}_i$.

For space-based detectors we must also choose a way to map between time and frequency. The limits of the SNR integral \eqref{eq:SNR} and the antenna patterns (which for LISA are functions of time) depend on this mapping. For multiband SOBH binaries we use the leading-order PN relation~\cite{Peters:1964zz,Berti:2004bd,Gerosa:2019dbe}
\begin{equation}\label{eq:pn_time}
f(t_{\MERG}) = \frac{5^{3/8}}{8 \pi} \left(\mathcal{M}\right)^{-5/8} t^{-3/8}_{\MERG}\,,
\end{equation}
where again $t_{\MERG}$ is the time before merger. For massive black hole (MBH) binaries, observed by LISA only through merger, this PN approximation is insufficient, so we use instead~\cite{Chamberlain:2018snj,Cornish:2020vtw}
\begin{equation}\label{eq:time_phase_deriv}
t_{\MERG}  = \frac{1}{2\pi} \frac{d \phi}{df}\,,
\end{equation}
where $\phi$ is the GW Fourier phase. When calculating detection rates, we will invert these relations numerically as needed.

\begin{figure*}[t]
\includegraphics[width=\linewidth]{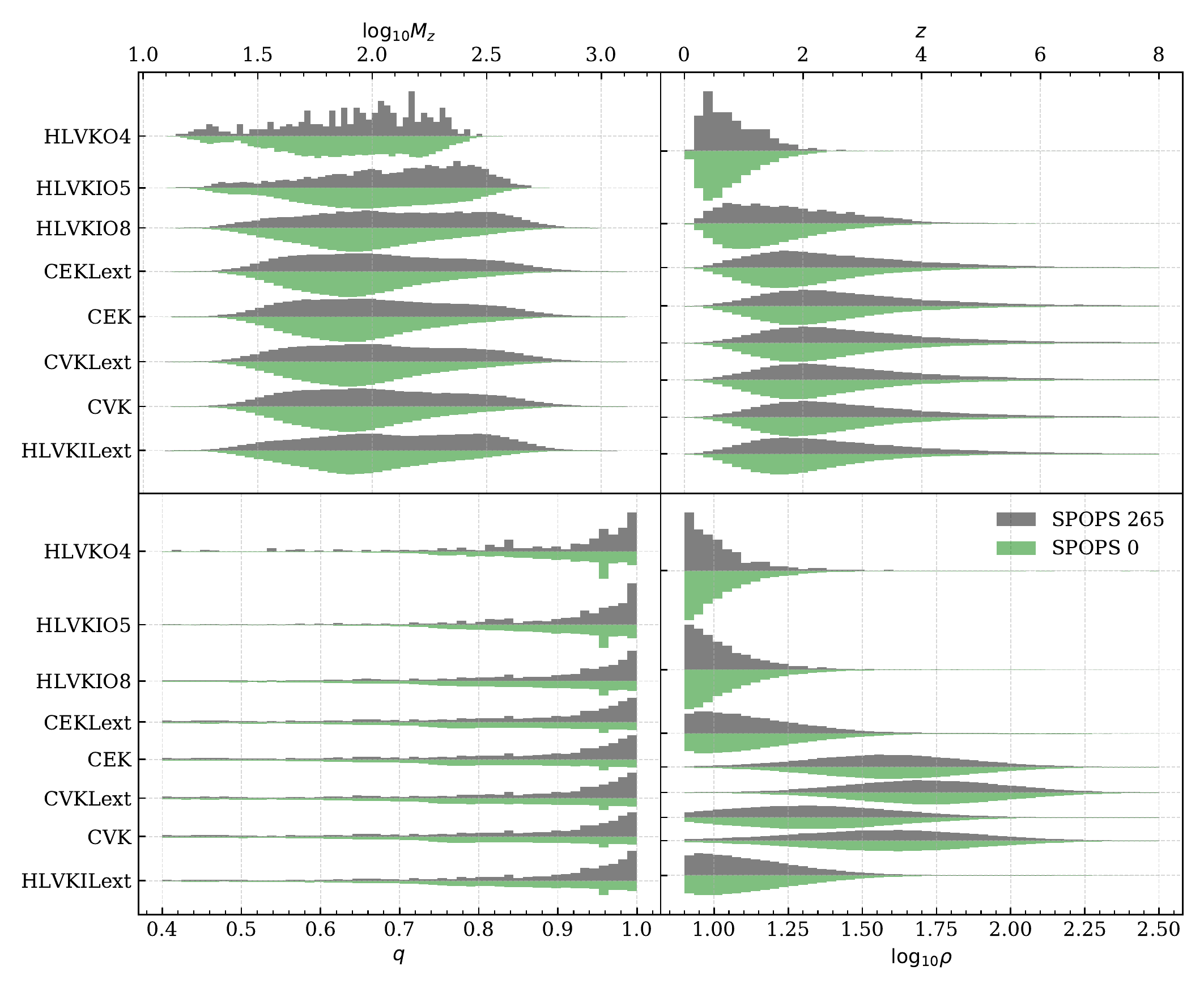}
\caption{
Distributions of the different source properties detected by each network. 
For each detector network, labeled across the y-axis, we plot the distribution of the total detector-frame mass $M_z=M(1+z)$, mass ratio $q=m_2/m_1<1$, redshift $z$, and SNR $\rho$ in log-space (base 10). 
Each plot is split, with the upper (grey) half coming from the $\sigma=265$\,km/s SPOPS simulations, and the lower (green) half coming from the $\sigma=0$\,km/s simulations.
}\label{fig:source_properties} 
\end{figure*}
\begin{figure*}
\includegraphics[width=\linewidth]{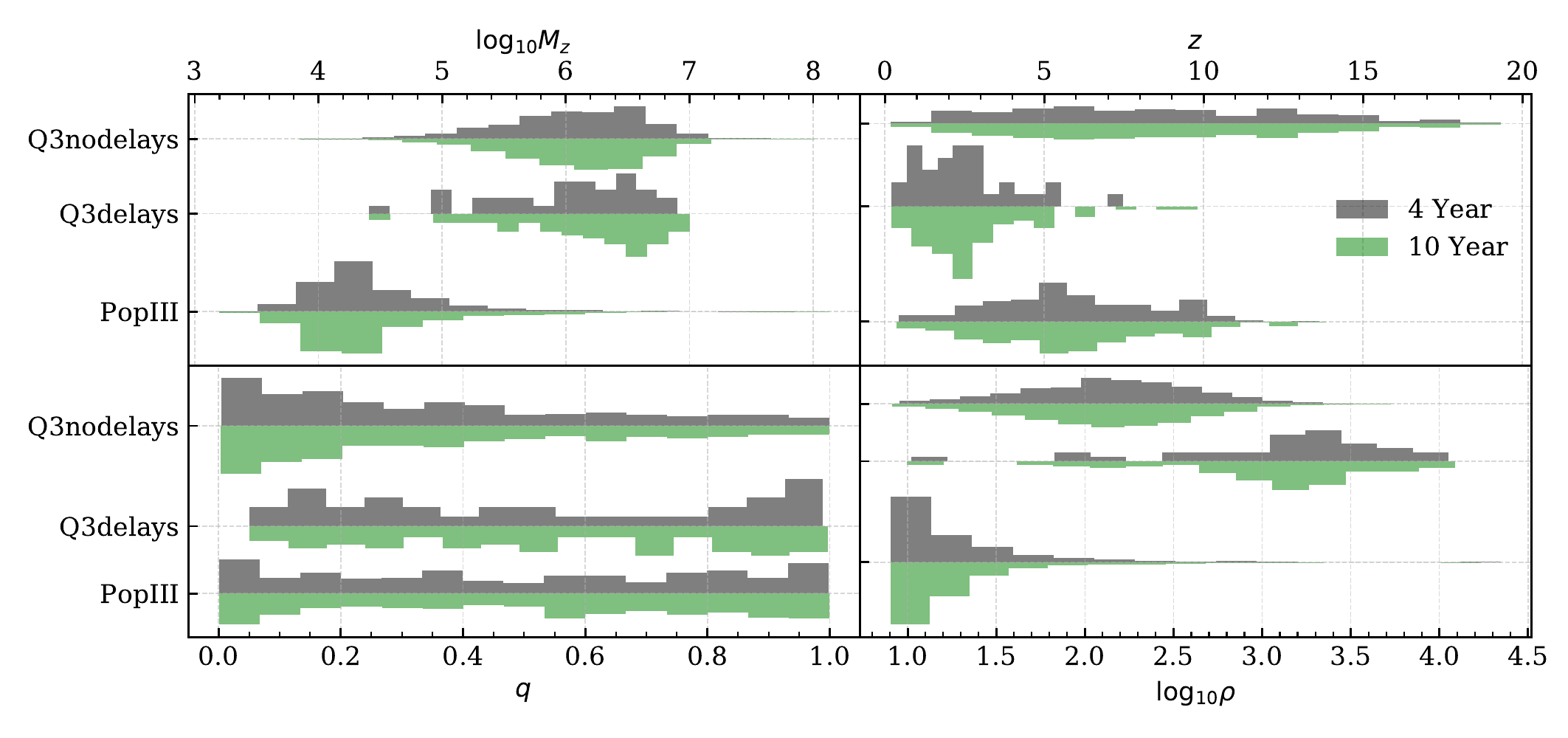}
\caption{
Distributions of the different MBH binary source properties detected by LISA. 
For each MBH binary simulations, labeled across the y-axis, we plot the distribution of the total detector-frame mass $M_z=M(1+z)$, mass ratio $q=m_2/m_1<1$, redshift $z$, and SNR $\rho$ in log-space (base 10). 
Each plot is split in two, with the upper (grey) half corresponding to a ``nominal'' four-year LISA mission, and the lower (green) half corresponding to an extended ten-year mission.
}\label{fig:source_properties_SPACE}
\end{figure*}

\section{Population Simulations}\label{sec:methodology}

A key ingredient of our work is the use of astrophysically motivated BBH population models (Sec.~\ref{sec:pop_synth}). 
Our methodology for computing detection rates and for creating synthetic catalogs from the models is explained in Sec.~\ref{sec:rates} and in Sec.~\ref{sec:catalog_creation}, respectively.

\subsection{Population Models}\label{sec:pop_synth}

For ease of comparison with previous work, we use the SPOPS catalogs~\cite{Gerosa:2018wbw} for SOBH binaries (Sec.~\ref{sec:spops}) and the MBH binary merger catalogs used in Ref.~\cite{Klein:2015hvg} (Sec.~\ref{sec:mbh}).

\subsubsection{Stellar Mass Simulations}\label{sec:spops}

We use the public SPOPS catalog of population synthesis simulations~\cite{Gerosa:2018wbw} in an effort to accurately capture the full spin orientations of the binaries at merger. 
The SPOPS catalog uses multiscale solutions of the precessional dynamics~\cite{Kesden:2014sla,Gerosa:2015tea} computed through the public code \software{PRECESSION}~\cite{Gerosa:2016sys} to quickly evolve the binary's spin orientations in time until the binary is about to merge. 

The catalog is parameterized by three different variables: the strength of the BH natal kicks, the BH spin magnitudes at formation, and the efficiency of tidal alignment~\cite{Gerosa:2018wbw}. 
In this model, natal kicks are caused by asymmetric mass ejection during core collapse, imparting a torque on one of the constituents of the binary, while the tidal alignment reflects spin-orbital angular momentum coupling through tidal interactions that can realign the spin vectors with the orbital angular momentum vector (see Ref.~\cite{Gerosa:2018wbw} for further details).

Following Ref.~\cite{Gerosa:2019dbe}, we choose to vary only one parameter of these models while keeping the others fixed. 
More specifically, we consider a uniform distribution in spin magnitude and the most realistic ( ``time'') prescription for tidal alignment of Ref.~\cite{Gerosa:2019dbe}, while varying the natal kick.
To estimate lower and upper constraints on the rates given uncertainties in our population modelling, we use the two most extreme natal kick models, corresponding to $\sigma=0$\,km/s and $\sigma=265$\,km/s, where $\sigma$ is the one-dimensional dispersion of the Maxwellian distribution the kicks are drawn from. 
The zero-kick scenario results in a lack of precessional effects and the highest detection rates for all detectors, while the $\sigma=265$\,km/s choice corresponds to a soft upper bound on the size of the kicks, which imparts the largest spin tilts and results in the lowest detection rate.
The two chosen values of $\sigma$ result in optimistic and pessimistic bounds on our projected constraints, and at the same time they provide a useful comparison between highly precessing systems and nonprecessing systems. 

\subsubsection{Massive Black Hole Simulations}\label{sec:mbh}

To model MBH binary populations, we adopt the semianalytical models of early Universe BH formation~\cite{Barausse:2012fy,Sesana:2014bea,Antonini:2015cqa} used in the LISA parameter estimation survey of Ref.~\cite{Klein:2015hvg}. 
As in that work, we focus on three populations models, characterized by different BH seeding mechanisms and different assumptions on the time delay between BH mergers and the mergers of their host galaxies. 
These population models are denoted as
\begin{enumerate}
\item PopIII -- seeds are produced from the collapse of population III stars in the early Universe (a light-seed scenario);
\item Q3delays -- seeds are produced from the collapse of a protogalactic disk (heavy-seed scenario), and there are delays between galaxy mergers and BH mergers; 
\item Q3nodelays -- seeds are produced from the collapse of a protogalactic disk (heavy-seed scenario), and there are no delays between galaxy mergers and BH mergers.
\end{enumerate}
These three models embody two seed formation mechanisms, with two models representing optimistic and pessimistic heavy-seed scenarios. The difference between PopIII simulations with and without delays is less than a factor of two, so, following Ref.~\cite{Klein:2015hvg}, we consider only the more conservative estimate, in which delays are incorporated.

\subsection{Detection Rate Calculations}\label{sec:rates}

With population synthesis simulations at our disposal, we can now estimate expected detection rates for a given detector network. 
This involves taking a model for our Universe that predicts a certain rate of merging BBHs per comoving volume, and filtering the model through the lens of a particular detector configuration and sensitivity.
The detection rate $r$ for a given network follows from the following relation~\cite{Gerosa:2019dbe,Belczynski:2015tba}:
\begin{equation}\label{eq:rate}
r = \iint dz \; d\vec{\lambda} \; \mathcal{R}(z) \; p(\vec{\lambda}) \; \frac{dV_{c}(z)}{dz} \; \frac{1}{1+z} \; p_{\DET}(\vec{\lambda} , z)\,,
\end{equation}
where $z$ is the cosmological redshift, $\mathcal{R}$ is the intrinsic merger rate (a function of the redshift), $p$ is the probability of a binary forming and merging given a set of intrinsic source parameters $\vec{\lambda}=\vec{\lambda}_{D}$ (discussed in Sec.~\ref{subsec:wf-pop}), and $dV_{c}/dz$ is a shell of comoving volume $V_{c}$ at redshift $z$.

The quantity $p_{\DET}$ is the probability of a binary being detected by a given detector network with some threshold SNR, as discussed in Sec.~\ref{sec:detector_prob}. 
The type of detector network affects the quantity $p_{\DET}$ only, while the other terms in the integral above depend only on information contained in the population simulation.
For this study, we have used a threshold SNR of 8 for terrestrial and space detections, while for multiband detections we require the terrestrial SNR and the LISA SNR to both be above $8$ independently.
Because of the intrinsic difference in the duration of signals observed by space detectors and terrestrial networks, we treat the calculation of $p_{\DET}$ slightly differently between the two cases, as discussed in Sec.~\ref{sec:T_pdet} for terrestrial detectors, and in Sec.~\ref{sec:S_pdet} for space-based detectors.

For all binaries, we evaluate the integral in Eq.~\eqref{eq:rate} through a large population of binary systems that are evolved to the point of becoming BBHs, and are weighted according to the probability that a binary of this type would actually be found in the Universe given some population model. 
This probability is comprised of factors like the star formation rate (SFR), cosmological evolution of the metallicity, the distribution of masses for these stellar populations, etc.; 
the continuous equation in Eq.~\eqref{eq:rate} then becomes a discrete sum
\begin{equation}\label{eq:rate_sum}
r = \sum_i r_i \; p_{\DET}(\vec{\lambda}_i)\,,
\end{equation}
where the index $i$ refers to samples in the simulation, $r_i$ is the intrinsic merger rate, which depends on parameters like the SFR and the mass distribution, and $p_{\DET}(\vec{\lambda}_{i})$ is the detection probability evaluated for the source parameters of the particular sample. This detection probability is $p_{\DET,\TERR}$ when considering a terrestrial network only, $p_{\DET,\SPACE}^{\SOBH}$ when considering multiband events, or $p_{\DET,\SPACE}^{\MBH}$ when considering MBH binaries detectable only by LISA. 

The intrinsic merger rate $r_i$ varies depending on the catalog used. 
For the case of the SPOPS simulations, we utilized the original \software{StarTrack} data at the foundation of each SPOPS catalog (cf. Ref.~\cite{Belczynski:2015tba} for details) to construct the intrinsic merger rate in Eq.~\eqref{eq:rate}.
For MBH catalogs, the intrinsic merger rate $r_{i}$
becomes~\cite{Klein:2015hvg}
\begin{equation}\label{eq:MBHB_rate}
r_i = 4 \pi W_{{\rm PS},i} \left(\frac{D_L (z_i)}{1+z_i}\right)^2 \,,
\end{equation}
as outlined in the data release~\cite{Klein:2015hvg,MBH_data_release}. 
The parameter $W_{{\rm PS},i}$ is the weight on the Press-Schechter mass function divided by the number of realizations~\cite{Barausse:2012fy}.

\subsection{Synthetic Catalog Creation}\label{sec:catalog_creation}

Calculating the BBH detection rate only gets us half-way to our end goal. 
Once we have the number of mergers we expect to detect for each network and simulated population, we still need to synthesize BBH catalogs to use for the later Fisher analysis in this paper. 

To create these synthetic catalogs, we sample directly from the population simulations, using Monte Carlo rejection sampling.
The probability of accepting a sample is based on the intrinsic merger rate $r_i$ in Eq.~\eqref{eq:rate_sum}, evaluated for a single simulation entry, which comes directly from the simulation data itself.
This gives a distribution of sources that reflects the expected BBH distributions for each evolution prescription.
With a distribution of ``intrinsic'' mergers in this realization of the Universe, we assign any remaining parameters according to reasonable distributions. 
For sky-location and orientation, this distribution is uniform in $\alpha$, $\sin\delta$, $\cos \theta_{\rm L}$, and $\phi_{\rm L}$. 

For the binary's merger time, we use a uniform distribution in GMST for the terrestrial networks, which impacts the orientation of the terrestrial network at the time of merger. 
This effect is completely degenerate with the right ascension of the binary, which is also randomly uniform in $\alpha$.
We use a similar prescription for MBH binaries, where the signal duration is typically shorter than the observation period. 
We employ a uniform distribution in time from $0$ to $T_{\OBS}$, which again translates to a uniform distribution in detector orientation (random position of LISA in its orbit).

Candidates for multiband detection are more nuanced. 
The signal is typically detectable for much longer than the observation period, and the frequency-time relation is nonlinear because of the familiar chirping behavior of GW signals.
For this class of sources, we randomly assign a signal starting time, which has a power-law relation with the starting frequency: cf. Eq.~\eqref{eq:pn_time}.
In this case, the position of the binary in time not only affects the orientation of LISA, but also the initial and final frequencies of the signal.
This assignment of time is important, as assigning a uniformly random initial frequency would create a bias towards seeing sources close to merger.

Once the full parameter vector has been specified, we proceed to calculate the SNR for the source in question. 
Sources meeting the SNR threshold requirements are retained in the final catalog. 
This process is repeated as necessary until we have a catalog of sources that matches the number of BBHs predicted by our rate calculations in Sec.~\ref{sec:rates}.

There are some drawbacks to this scheme.
If this process is repeated enough times, sources in the simulation will begin to be reused, as there are a fixed number of possible sources to draw from.
For this study, however, these effects are negligible, as the number of the sources in the simulations is larger than any single catalog we construct.
Furthermore, the effects will be further mitigated by randomly assigning the rest of the parameter vector not coming from the simulation, which will imbue at least slightly different properties to each source, even if one were reused. 

\begin{figure*}[t]
\includegraphics[width=\linewidth]{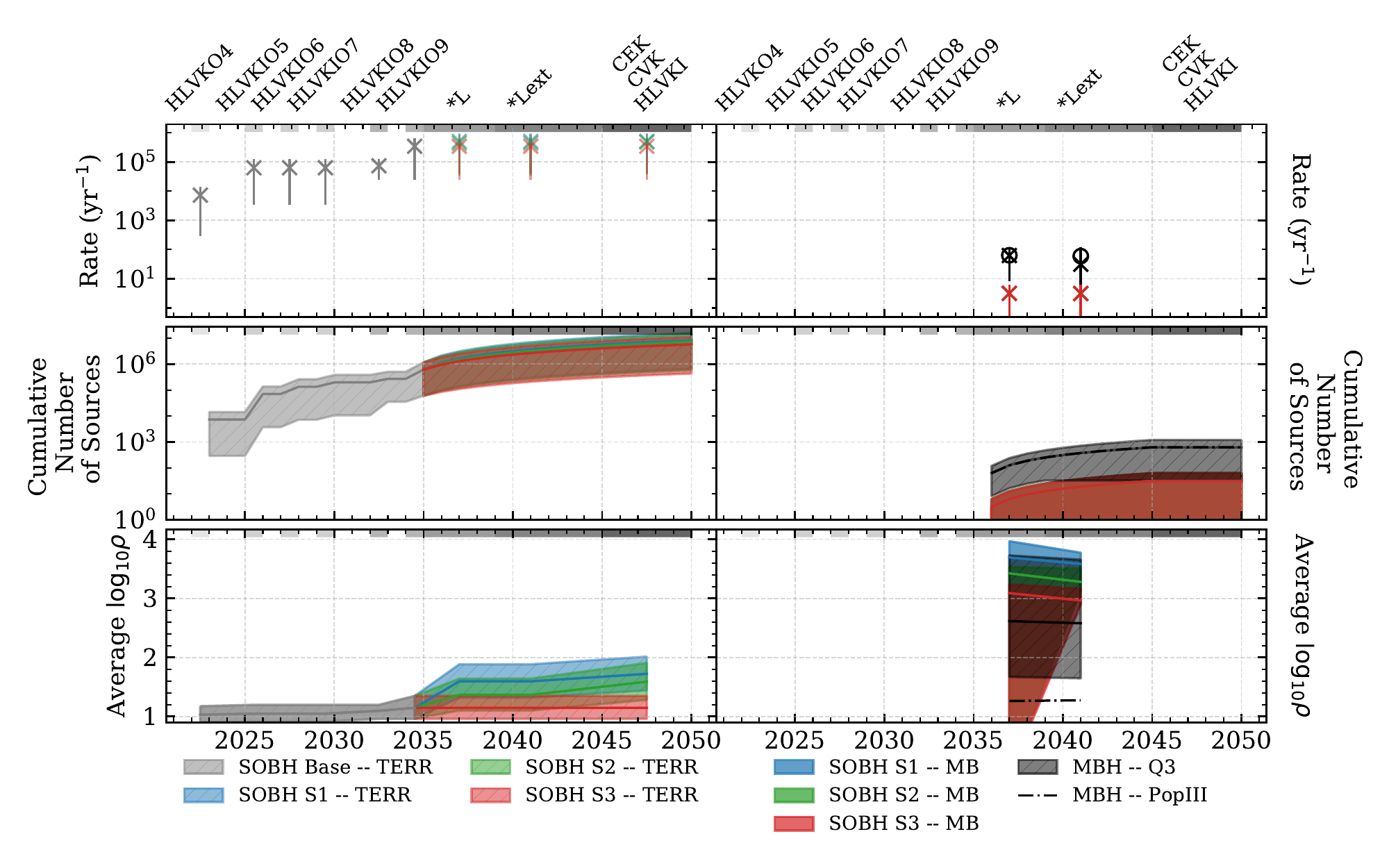}
\caption{Properties of detected merger events for various detector networks and population models. The left panels refer to terrestrial-only sources, while MBHs and multiband sources are shown on the right.
The points and thick lines show the mean values, while the shaded regions and error bars encompass the optimistic and pessimistic scenarios.
The assumed detector network is shown in the top x-axis (using the notation of Table~\ref{tab:timeline}), while the corresponding years are shown on the bottom x-axis. 
The top panels show the rates of detected mergers for each class of sources; circles refer to the PopIII MBH population.
The middle panels show the cumulative number of observed sources: here the three different multiband scenarios are identical, as the choice of terrestrial network has little impact on the number of multiband sources we can detect~\cite{Gerosa:2019dbe}.
The bottom panels show the average $\log_{10}$SNR. Here the lower (upper) bounds correspond to subtracting (adding) the standard deviation to the mean value of the most pessimistic (optimistic) scenario.}
\label{fig:sources_per_year}
\end{figure*}
\begin{table}[t]
\begin{tabularx}{1\linewidth}{ @{\hspace{\tabcolsepcustom}}Y  @{\hspace{\tabcolsepcustom}}Y  @{\hspace{\tabcolsepcustom}}Y }
\hline 
\hline 
\multicolumn{3}{c}{SOBH Rates (yr${}^{-1}$)} \\ 
\hline 
Network & \makecell{SPOPS 0 \\(T, MB)} & \makecell{SPOPS 265 \\(T, MB)}  \\ 
\hline  
\hline
HLVKO4 & ($1.43\times 10^{4}$,0) & ($2.90\times10^{2}$,0) \\ 
HLVKIO5-O7 & ($1.22\times10^{5}$,0) & ($3.43\times10^{3}$,0) \\ 
HLVKIO8-O9 & ($6.60\times10^{5}$,0) & ($2.48\times10^{4}$,0) \\ 
\hline
\multicolumn{3}{c}{Scenario 1} \\ 
\hline
CEKL & ($9.70\times10^{5}$,2.58) & ($3.96\times10^{4}$,0.0854) \\ 
CEKLext & ($9.70\times10^{5}$,6.24) & ($3.96\times10^{4}$,0.210) \\ 
CEK & ($9.72\times10^{5}$,0) & ($3.97\times10^{4}$,0) \\ 
\hline
\multicolumn{3}{c}{Scenario 2} \\ 
\hline
CVKL & ($8.36\times 10^{5}$,2.58) & ($3.36\times10^{4}$,0.0854) \\ 
CVKLext & ($8.36\times 10^{5}$,6.24) & ($3.36\times10^{4}$,0.210) \\ 
CVK & ($9.26\times 10^{5}$,0) & ($3.77\times10^{4}$,0) \\ 
\hline
\multicolumn{3}{c}{Scenario 3} \\ 
\hline
HLVKIL & ($6.60\times 10^{5}$,2.58) & ($2.48\times10^{4}$,0.0854)  \\ 
HLVKILext & ($6.60\times 10^{5}$,6.24) & ($2.48\times10^{4}$,0.210) \\ 
HLVKI+ & ($6.60\times 10^{5}$,0) & ($2.48\times10^{4}$,0) \\ 
\hline
\multicolumn{3}{c}{} \\ 
\hline
\hline
\multicolumn{3}{c}{MBH Rates (yr${}^{-1}$)} \\ 
\hline  
Network & PopIII & Q3 (delay,nodelay)  \\ 
\hline  
\hline
LISA & 62.5 & (8.11,119.1) \\ 
\hline
\hline
\end{tabularx}
\caption{Detection rates for the detector networks and population models examined in this study. For SOBH populations, the first number in the parentheses is the detection rate for the terrestrial-only network (neglecting LISA), while the second number is the detection rate for multiband events seen in both the terrestrial network and LISA. For MBH populations, we show the detection rate for LISA for the PopIII, light-seeding scenario, as well as for the Q3, heavy-seeding scenario. In the case of Q3, the first number in parentheses corresponds to delayed mergers (Q3delays) and the second number to the nondelayed version (Q3nodelays).}
\label{tab:rates}
\end{table}

To recap, our process can be broken down into the following steps:
\begin{enumerate}
\item Perform rejection sampling on the simulation entries according to the probability of merging, neglecting detector selection effects.
\item Keep the ``successful'' events, and randomly draw the rest of the requisite parameters according to their individual distributions.
\item Calculate the SNR for the given detector network. If the binary meets the threshold requirements, keep the source in the final catalog.
\end{enumerate}

The source properties of the various \emph{detected} catalogs are shown in Fig.~\ref{fig:source_properties} for the SOBH populations, and in Fig.~\ref{fig:source_properties_SPACE} for the MBH populations targeted by LISA. 
Both figures show the distributions of the redshifted total mass $M_z$, the mass ratio $q=m_2/m_1<1$, the redshift $z$, and the SNR $\rho$ of the detected populations of sources for different detector configurations and population models.
For the SOBH sources shown in Fig.~\ref{fig:source_properties}, the y-axis labels correspond to different detector combinations, while the upper (grey) and lower (green) histograms correspond to the two different kick magnitudes ($\sigma=265$\,km/s and $\sigma=0$\,km/s) chosen to bracket SOBH population models.

In the LISA SMBH case of Fig.~\ref{fig:source_properties_SPACE}, the same properties are plotted for the three populations models and for a four-year and ten-year LISA mission.
Note that the y-axis label now corresponds to different population models, and each half of the violin plot corresponds to different mission durations: the upper (grey) half corresponds to the ``nominal'' four-year LISA mission, and the lower (green) half corresponding to an extended ten-year mission.

The detection rates, cumulative detected sources, and average SNR for each class of sources are shown in Fig.~\ref{fig:sources_per_year}, where sources are broken down into 4 distinct categories: 
\begin{itemize}
\setlength\itemsep{0.1em}
\item [(i)] ``SOBH - TERR'': SOBH candidates detected only by a terrestrial network;
\item [(ii)] ``SOBH - MB'': SOBH candidates detected by both a terrestrial network and LISA (multiband);
\item [(iii)] ``MBH - PopIII'': MBH sources from the PopIII model (light seeds);
\item [(iv)] ``MBH - Q3'': MBH sources from both Q3 (heavy seeds) models, with shaded bands indicating the range of uncertainty on delays between galaxy mergers and BH mergers.
\end{itemize}
The year is shown across the bottom x-axis, while the detector network timeline is shown across the top x-axis using the acronyms defined in Table~\ref{tab:timeline}.
The solid lines and markers represent the mean values of the different quantities when considering each population model and optimistic/pessimistic detector configurations. 
The error bars and shaded regions represent the most optimistic and most pessimistic scenarios, except in the case of the SNR in the third panel, where the upper and lower bounds are the optimistic (pessimistic) average plus (minus) the standard deviation of the optimistic (pessimistic) distribution.
There is no error for the PopIII model, as we only have one iteration of this model and only one noise curve for LISA.
The detection rates for SOBHs and MBHs in the different scenarios are also listed in Table~\ref{tab:rates}.

Roughly speaking, the power of a detector network to reveal new physics comes from a combination of (i) the number of sources the network can detect, and (ii) the typical quality of each signal (as measured by the SNR). 
Figure~\ref{fig:sources_per_year} attempts to capture the zeroth-order difference between each detector configuration and population model in these two aspects. The punchline is that although LISA will be able, on average, to see events with much larger SNR, these are just a few compared to the abundant number of sources that ground-based detectors will observe (albeit at typically lower SNR). The precision of GR tests scales as $\rho^{-1}$ and it is approximately proportional to $\sqrt{N}$ for $N$ events~\cite{Berti:2011jz}, therefore it is not immediately obvious which set of observations will be best at testing GR. With our catalogs this question can be answered quantitatively. As we discuss below, ground-based and space-based detectors are complementary to each other. 

\section{Parameter Estimation}\label{sec:stat}

In this section we describe the statistical methods we will use to carry out projections on the strength of tests of GR in the future, as well as our waveform model and the numerical implementation.

\subsection{Basics of Fisher Analysis}

The backbone of this work is built on the estimation of the posterior distributions that might be inferred based on our synthetic signals. Given a loud signal with a large enough SNR, the likelihood of the data, i.e., the probability that one would see a data set $d$ given a model with parameters $\vec{\theta}$, can be expanded about the maximum likelihood (ML) parameters $\vec{\theta}_{\ML}$.  This expansion taken out to second order results in the following approximate likelihood function (where we focus on a single detector for the moment)~\cite{Poisson:1995ef,Berti:2004bd}:
\begin{equation}\label{eq:approx_likelihood}
\mathcal{L} \propto \exp \left[ -\frac{1}{2} \Gamma_{ij} \Delta \theta^i \Delta \theta^j \right]\,,
\end{equation}
where $\Delta \theta^i = \theta^i_{\ML} - \theta^i$ are deviations from the ML values, and $\Gamma_{ij}$ is the Fisher information matrix
\begin{equation}
\Gamma_{ij} = \left( \partial_{i} h | \partial_{j} h \right) |_{\ML} \,.
\end{equation}
As before, $h$ is the template response function, and the noise-weighted inner product is given by 
\begin{equation}\label{eq:inner_product}
(A|B) = 4 \Re \left[ \int \frac{ \tilde{A} \; \tilde{B}^{\ast}}{S_n (f)} df \right]\,,
\end{equation}
with $S_{n}(f)$ the noise power spectral density.  By truncating the expansion at second order, we have effectively represented our posterior probability distribution as a multidimensional Gaussian with a covariance matrix given by $\Sigma^{ij} = \left(\Gamma^{-1}\right)^{ij}$. The variances of individual parameters can then be read off to be $\sigma^i = \sqrt{\Sigma^{ii}}$, where index summation is not implied.

\begin{table*}[t]
\begin{tabular*}{15 cm}{c|c|c|c|c|c|c}
\hline \hline
\makecell{Theory or \\ physical process} & \makecell{Physical \\ modification} & G/P & \makecell{PN \\ order} & $\beta$ & \makecell{Theory \\ parameter}& $b$ \\ \hline \hline
\makecell{Generic dipole \\ radiation} & \makecell{Dipole \\ radiation} & G & -1 &\eqref{eq:dipole_beta} & $\delta\dot{E}$ &-7  \\ \hline 
\makecell{Einstein-dilaton\\Gauss-Bonnet}& \makecell{Dipole \\ radiation} & G & -1 &\eqref{eq:EdGB_beta}&$\sqrt{\alpha_{\EdGB}} $&-7 \\ \hline 
\makecell{Black Hole \\Evaporation} & \makecell{Extra \\ dimensions}& G & -4 &  \eqref{eq:BHE_beta} & $\dot{M}$ & -13 \\ \hline
\makecell{Time varying $G$} & LPI & G & -4 & \eqref{eq:Gdot_beta}&$\dot{G}$ & -13  \\ \hline
\makecell{Massive \\ Graviton} & \makecell{Nonzero \\ graviton mass}  & P & 1 &\eqref{eq:MG_beta}&$m_g$  &-3  \\ \hline
\makecell{dynamical \\Chern-Simons} & \makecell{Parity \\ violation} &G & 2 & \makecell{ \eqref{eq:dCS_beta}} &$\sqrt{\alpha_{\dCS}} $  &-1 \\ \hline
\makecell{Noncommutative\\gravity} & \makecell{Lorentz \\ violation} & G & 2 & \eqref{eq:NC_beta} &$\sqrt{\Lambda}$ &-1 \\ 
  \hline \hline
\end{tabular*}
\caption{
A summary of the theories examined in this work (adapted and updated from~\cite{Chamberlain:2017fjl,Carson:2019kkh}). 
The columns (in order) list the theory in question (unless a generic deviation is being examined), the physical interpretation of the modification, the way the modification is introduced into the waveform, the PN order at which the modification is introduced, the equation specifying the ppE-theory mapping, and the $b$ parameter in the ppE framework. 
The practical ramifications between ``generation'' vs ``propagation'' effects relates to how the modification is introduced into the waveform, as explained in Appendix~\ref{sec:imr_vs_ins}.
}
\label{tab:theory}
\end{table*}

In an attempt to capture the hard boundaries on the spin components (the dimensionless spin magnitudes $|\chi_i|$ and in-plane spin component $\chi_p$ in GR should not exceed 1), we incorporate a Gaussian prior on these two parameters with a width of $1$. We do so by adding to the Fisher matrix diagonal terms of the form~\cite{Cutler:1994ys,Poisson:1995ef,Berti:2004bd}
\begin{equation}
\Gamma_{ij} \to \Gamma_{ij} + \Gamma_{ij}^0 \,,
\end{equation} 
where $\Gamma_{ii}^0$ represents our prior distribution and is given by 
\begin{equation}
\Gamma^0_{ij} =  \delta_{\chi_1, \chi_1} + \delta_{\chi_2, \chi_2}+ \delta_{\chi_p, \chi_p}\,.
\end{equation}

In the case of multiple observations for a single source, we simply generalize the above results through sums. For example, the likelihood for a single event observed with $N$ detectors can be expanded quadratically via
\begin{equation}
\mathcal{L} \propto \exp \left[ -\frac{1}{2} \Delta \theta^i \Delta \theta^j \sum_k^N \Gamma_{ij,k} \right]\,,
\end{equation}
where the subscript $k$ labels the $k$-th detector, and we have assumed that the parameters $\vec{\theta}$ are globally defined.
This gives the final covariance matrix
\begin{equation}\label{eq:total_cov}
\Sigma^{ij} = \left( \bigl(\sum_k^N \Gamma_k + \Gamma^0\bigr)^{-1}\right)^{ij}\,.
\end{equation}
To improve readability, additional details on the calculation of the Fisher matrix are given in Appendix~\ref{app:Fisher}.

\subsection{Waveform Model for the Fisher Analysis}

For the Fisher studies carried out in this paper, we model binary merger waveforms using the phenomenological waveform model \software{IMRPhenomPv2}~\cite{Hannam:2013oca,Khan:2015jqa,Husa:2015iqa}, which allows us to capture certain spin precessional effects from inspiral until merger. The software used in this work was predominantly written from scratch, but the software library \software{LALSuite}~\cite{lalsuite} was used for comparison and to verify our implementation. For the actual parameter estimation calculation with LISA, we rescale the sensitivity curve to remove the sky-averaging numerical factor, and we account for the geometric factor of $\sqrt{3}/2$ manually in the LISA response function (``LISA -- non-sky-averaged'' in Fig.~\ref{fig:SN}), following Ref.~\cite{Tanay:2019knc}.

To fully specify the waveform produced by the \software{IMRPhenomPv2} template in GR, we need a 13-dimensional vector of parameters:
\begin{equation}
\vec{\lambda}_{\text{Pv2},\text{GR}} = \left[\alpha,\delta,\theta_{\rm L},\phi_{\rm L}, \phi_{\text{ref}}, t_{c,\text{ref}}, D_L, \mathcal{M}, \eta, \chi_1, \chi_2, \chi_\text{p}, \phi_\text{p}\right]. 
\end{equation}
The first 11 parameters are the same as those introduced for the \software{IMRPhenomD} model in Sec.~\ref{subsec:wf-pop}. The parameters $\chi_\text{p}$ and $\phi_\text{p}$ define the magnitude and direction of the in-plane component of the spin, defined as~\cite{LIGOScientific:2018mvr}
\begin{equation}\label{eq:chip}
\chi_\text{p} = \frac{1}{B_1 m_1^2} \text{max}\left( B_1 S_{1 \perp}, B_2 S_{2 \perp} \right) \,,
\end{equation}
where $B_1 = 2 + 3 q / 2$, $B_2 = 2+ 3/ ( 2q )$, $q = m_2/m_1 < 1$ is the mass ratio, and $S_{i\perp}$ is the projection of the spin of BH $i$ on the plane orthogonal to the orbital angular momentum $\mathbf{L}$.

This \software{IMRPhenomPv2} is then deformed through parameterized post-Einsteinian corrections to model generic, theory-independent modifications to GR~\cite{Yunes:2009ke,Cornish:2011ys,Sampson:2013lpa,Chatziioannou:2012rf}. We worked with deformations of two types:
\begin{align}
\tilde{h}_{\text{gen}}(\vec{\lambda}_{\text{Pv2}},\beta ) &= 
\begin{cases} 
\tilde{h}_{\text{GR}} e^{i \beta \left(\mathcal{M} \pi f \right)^{b/3}} & f<0.018m \\ 
\tilde{h}_{\text{GR}} &  0.018m<f \,,\\ 
\end{cases}
\label{eq:ppE1}
\\
\tilde{h}_{\text{prop}}(\vec{\lambda}_{\text{Pv2}},\beta ) &= 
\tilde{h}_{\text{GR}} e^{i \beta \left(\mathcal{M} \pi f \right)^{b/3}}\,,
\label{eq:ppE2}
\end{align}
where the first waveform $h_{\text{gen}}$ represents deviations from GR caused by modified generation mechanisms, and $h_{\text{prop}}$ represents deviations from GR caused by modified propagation mechanisms.
Details (including the motivation for these implementations, and the disparity of the results between the two types of deviations) are discussed in Appendix~\ref{sec:imr_vs_ins}.
As outlined there, differences are minor, and therefore from now on we will focus on the propagation mechanism, unless otherwise specified.
The parameter $\beta$ controls the magnitude of the deformation, and $b$ controls the type of deformation considered. 
The ppE version of the \software{IMRPhenomPv2} model is then controlled by the parameters
\begin{equation}
\vec{\lambda}_{\text{Pv2},\PPE} = \vec{\lambda}_{\text{Pv2},\text{GR}} \cup \{\beta\}. 
\end{equation}
Recall that, in PN language~\cite{Blanchet:2013haa}, a term in the phase that is proportional to $\left( \pi \mathcal{M} f\right)^{b/3}$ is said to be of $(b+5)/2$ PN order. The waveform model above is identical to the \software{gIMR} model coded up in LAL, and used by the LVC when performing parameterized PN tests of GR on GW data.  

The main power of the ppE approach is its ability to map the ppE deformations to known theories of gravity. Table~\ref{tab:theory} presents the mapping between $(\beta,\,b)$ and the coupling constants in various theories of gravity  (see Appendix~\ref{sec:theories} for a more detailed review of these mappings).

This table makes it clear then that ppE deformations are not false degrees of freedom, in the language of~\cite{Chua:2020oxn}. Once a constraint is placed on $\beta$, one can easily map it to a constraint on the coupling constants of a given theory through Table~\ref{tab:theory}. This reparameterization is typically computationally trivial, and therefore it saves significant resources by reusing generic results, instead of repeating the analysis for every individual theory.

\subsection{Numerical Implementation}

Common methods for calculating the requisite derivatives for the Fisher matrices typically involve either symbolic manipulation software, such as \software{Mathematica}~\cite{Mathematica}, or the use of numerical differentiation based on a finite difference scheme.
The calculation of the derivatives is always followed by some sort of numerical integration, which can be based on a fairly simple method such as Simpson's rule, or some more advanced integration algorithm that might appear prepackaged in \software{Mathematica}.

All of these methods have their respective benefits: symbolic manipulation and complex integration algorithms provide the most accuracy, while numerical differentiation and simpler integration schemes are typically much faster.
All methods also come with their respective drawbacks.
The maximally accurate method of adaptive integration and symbolic differentiation in \software{Mathematica} can be computationally taxing, while the fully numerical approach can be prone to large errors if the stepsizes are not tuned correctly, both for the differentiation with respect to the source parameters $\vec{\theta}$, as well as for the frequency spacing in the Fisher matrix integrals.
On top of these aspects, using a program like \software{Mathematica} can be cumbersome at times, as interfacing with lower-level (or even scripting) languages adds an extra layer of complexity.

A combination of the two extremes implemented in one low-level language would be ideal, and it is the route chosen for this work.
While symbolic manipulation is not available in the language that we chose (\software{C++}), we instead implemented an automatic differentiation (AD) software package natively written in \software{C}/\software{C++}: \software{ADOL-C}~\cite{10.1145/229473.229474}.
The basic premise of AD (as implemented in \software{ADOL-C}) is to use operator-overloading to perform the chain-rule directly on the program itself.
By hard-coding a select number of derivatives on basic mathematical functions and operations (such as trigonometric functions, exponentials, addition, multiplication, etc.) and tracing out all the operations performed on an input parameter as it is transformed into an output parameter, \software{ADOL-C} can stitch together the derivative of the original function.
This results in derivatives that are exact to numerical precision.
As no final, mathematical expression is output, this does not exactly constitute symbolic differentiation, but perfectly fulfills our requirements.

To complete the Fisher calculation, we take our exact derivatives (to floating-point error) and integrate them with a Gaussian quadrature scheme based on Gauss-Legendre polynomials, as in Ref.~\cite{Berti:2004bd}.
To calculate the weighting factors and the evaluation points, we have implemented a modified version of the algorithm found in Ref.~\cite{10.5555/1403886}.
While this typically incurs a high computational cost to calculate the weights and abscissas, we mitigate this fact by doing the calculation only once, and reusing the results for each Fisher matrix. 
This results in integration errors orders of magnitude lower than a typical ``Simpson's rule'' scheme, with the same computational speed per data point.

\section{Tests of General Relativity}
\label{sec:results}

In this section we summarize the main results of the analysis described above. 
We begin with the constraints on generic modifications as a function of time for each population and network (Sec.~\ref{sec:general_mod}).
Next, we translate these into constraints on specific theories (Sec.~\ref{sec:specific_theories}, and in particular Table~\ref{tab:theory}).

\subsection{Constraints on Generic Modifications}\label{sec:general_mod}

\begin{figure*}
\includegraphics[width=\textwidth]{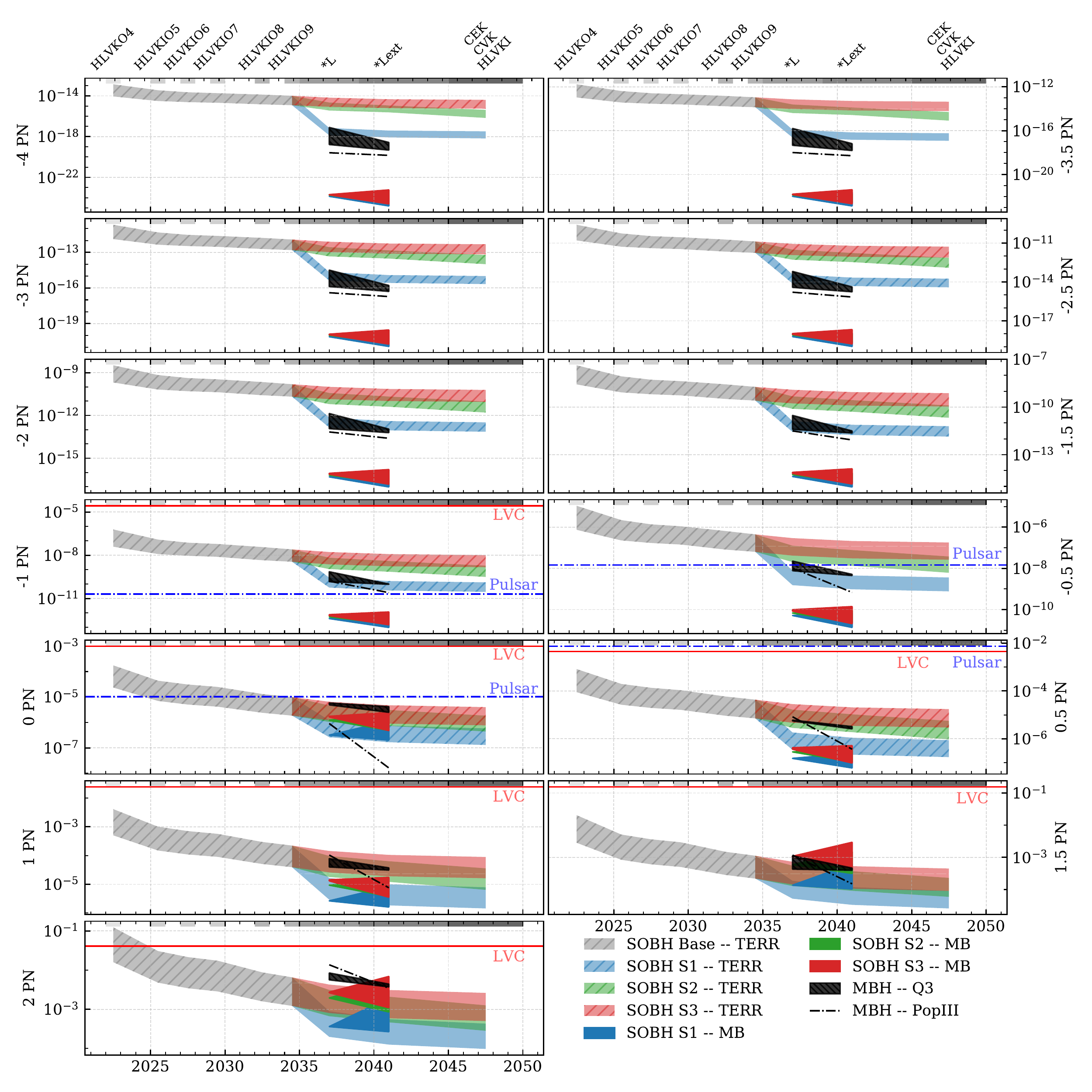}
\caption{
Constraints on modifications to GR at various PN orders as a function of time. 
The colors represent different classes of populations (including SOBH terrestrial-only sources, SOBH multiband sources, MBH sources from the Q3 heavy-seed scenario, and MBH sources from the light-seed PopIII scenario). The bands in all of these scenarios -- except for PopIII -- correspond to astrophysical uncertainties: kick velocities $\sigma=265$\,km/s and $\sigma=0$\,km/s give the upper and lower bounds for SOBHs, while the inclusion of delays affects Q3 scenarios.
Greyscale patches at the top of each panel correspond to the observation period for each network, labeled across the top. 
Multiband sources and MBHs yield strong constraints at negative PN orders. 
Terrestrial-only SOBH sources begin to contribute substantially at positive PN orders for all detector networks, with the optimistic scenario S1 yielding the best constraints.
We overlay as horizontal lines the most stringent current bounds, where available and competitive, from pulsars~\cite{Nair:2020ggs} and LVC observations of GWs~\cite{LIGOScientific:2019fpa}.
}\label{fig:SMBH_time}
\end{figure*}

Let us begin by showing in Fig.~\ref{fig:SMBH_time} the projected strength of constraints on modifications at various PN orders (shown in different panels) as a function of time.
Detector scenarios are labeled at the top, and the various astrophysical population classes are separated to facilitate visual comparisons. 
Recall from Sec.~\ref{sec:detector_networks} that we consider three detector scenarios (S1, S2, and S3) bracketing funding uncertainties in the development of the future detector network.
 The source classes include the following: 
\begin{itemize}
\setlength\itemsep{0.1em}
\item[(i)] SOBH - TERR: SOBH populations as seen by only terrestrial networks; 
\item[(ii)] SOBH - MB: SOBH events observed by both terrestrial networks and LISA; 
\item[(iii)] MBHs: heavy-seed (Q3) and light-seed (PopIII) scenarios as seen by LISA.
\end{itemize}
When relevant, the error estimates shown in the figures below come from the different versions of the population model (i.e. SPOPS 265 vs SPOPS 0 and Q3delays vs Q3nodelays), as well as marginalization over the different estimates of the noise curves (i.e. the ``high'' and ``low'' sensitivity curve for Virgo and the ``128Mpc'' and ``80Mpc'' curves for KAGRA). 
The uncertainties correspond to the minimum and maximum bounds from all the combinations we studied at that point in the timeline. 

Figure~\ref{fig:SMBH_time} is one of the main results of this paper. It allows us to draw many conclusions, itemized below for ease of reading\footnote{Throughout this analysis, the $0^{\text{th}}$ PN order in the GW phase refers to the first (often called ``Newtonian'') term in the GR series, which is proportional to $v^{-5} \propto f^{-5/3}$. Consistently, negative (positive) PN orders identify modifications entering in at lower (higher) powers of $v$, relative to this leading-order term.}:
\begin{itemize}
\item [(i)] {\bf{Multiband sources yield the best constraints at negative PN orders.}}  This is expected from prevous work~\cite{Barausse:2016eii,Chamberlain:2017fjl}: the long, early (almost monochromatic) inspiral signals coming from LISA observations stringently constrain deviations at low frequencies.
\item [(ii)] {\bf{LISA MBH observations do better than terrestrial SOBH observations at negative PN orders.}} Constraints coming from the large-SNR MBH populations outperform the terrestrial networks at negative PN order, despite the large number of expected SOBH sources in the terrestrial network.
\item [(iii)] {\bf{Terrestrial SOBH observations can do slightly better than LISA MBH observations at positive PN orders.}} Positive PN order effects can be constrained better when the merger is in band. The terrestrial networks begins to benefit from the millions of sources in the SOBH catalogs, but the extremely high-SNR sources in the MBH catalogs mean that LISA constraints are still competitive with terrestrial constraints. 
\item [(iv)] {\bf{Terrestrial network improvements make a big difference at negative PN orders.}} The different terrestrial network scenarios are widely separated for the negative PN effects, with the most optimistic S1 scenario vastly outperforming the S2 and S3 scenarios. This conclusions is robust with respect to astrophysical uncertainties in the population models.  
\item [(v)] {\bf{Network improvements are less relevant at higher PN order.}} In this case the three different scenarios overlap considerably (but the S1 scenario maintains a clear edge over the other two).
\end{itemize}

To understand some of these features, it can be illuminating to model the scaling behavior of bounds at different PN orders with respect to various source parameters. 
Below we consider an analytical approximation that can reproduce most of the observed features. We first model constraints on individual sources, and then fold in the enhancement achieved by stacking multiple events. 

\subsubsection{Analytical scaling: individual sources}\label{sec:ind_scaling}

A good first approximation is to ignore any covariances between parameters by treating the Fisher matrix as approximately diagonal, so that the bounds on the generic ppE parameter $\beta$ is roughly
\begin{equation}
\sigma_{\beta\beta} \approx \left(\frac{1}{\Gamma_{\beta \beta}}\right)^{1/2} = \left[  4 \Re \int_{f_{\text{low}}}^{f_{\text{high}}} \frac{(\pi \mathcal{M} f)^{2b/3} \left|\tilde{h}\right|^{2}}{S_n(f)} df\right]^{-1/2}\,,
\end{equation}
where $f_{\text{low}}$ and $f_{\text{high}}$ are the lower and upper bounds of integration. This expression can be simplified further by assuming white noise, so that $S_n(f)=S_0$ is constant, and by ignoring PN corrections to the amplitude, i.e. $|\tilde h|=A f^{-7/6}$, where $A\propto \mathcal{M}^{5/6}/D_{L}$ is an overall amplitude (see e.g.~\cite{Maggiore:1900zz}). This leads to
\begin{equation}
\sigma_{\beta\beta} \approx \left[  \frac{6 A^2}{S_0} \frac{\left( f_{\text{low}}^{2(b-2)/3}- f_{\text{high}}^{2(b-2)/3}\right) \left(\pi \mathcal{M}  \right)^{2 b /3}}{2-b} \right]^{-1/2}\,,
\end{equation}
as long as $b \neq 2$.
We can further simplify the expression for $\sigma_{\beta \beta}$ by using the fact that, within the same approximations, the SNR scales like
\begin{align}
\rho^2  	= 4 \Re\left[ \int_{f_{\text{low}}}^{f_{\text{high}}} \frac{h h^{\ast}}{S_n(f)} df \right] \approx \frac{3 A^2}{S_0} \left(f_{\text{low}}^{-4/3} - f_{\text{high}}^{-4/3}\right)\,,
\end{align}
which then leads to 
\begin{equation}
\sigma_{\beta\beta} \approx  \frac{(\pi \mathcal{M})^{-b/3}}{\rho}  \left[ \left(1-\frac{b}{2}\right) \frac{f_{\text{low}}^{-4/3} - f_{\text{high}}^{-4/3}  }{f_{\text{low}}^{2(b-2)/3} - f_{\text{high}}^{2(b-2)/3} }\right]^{1/2}
\end{equation}
Assuming the higher frequency cutoff to be at the Schwarzschild ISCO, so that $f_{\text{high}} = f_{\rm ISCO} = 6^{-3/2} \eta^{3/5}/(\pi {\cal{M}})$, and expanding to leading order in the small quantity $\pi {\cal{M}} f_{\text{low}} \ll 1$, we finally obtain the approximate scaling
\begin{align}
\label{eq:single_source_scaling}
\sigma_{\beta\beta} &\approx \left[ 6^{b-2} \left(\frac{b}{2}-1\right)\right]^{1/2} \frac{\left(\pi {\cal{M}} f_{\text{low}}\right)^{-2/3}}{\eta^{(b-2)/5}\rho}  \,,  \quad b > 2\,,
  \\
\sigma_{\beta\beta} &\approx  \left(1-\frac{b}{2}\right)^{1/2} \frac{\left(\pi {\cal{M}} f_{\text{low}}\right)^{-b/3} }{\rho} \,,  \quad b < 2\,.
\label{eq:single_source_scaling2}
\end{align}
The expressions above do not apply to the case $b=2$, as the integration would lead to a logarithmic scaling. Recall that $b>2$ corresponds to PN orders higher than $3.5$.

As expected, all bounds on generic ppE parameters approximately scale as the inverse of the SNR, regardless of the PN order at which they enter. What is more interesting is that they also scale with the chirp mass as $\mathcal{M}^{-b/3}$ when $b < 2$, or as ${\cal{M}}^{-2/3}$ when $b>2$. 
For a single event, we then have the ratio
\begin{align}\label{eq:ratio_scaling}
\frac{\sigma_{\beta\beta}^{\text{TERR}}}{\sigma_{\beta\beta}^{\text{MBH}}} &\approx   
\frac{\rho^{\text{MBH}}}{\rho^{\text{TERR}}} 
\left(\frac{{\cal{M}}^{\text{TERR}}}{{\cal{M}}^{\text{MBH}}}\right)^{-b/3} 
\left(\frac{f_{\text{low}}^{\text{TERR}}}{f_{\text{low}}^{\text{MBH}}}\right)^{-b/3}\,,
\end{align}
for $b < 2$. Since ${\rho^{\text{MBH}}}/{\rho^{\text{TERR}}} \sim 10^{2}$, ${{\cal{M}}^{\text{TERR}}}/{{\cal{M}}^{\text{MBH}}} \sim 10^{-4}$
and ${f_{\text{low}}^{\text{TERR}}}/{f_{\text{low}}^{\text{MBH}}} \sim 10^{5}$,
we conclude that the ratio
${\sigma_{\beta\beta}^{\text{TERR}}}/{\sigma_{\beta\beta}^{\text{MBH}}} \approx 10^{3 - b/3}$.
This ratio is large (favoring MBH sources) when $b$ is negative and large, i.e. at highly negative PN orders, and slowly transitions to favor terrestrial, SOBH sources at positive PN orders, explaining the observations in items (ii) and (iii) above. The ratio degrades by approximately four orders of magnitude between -4 PN and 2 PN, in favor of the terrestrial network, and in agreement with Fig.~\ref{fig:SMBH_time}. This scaling with $b$ holds true regardless of the typical SNRs of the sources, as the ratio of SNRs depends on the ratio of the chirp masses of the sources, but not on the PN order. 

\begin{figure}
\includegraphics[width=\linewidth]{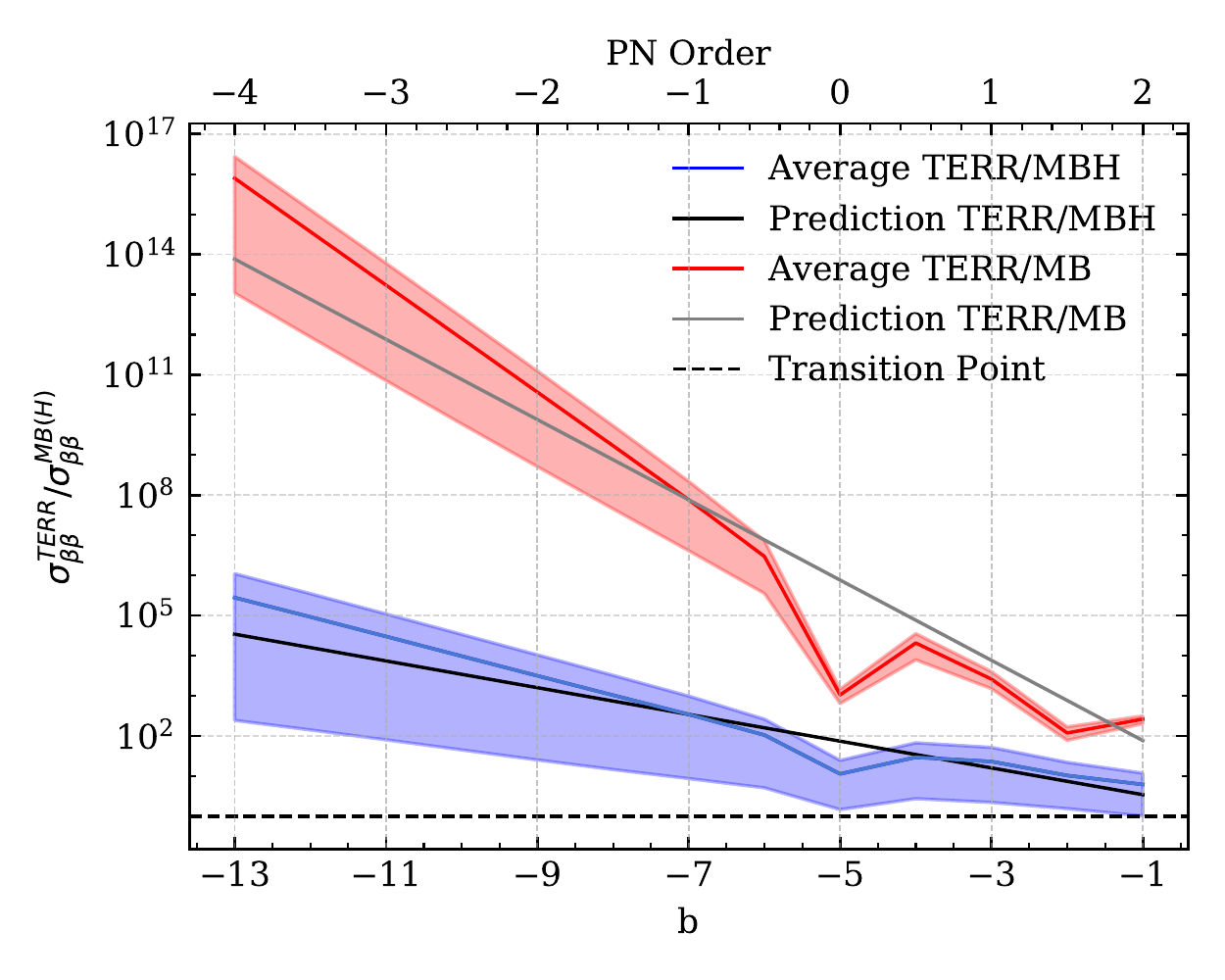}
\caption{
Scaling relations discussed in Sec.~\ref{sec:ind_scaling}.
The ratio $\sigma_{\beta\beta}^{\text{TERR}}/\sigma_{\beta\beta}^{\text{MBH}}$, calculated from the full Fisher simulations including the realistic noise curves shown in Fig.~\ref{fig:SN} and the \software{IMRPhenomPv2} waveform, is shown in blue.
The empirically measured trend is derived from averaging the constraints from each terrestrial network and each population model, then calculating the ratios of every combination of terrestrial network and SOBH model against each MBH heavy-seeding model.
The blue line shows the mean ratio, and the blue shaded region is the area bounded by the maximum and minimum ratios.
The red line and the red shaded region refer instead to the ratio between the terrestrial-only constraints and the multiband constraints, i.e. $\sigma_{\beta\beta}^{\text{TERR}}/\sigma_{\beta\beta}^{\text{MB}}$.
For this class of sources, we calculate the ratio for each population model and detector network, one at a time. 
That is, the terrestrial-only constraints from the S1 network derived from the SPOPS 265 model are compared against the multiband constraints from the S1 network and the SPOPS 265 model.
The trends predicted analytically in the text are shown in black and grey for MBH and multiband sources, respectively.
The trend lines we show for our predictions have been shifted along the y-axis to better compare the with the data.
}\label{fig:source_class_scaling}
\end{figure}

Let us now consider the scaling of the bounds with PN order in more detail. Figure~\ref{fig:source_class_scaling} shows an averaged ratio $\sigma_{\beta\beta}^{\text{TERR}} / \sigma_{\beta\beta}^{\text{MBH}}$ computed from the full numerical simulations of Fig.~\ref{fig:SMBH_time} (solid blue line), together with the prediction in Eq.~\eqref{eq:ratio_scaling} that the ratio should scale as $\propto 10^{-b/3}$ (solid black line).
The numerical results (blue line, with an ``uncertainty'' quantified by the shaded blue region) were computed as follows. We first averaged the constraints for each population model at each PN order and for each detector network that concurrently observes with LISA; this allowed us to isolate the effect of the combination of source class and detector, neglecting the sometimes significant contribution from stacking. Ratios of the averaged quantities were then calculated for each combination of SOBH model (SPOPS 0 and SPOPS 265) and heavy-seeding MBH model (Q3delays and Q3nodelays) and for each detector network -- the CEKLext, CVKLext, and HLVKILext (optimistic and pessimistic) configurations -- resulting in $16$ combinations in all at each PN order, assuming an extended ten-year LISA mission duration.
The average of these combinations is shown as the solid blue line in Fig.~\ref{fig:source_class_scaling}, and the region bounded by the minimum and maximum ratios is shown shaded in blue. Observe that the scaling of Eq.~\eqref{eq:ratio_scaling} is consistent with the averaged ratio in the entire domain; the small dip at $b=-5$ (or $0$PN order) is due to degeneracies with the chirp mass, which the scaling relation does not account for.  

The relation $\sigma_{\beta\beta}^{\text{TERR}}/\sigma_{\beta\beta}^{\text{MBH}}$ can be pushed further by comparing multiband sources against the rest of the SOBH sources detected \emph{only} by the terrestrial network.
For these two classes of sources, the masses would be comparable. Let us focus on the impact of the early inspiral observation. The ratio of the SNRs in the LISA band
is of ${\cal{O}}(1)$ for typical sources, so we will neglect it for now. 
Typical initial frequencies, however, are quite different, with multiband sources having initial frequencies of about $10^{-2}$Hz for SOBH sources that merge within several decades in the terrestrial band. 
This makes the ratio $f_{\text{low}}^{\text{TERR}}/f_{\text{low}}^{\text{MB}}\sim 10^{3}$, and thus, the constraining power of multiband sources relative to that of terrestrial-only sources is approximately $\sigma_{\beta\beta}^{\text{TERR}}/\sigma_{\beta\beta}^{\text{MB}} \sim 10^{-b}$, which explains the scaling observed in item (i) above.
In Fig.~\ref{fig:source_class_scaling} we show the averaged ratio measured from our full simulations including the noise curves shown in Fig.~\ref{fig:SN} and the \software{IMRPhenomPv2} waveform (solid red line) as well as the $10^{-b}$ scaling derived from Eq.~\eqref{eq:ratio_scaling}  (solid gray line). Again, we average the constraints from each population model at each PN order, assuming a ten-year LISA mission duration. However we do not consider every combination of population models and detector networks, but instead compare the multiband constraints from each network and SOBH model
against the terrestrial-only constraints from the same combination of terrestrial network and SOBH model. That is, we compare S1 terrestrial-only constraints derived from the SPOPS 265 model against the multiband constraints with the S1 network and from the SPOPS 265 model, repeating the procedure for each terrestrial network and population model. This yields 8 different combinations of population models and networks. The red line shows the average ratio for all the combinations considered, and the red-shaded region shows the area bounded by the maximum and minimum ratios. The simple analytical scaling reproduces the numerics quite well at negative PN orders, where the contribution to the constraint on the ppE parameter primarily comes from LISA observations. At positive PN orders the scaling relation breaks down for two main reasons: (i) our scaling relation neglects covariances, and (ii) the dominant source of information is no longer LISA's observation of the early inspiral, but the signal from the merger-ringdown seen by the terrestrial network.

\begin{figure}[t]
\includegraphics[width=\linewidth]{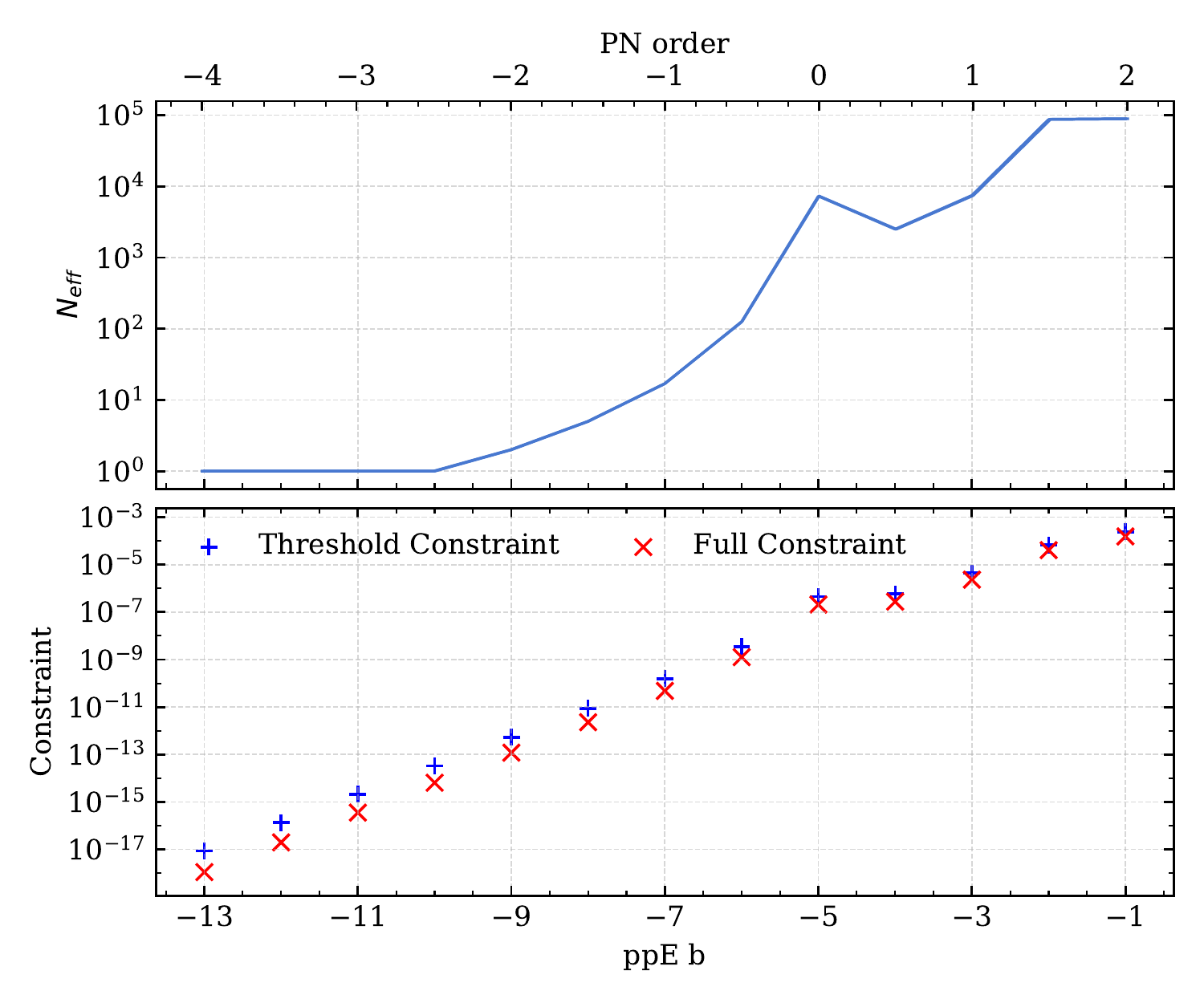}
\caption{
Empirically determined values of $N_{\text{eff}}$ for the CEK (Scenario 1) network and the SPOPS 0 catalog, derived from our full Fisher analysis, including the noise curves shown in Fig.~\ref{fig:SN} and the \software{IMRPhenomPv2} waveform. 
The parameter $N_{\text{eff}}$ is defined as the number of sources needed from the full catalog in order to achieve a threshold constraint $\sigma_{\beta,\THRESH}$, using the most constraining sources first.
Here we choose $\log_{10} \sigma_{\beta,\THRESH} = 0.95 \log_{10} \sigma_{\beta}$, where $\sigma_{\beta}$ is the cumulative bound from the full Fisher analysis for the entire catalog. 
The values of the threshold constraint (blue $+$ signs) are shown alongside the full constraint (red $\times$ signs) in the lower panel.
The number of required sources grows exponentially as a function of PN order: large catalogs benefit positive PN orders, but they are not as important for highly negative PN orders.
}\label{fig:Neff_pn}
\end{figure}

\begin{figure*}[t]
\includegraphics[width=\linewidth]{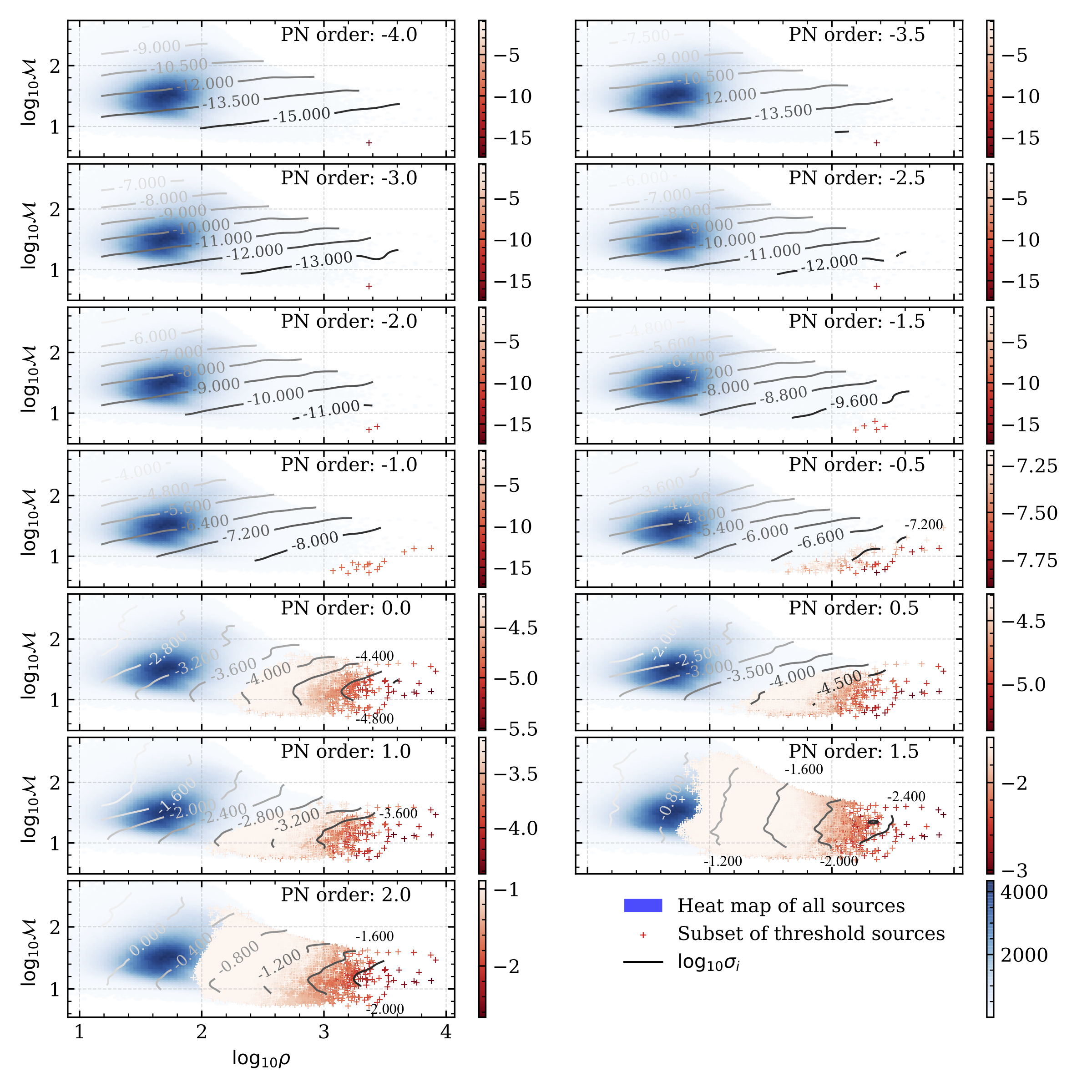}
\caption{
Three different distributions in the $\log_{10} \mathcal{M}-\rho$ plane for the CEK network and the SPOPS 0 population model.
The blue heat map shows the distribution of the sources directly in the $\log_{10} \mathcal{M}-\rho$ plane, and it is the same for all PN orders.
The black contours show the constraints from individual sources.
The red scatter plots show the sources needed to obtain a threshold cumulative constraint $\log_{10} \sigma_{\beta,\THRESH} = 0.95 \log_{10} \sigma_{\beta}$, where the shade of red indicates the strength of the individual bounds (in log base 10).
We utilized a $2\sigma$ gaussian filter over the data to smooth out the noise and create more easily interpretable contour plots.
In conjunction with Fig.~\ref{fig:Neff_pn}, the growing number of scatter points as a function of PN order illustrates the increasing dependence of the cumulative constraint on the size of the source catalog.
Furthermore, the relation between chirp mass, SNR, and individual bound can be seen to shift significantly between positive and negative PN orders, agreeing with the commonly held intuition that lower-mass sources are better for constraining negative PN effects.
In more detail, the negative PN orders benefit highly from low-mass systems, with slight dependence on SNR, while positive PN order effects depend much more strongly on the SNR and have more minimal dependence on the chirp mass.
Finally, the range of individual bounds ($\sim 4$ orders of magnitude at negative PN orders and $\sim 2$ orders of magnitude at positive PN orders) helps to explain the different scaling relations between the cumulative bounds and the total number of sources.
}\label{fig:Neff_distribution}
\end{figure*}

\subsubsection{Analytical scaling: multiple sources}\label{sec:multiple_source_scaling}

Our analysis above helps to elucidate some of the trends observed in our numerical simulations by examining individual sources, but it fails to capture the power of combining observations to enhance constraints on modified theories of gravity.
Especially when considering terrestrial networks, this element is critical in predicting future constraints, and it is connected with our observations (iv) and (v) in the previous list.

To fully explore this facet of our predictions, we try to isolate the impact of the total number of sources on the final, cumulative constraint for a given network. 
As shown in Eq.~\eqref{eq:PPEcombined} of Appendix~\ref{app:Fisher}, the combined constraint from an ensemble of simulated detections is 
\begin{equation}
\sigma_{\beta}^2 = \left( \sum_i^{N} \frac{1}{\sigma_{\beta,i}^2} \right)^{-1}\,,
\end{equation}
where $\sigma_{\beta,i}$ is the variance on $\beta$ of the $i$-th source marginalized over the source-specific parameters, including all detectors and priors, and $N$ is the total number of sources in the ensemble.
The effect of the population on all the different combinations of detector networks and PN orders can be summarized by the distribution in $\sigma_{\beta,i}$, and we find empirically that they all lie somewhere in the spectrum bounded by the following extreme scenarios:
\begin{itemize}
\item [(a)] all the constraints contribute more or less equally, 
\item [(b)] the total constraint is dominated by a single (or a few) observations.
\end{itemize}
When the covariances are all approximately equal, the sum above reduces to $\sigma_{\beta} \approx \sigma_{\beta,i} / \sqrt{N}$, but when one constraint (say $\sigma_{\beta,\text{strongest}}$) dominates the ensemble, the sum reduces to $\sigma_{\beta} \approx \sigma_{\beta,\text{strongest}} $.
Naturally, in the case where all sources are more or less equally important, the power of large catalogs is maximized, and one would expect terrestrial networks observing hundreds of thousands to millions of sources to outperform networks with smaller populations, such as MBHs and multiband sources (everything else being equal).
When one observation dominates the cumulative bound because of loud SNR or source parameters that maximize the constraint, then large catalogs are not as important.

In an attempt to quantify this effect, we can ask the following question: what is the minimum number of sources we can retain and still achieve a similar constraint on $\beta$?
To answer this question, we take all the variances calculated with our Fisher analysis for a given population model and detector network, and order them according to the strength of the constraint from each individual source. 
With some threshold constraint set, we can work our way down the list, calculating the cumulative bound for the ``best'' $N'$ sources at a time. 
We define $N_{\text{eff}}$ as the value of $N'$ such that our threshold constraint is achieved. 
Comparing the values of $N_{\text{eff}}$ at each PN order for a single population model and network provides useful insights into how generic constraints benefit from the catalog size.

The upper panel of Fig.~\ref{fig:Neff_pn} shows the values of $N_{\text{eff}}$ calculated using the results from our full Fisher analysis, including the noise curves shown in Fig.~\ref{fig:SN} and the \software{IMRPhenomPv2} waveform, for the CEK network with the SPOPS 0 population model and a threshold constraint of $\log_{10}\sigma_{\beta,\THRESH} = 0.95\log_{10}\sigma_{\beta}$. A pronounced trend is evident: positive PN orders require up to $\sim 10^5$ sources to retain a constraint equal to our threshold value, while the most negative PN effects only require a single, highly favorable source to reach the threshold value.
The lower panel of Fig.~\ref{fig:Neff_pn} merely shows the value of the full numerical constraint (red $\times$ signs) compared with our value of the threshold constraint (blue $+$ signs): by our own definition, the threshold constraint captures most (i.e.~95\%) of the full constraint. 

Figure~\ref{fig:Neff_distribution} shows several different facets of the data relevant to the analysis of Fig.~\ref{fig:Neff_pn}.
For each PN order, we have plotted three different quantities: (i) a heat map of all the sources in the catalog in the $\log_{10} \mathcal{M}-\rho$ plane (shown in blue), which is the same for all PN orders, (ii) the contours showing the strength of the individual constraints from each source for the entire catalog (in black), and (iii) the subset of sources required to meet the threshold constraint $\sigma_{\beta,\THRESH}$ (in red), where the shade corresponds to the strength of the individual bounds.

Several interesting conclusions can be drawn from this figure. 
First, the relation between the constraint, the SNR, and the chirp mass changes as a function of PN order. 
The highly positive PN orders benefit highly from loud sources, with only a slight preference for the lower mass systems (if at all), while highly negative PN effects benefit greatly from low-mass systems, with a slight preference for louder sources.
This agrees with our intuition about low-mass systems being most important for negative PN effects: in Eq.~\eqref{eq:single_source_scaling} the chirp mass is raised to the $-b/3$ power, significantly enhancing the impact of low-mass systems for negative PN effects, while minimizing their impact for positive PN effects (assuming $b<2$). As these figures are constructed from our fully numerical data, these trends take into account the nonlinear relation between SNR and chirp mass, as these are not independent parameters when considering realistic population models. 
Reasonably accurate population models are important in studies of this type, as bounds can be significantly altered by changing the distributions of source properties.

A second observation one can draw from Fig.~\ref{fig:Neff_distribution} relates to the change in the relation between SNR and individual constraints, which explains why the constraining-power gap between the different terrestrial network scenarios closes at positive PN orders (items (iv) and (v) from above).
The relaxation in the SNR-constraint correlation at high positive PN orders means that the huge boost in SNR from utilizing 3g detectors, as compared to a 2g only network, has only a moderate impact on the cumulative bound, \emph{if} the 2g network is sensitive enough to observe a comparable number of sources to the 3g network. 
In the case of the Voyager network (HLVKI+), the much lower average SNR (shown in Fig.~\ref{fig:source_properties} and Fig.~\ref{fig:sources_per_year}) hinders the network's capability greatly at negative PN orders, but only minimally at positive PN orders, as compared with the CEK or CVK networks shown in Fig.~\ref{fig:SMBH_time}.
This is because the total number of sources observed in each scenario is comparable with Scenario 3, only differing by $\sim 30\%$, and allowing HLVKI+ to maintain competitive constraining power through comparably sized catalogs.

A third observation that we can make about Fig.~\ref{fig:Neff_distribution} is that the range in individual bounds is also clearly PN-order dependent. The most negative PN corrections change by $\sim 4$ orders of magnitude, while the most positive PN corrections only change by $\sim 2$ orders of magnitude. 
This change in constraint range lends credence to the interpretation outlined above. When constraints are clustered closer together and contribute equally, the cumulative constraint scales strongly with the number of sources. The opposite is true when the clustering is weaker and one constraint dominates over the whole ensemble.
The analysis performed here, coupled with that done in Sec.~\ref{sec:ind_scaling}, further clarifies the trend observed in items (ii) and (iii). 
The combination of the individual source scaling favoring LISA at negative PN orders is enhanced by the significant benefit from large catalogs for terrestrial networks for positive PN orders.

\subsection{Specific Theories}\label{sec:specific_theories}

We can now recast the constraints on generic ppE parameters from Sec.~\ref{sec:general_mod} into constraints on relevant quantities in a variety of specific modified gravity theories. 
We list and categorize these theories in Table~\ref{tab:theory}.

We will utilize the scaling analysis outlined in the previous section, with the additional step 
\begin{equation}
\Gamma_{\text{theory}} = \mathcal{J}^T \cdot \Gamma_{\PPE} \cdot \mathcal{J}\,,
\end{equation}
where $\mathcal{J}$ is the Jacobian $\partial \vec{\theta}_{\PPE}/\partial \vec{\theta}_{\text{theory}}$ of the transformation, and ${(\cdot)}^T$ is the transpose operation. 
In our case, the Jacobian is diagonal. This is because the off-diagonal components are all proportional to the theory-specific modifying parameter;
as we inject with GR models, these are always set to zero for any specific beyond-GR theory.
We can then write
\begin{equation}
\Gamma_{\alpha_\text{theory}\alpha_\text{theory}} = \left(\frac{\partial \beta}{\partial \alpha_{\text{theory}}}\right)^2 \Gamma_{\beta\beta}\,,
\end{equation}
where $\beta$ is the generic ppE modification at the corresponding PN order for a given theory, and $\alpha_{\text{theory}}$ is the theory-specific modifying parameter. 
The interested reader can find the mappings $\beta(\alpha_{\text{theory}})$ between each theory and the ppE formalism, and more in-depth explanations of their motivations, in Appendix~\ref{sec:theories}.

This mapping between ppE constraints and theory-specific constraints changes the scaling relations between the theory-specific bound and different source parameters, with many of the conclusions made by examining the generic constraints changing quite drastically. This is because the Jacobian typically depends on source parameters, like $\mathcal{M}$, $\eta$, $\chi_1$, and $\chi_2$, and this can strongly enhance the constraining power of one population of BBHs over another. No general trend can be ascertained across multiple modified theories since each coupling is different, so we will examine each theory in turn. As we will see, constraints on different theory-specific parameters scale differently with SNR, chirp mass, etcetera, impacting how the cumulative bound improves with stacking and how dependent the bound is on small numbers of loud sources.
To examine this in more detail, we will focus on a single detector network (HLVKIO8) with a single population model (SPOPS 0) to try and isolate the pertinent effects for each theory. 

\begin{figure}[ht!]
\includegraphics[width=\linewidth]{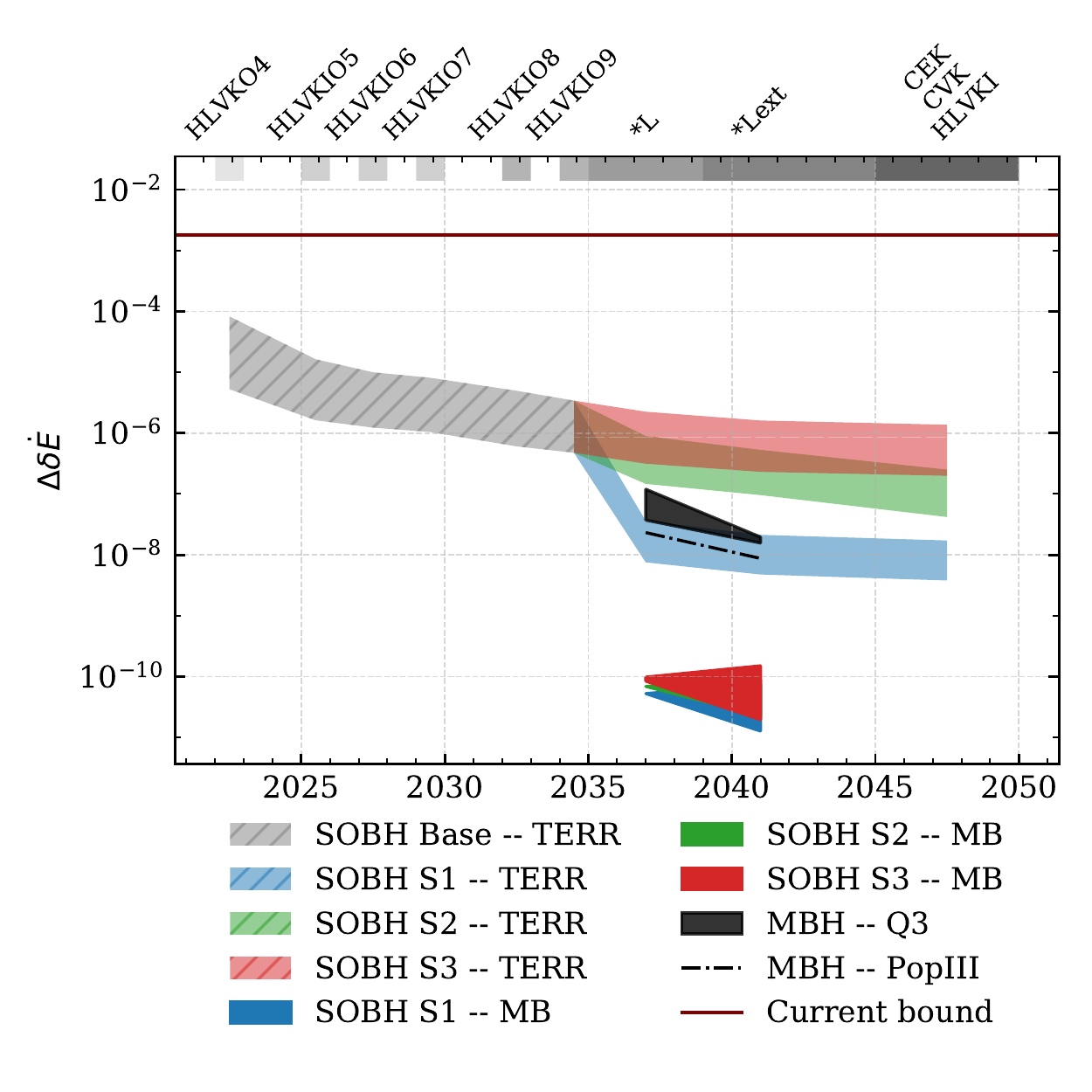}
\caption{
Projected cumulative constraints on generic dipolar radiation for the detector networks and population models examined in this paper. 
The multiband sources outperform all other source classes by at least $\sim 2$ orders of magnitude, with MBH sources and the most optimistic terrestrial scenario performing comparably. 
}\label{fig:dipole}
\end{figure}

\begin{figure*}
\includegraphics[width=\linewidth]{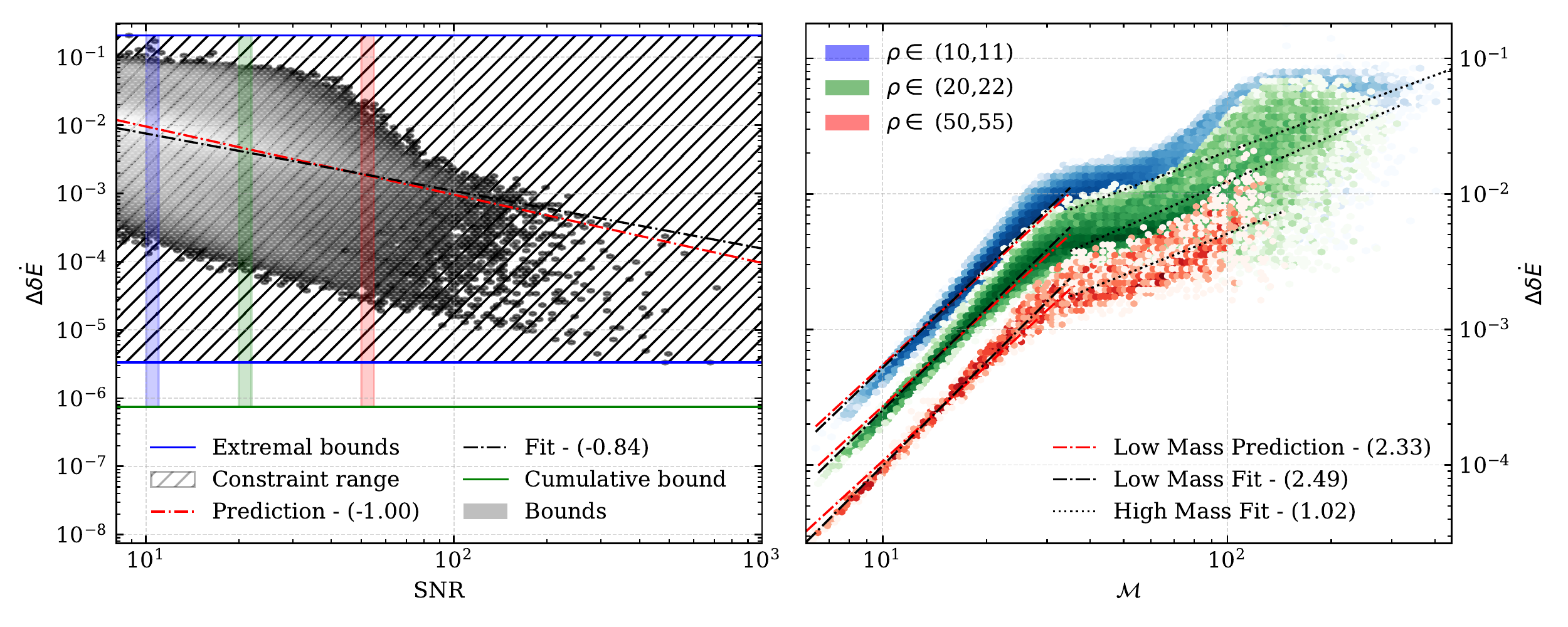}
\caption{
Result of the scaling analysis outlined in Sec.~\ref{sec:res_dipole} performed on the data synthesized with the HLVKIO8 network and the SPOPS 0 population.
The left panel shows a heat map of the constraint on $\delta \dot{E}$ versus the SNR of the source. 
The solid blue lines correspond to the strongest and weakest single-source constraint, and the area between these two bounds is shown in hatching.
The cumulative bound from the entire catalog is shown as the solid green line.
The power-law fit to the data in the left panel is shown as the solid black curve, and our prediction for the scaling is shown as the solid red curve.
The right panel shows three distinct slices of the catalog, with ranges in SNR from 10 to 11 (blue), 20 to 22 (green), and 50 to 55 (red).
These ranges are highlighted in the left panel.
The right panel shows the density of the constraint versus the chirp mass, with empirical trends shown in black and predicted trends shown in red.
There is a noticeable transition point in the distribution, so low-mass and high-mass systems were analyzed separately. 
The powers used in all trend lines are shown in the legend.
For trend lines, the (logarithmic) offset for the predicted scaling relations has been adjusted to coincide with the empirically fit offset, to better compare the slopes of the trends.
Of particular interest is the strong trend relating the SNR and the bound, as well as the tight correlation between chirp mass and constraint for low-mass systems, which seems to taper off for high-mass systems.
}\label{fig:dipole_scaling}
\end{figure*}

\subsubsection{Generic Dipole Radiation}\label{sec:res_dipole}

Dipole radiation is absent in GR, since in Einstein's theory GWs are sourced by the time variation of the quadrupole moment of the stress-energy tensor. 
Therefore, any observation of dipole radiation would indicate a departure from GR.
Dipole radiation must be sourced by additional channels of energy loss, due to the presence of new (scalar, vector or tensor) propagating degrees of freedom. 
By the balance law, these new channels of energy loss affect the time variation of the binding energy $E$,
and therefore dipole effects generically enter the GW Fourier phase at $-1$ PN (to leading order)~\cite{Chamberlain:2017fjl}.
While many theories predict specific forms of dipole radiation, we can constrain any process leading to dipole radiation by the time rate of change of the binding energy, $\dot{E}$.

We we show in Appendix~\ref{sec:theories}, the Jacobian in this specific class of modifications scales as 
\be
\left(\frac{\partial \beta}{\partial \delta \dot{E}}\right)^{2} \propto  \eta^{4/5}\,,
\ee
where $\delta \dot{E} =\dot{E} - \dot{E}_{\GR}$ is the variation in $\dot{E}$ due to dipole radiation: see Eq.~\eqref{eq:dip_energy}.
This implies that the scaling relations found earlier for generic ppE modifications should not change much when we translate them into constraints on dipole radiation. 

These constraints are shown in Fig.~\ref{fig:dipole}.
As dipole radiation is a negative PN effect, multiband sources will contribute significantly, improving bounds by at least two orders of magnitude over any other detector network or population class.
LISA observations of MBH binaries are still highly competitive, outpacing the terrestrial-only network in all cases except the most optimistic detector schedule. 
Furthermore, the different terrestrial networks see a wide variation, as the difference between the typical SNRs between the networks are quite large. 
After thirty years of GW measurements, our models suggest an improvement of 3--9 orders of magnitude over existing constraints, depending on source populations and detector characteristics, but a 9-orders-of-magnitude improvement is only possible with multiband events. 
All of these trends are consistent with the analysis presented in Sec.~\ref{sec:ind_scaling}, with constraints on this negative PN order effect benefitting from the low initial frequency and low chirp masses of LISA multiband sources.
This is because dipole radiation approximately scales like a generic ppE modification in terms of SNR and chirp mass, meaning that most of the analysis from above is still valid in this case.

To better understand the numerical results presented in Fig.~\ref{fig:dipole}, we can look at our analytical approximation of $\Delta \delta \dot{E}$ using the methods from the previous section. 
After mapping the bound on the generic $\beta$ to $\delta \dot{E}$, expanding in $\epsilon = \mathcal{M} f_{\text{low}}$, and setting the upper frequency to the ISCO frequency, we have the approximation
\begin{equation}
\Delta\delta \dot{E} \approx \frac{ 112 \sqrt{2} }{ \eta^{2/5} } \frac{ \left(\mathcal{M}\pi f_{\text{low}}  \right)^{7/3} }{\rho} \,.
\end{equation}
Results related to this approximation are shown in Fig.~\ref{fig:dipole_scaling}.
The left panel shows a density map of the bounds on $\delta \dot{E}$ versus the SNR of the source, with a numerical fit overlaid showing the SNR scaling trend in black.
Our $1/\rho$ scaling prediction, shown in red, matches the numerics very well.

The right panel shows a density plot of the bound on $\delta \dot{E}$ versus chirp mass.
To isolate the impact of the chirp mass on the attainable bound on $\delta \dot{E}$, we restrict ourselves to thin slices in different ranges of SNR (the ranges are highlighted in the top panel). 
This is to insulate our results from the fact that the SNR typically scales with the mass, causing a nonlinear relationship between the mass, SNR, and constraint. 
To ensure that the scaling does not change for different ranges of SNR, we have separately analyzed three different ranges.
For lower mass systems, we see good agreement with the analytically predicted $\mathcal{M}^{7/3}$ scaling relationship, but around $\mathcal{M}\sim 30 M_{\odot}$ we see a sharp transition, and our approximations fail.

The impact of these different scaling relations can be seen in the range of constraints and the cumulative constraint shown in Fig.~\ref{fig:dipole_scaling}.
In the left panel, we have plotted the strongest and weakest constraint as solid blue lines, bounding the parameter space of single-source bounds.
The cumulative bound for this one network-population combination is shown as a green line, near the bottom of the panel. 
As is evident in the figure, the improvement of the cumulative bound over the most stringent bound is marginal. 
This can be explained by the huge range of single-source bounds, covering five orders of magnitude, consistent with the analysis performed in Sec.~\ref{sec:multiple_source_scaling}.

\begin{figure}
\includegraphics[width=\linewidth]{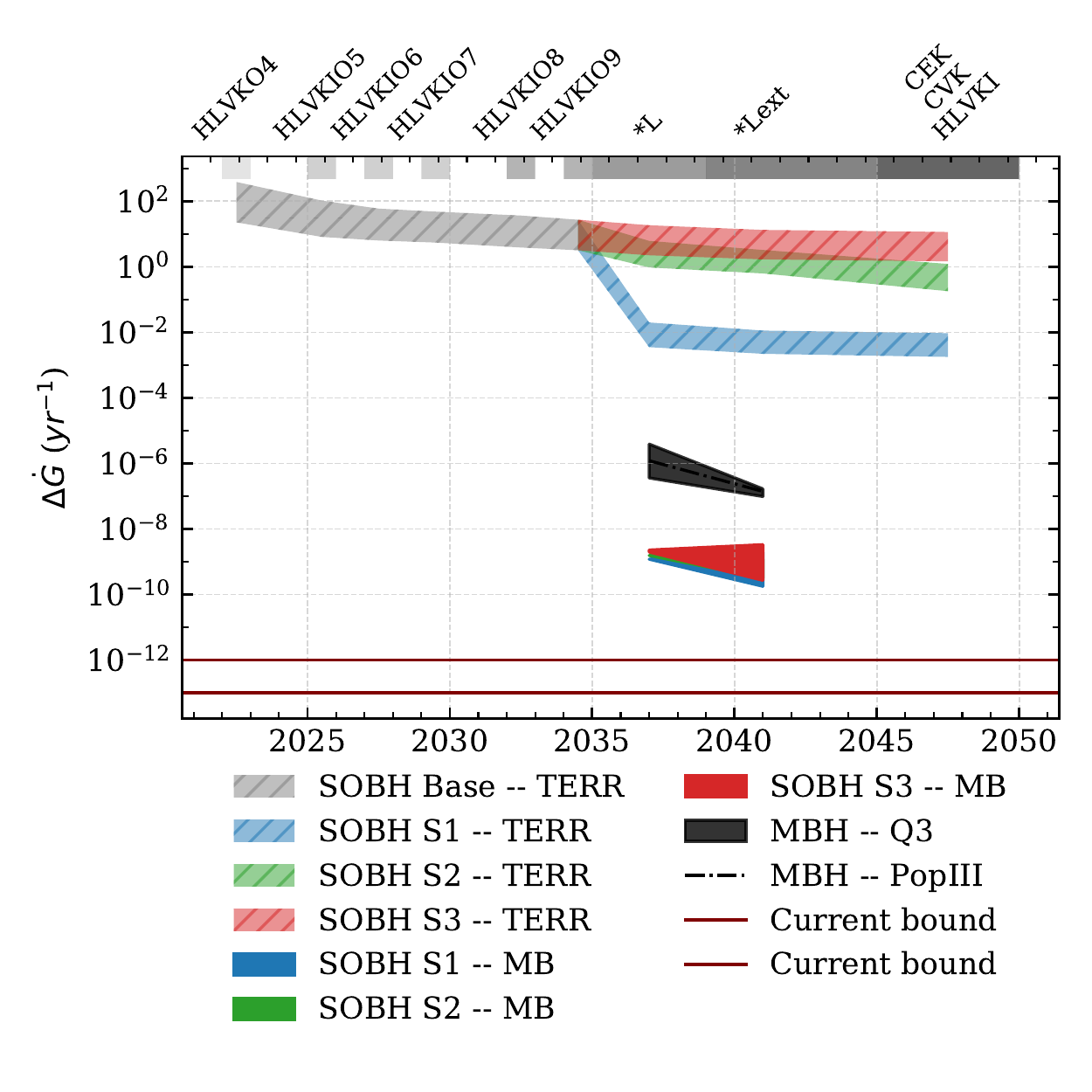}
\caption{
Projected cumulative constraints on the time derivative of the gravitational constant $\dot{G}$ for the detector networks and population models examined in this paper. 
Multiband sources outperform all other source classes by $\sim 1-2$ orders of magnitude, with MBH sources performing the next best.
SOBHs observed by the terrestrial network alone perform the worst, but with Scenario 1 outperforming Scenarios 2 and 3 due to the high SNR of the observations in the former network.
}\label{fig:varG}
\end{figure}

\begin{figure*}
\includegraphics[width=\linewidth]{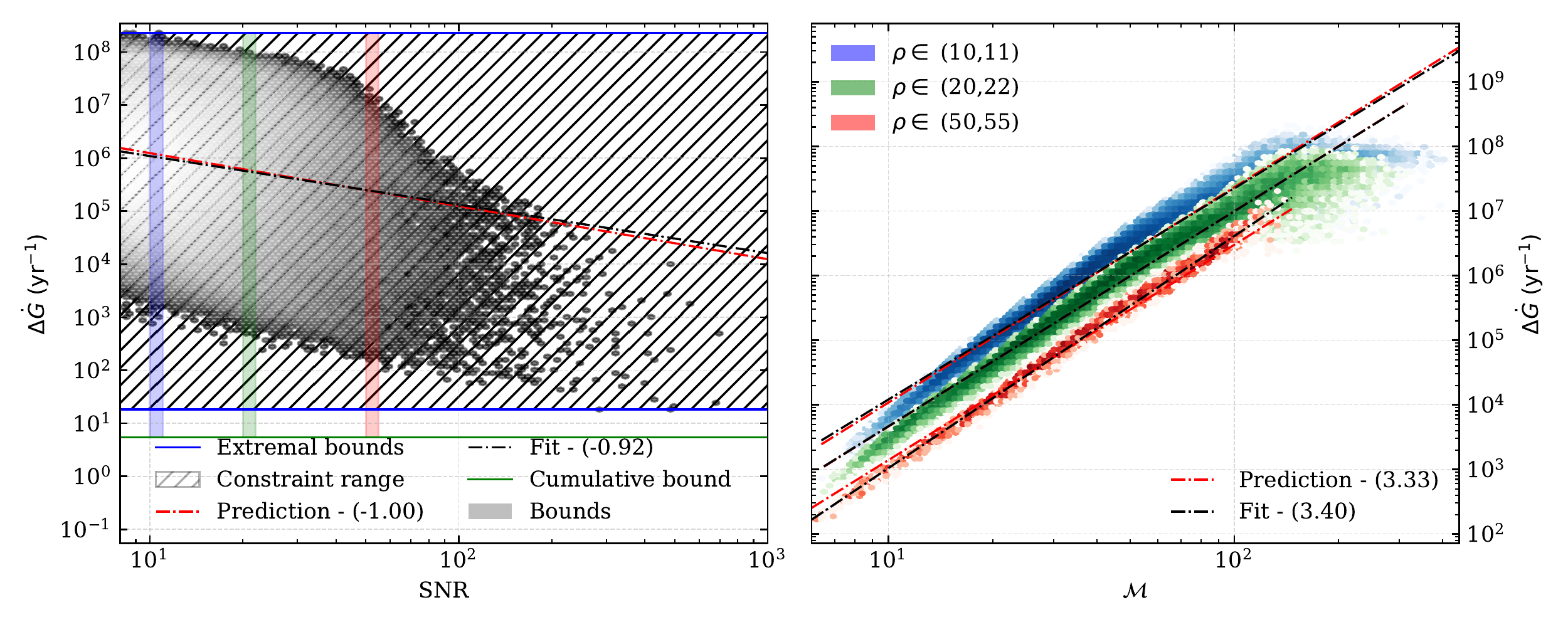}
\caption{
Result of the scaling analysis outlined in Sec.~\ref{sec:res_varg} performed on the data synthesized with the HLVKIO8 network and the SPOPS 0 population.
The plotting style is the same as in Fig.~\ref{fig:dipole_scaling}.
The left panel shows a heat map of the constraint on $\dot{G}$ versus the SNR of the source. 
The right panel shows the density of the constraint versus $\mathcal{M}$, with empirical trends shown in black and predicted trends shown in red.
Again, the strong trend relating the SNR and the bound agrees well with the prediction, and there seems to be a tight correlation between $\mathcal{M}$ and constraint, well approximated by our analysis in Sec.~\ref{sec:res_varg}.
}\label{fig:TVG_scaling}
\end{figure*}

\subsubsection{Local Position Invariance -- Variable G Theories}\label{sec:res_varg}

If the gravitational constant $G$ were time-dependent, we would observe anomalous acceleration in the inspiral of BBHs~\cite{Yunes:2009bv}.
At leading order, this affects the GW Fourier phase at $-4$ PN. 
From the transformation in Appendix~\ref{sec:theories}, the Jacobian to map from the generic ppE modification to the parameter $\dot{G}$ itself is 
\be
\left(\frac{\partial \beta}{\partial \dot{G}}\right)^{2} \propto  \left(\frac{\mathcal{M}}{1+z}\right)^2\,.
\ee
The mapping now includes a chirp mass-dependent factor, which can vary by orders of magnitude between source classes. 
From this scaling with chirp mass, and the fact that this modification enters at a highly negative PN order ($-4$PN), we expect that the best sources will be those that are seen at the widest separations (like multi-band sources) and have the largest chirp mass.

Our predictions for the constraints on $\dot{G}$ can be seen in Fig.~\ref{fig:varG}.
Multiband constraints again outperform all other source classes and detector configurations, as expected.
However, because the Jacobian is proportional to $\mathcal{M}^2$, MBH sources seen by LISA are not far behind. 
Comparatively, the terrestrial-only bounds trail significantly behind both of these source classes, by as much as three orders of magnitude. 
There is also a wide separation between the three different terrestrial-only observation scenarios. 
This suggests that the cumulative bound does not benefit too much from large catalogs, but instead is dominated by a small number of favorable observations.

A variable $G$ modification presents the first departure from our analysis on the scaling of generic results. MBH sources receive a sizeable benefit over the SOBH sources due to the Jacobian factor between parameters.
Consequently, constraints on this particular modification benefit greatly from the inclusion of LISA in the GW network, both in the form of multiband and MBH observations.

Even after thirty more years of GW detections with the most ideal networks, our models indicate that the bounds will still fall far short of the current constraints on $\dot{G}$ coming from cosmology. These constraints, however, are qualitatively different from those considered here. Cosmological constraints assume a Newton constant that is linearly dependent on time in the entire cosmological history of the Universe, i.e.~ that $G \to G(t) \sim G_{\rm BBN} + \dot{G}_{\rm BBN} t$, where $t$ is time from the Big Bang until today, and where $G_{\rm BBN}$ and $\dot{G}_{\rm BBN}$ are constants. Our $\dot{G}$ constraints only assume a linear time dependence \emph{near} the BBH merger, i.e.~ that $G \to G(t) \sim G_{t_{c}} + \dot{G}_{t_{c}} (t - t_{c})$ for $t < t_{c}$ where $t_{c}$ is the time of coalescence, $G_{t_{c}}$ and $\dot{G}_{t_{c}}$ are constants, and $G(t)$ relaxes back to $G_{t_{c}}$ in a few horizon light-crossing times. In our stacking analysis, we are implicitly assuming that $\dot{G}_{t_{c}}$ is the same for all sources in all catalogs. Therefore, it is not strictly fair to compare cosmological and GW bounds.

We can again repeat the analysis from Sec.~\ref{sec:general_mod} to better understand the relationship between the bound on $\dot{G}$ and various source parameters.
Making the approximations outlined in Sec.~\ref{sec:ind_scaling}, we can approximately rewrite the constraint on $\dot{G}$ as 
\begin{equation}
\Delta \dot{G} \approx \frac{32763}{5} \sqrt{ \frac{6}{5} } \frac{ \left( \pi \mathcal{M} f_{\text{low}} \right)^{13/3}(1+z) }{\mathcal{M} \rho}\,,
\end{equation}
where we obtain the expected extra dependence on the chirp mass from the Jacobian transformation.
Results pertinent to this approximation are shown in Fig.~\ref{fig:TVG_scaling}.
The left panel shows a heat map of the $\dot{G}$ constraints against the SNR for the sources in the HLVKIO8 network and the SPOPS 0 model.
The right panel shows a heat map of the constraint on $\dot{G}$ against the chirp mass, for different slices in the SNR.
Notably, the scaling of the constraint on $\dot{G}$ with respect to the chirp mass matches well with our prediction of $\mathcal{M}^{10/3}$, which differs from the generic constraint by a factor of $\mathcal{M}^{-1}$ due to the Jacobian factor.
Again, we see a large spread in the magnitude of the constraint, ranging over $\sim 6$ orders of magnitude.
This leads to a marginal improvement of the cumulative bound over the strongest bound from a single observation, further hampering the terrestrial-only networks, in agreement with our analysis in Sec.~\ref{sec:multiple_source_scaling}.
After accounting for the modified scaling due to the Jacobian, the scaling relations and techniques from Sec.~\ref{sec:general_mod} generally hold for predicting constraints on variable $G$ theories.

\begin{figure}
\includegraphics[width=\linewidth]{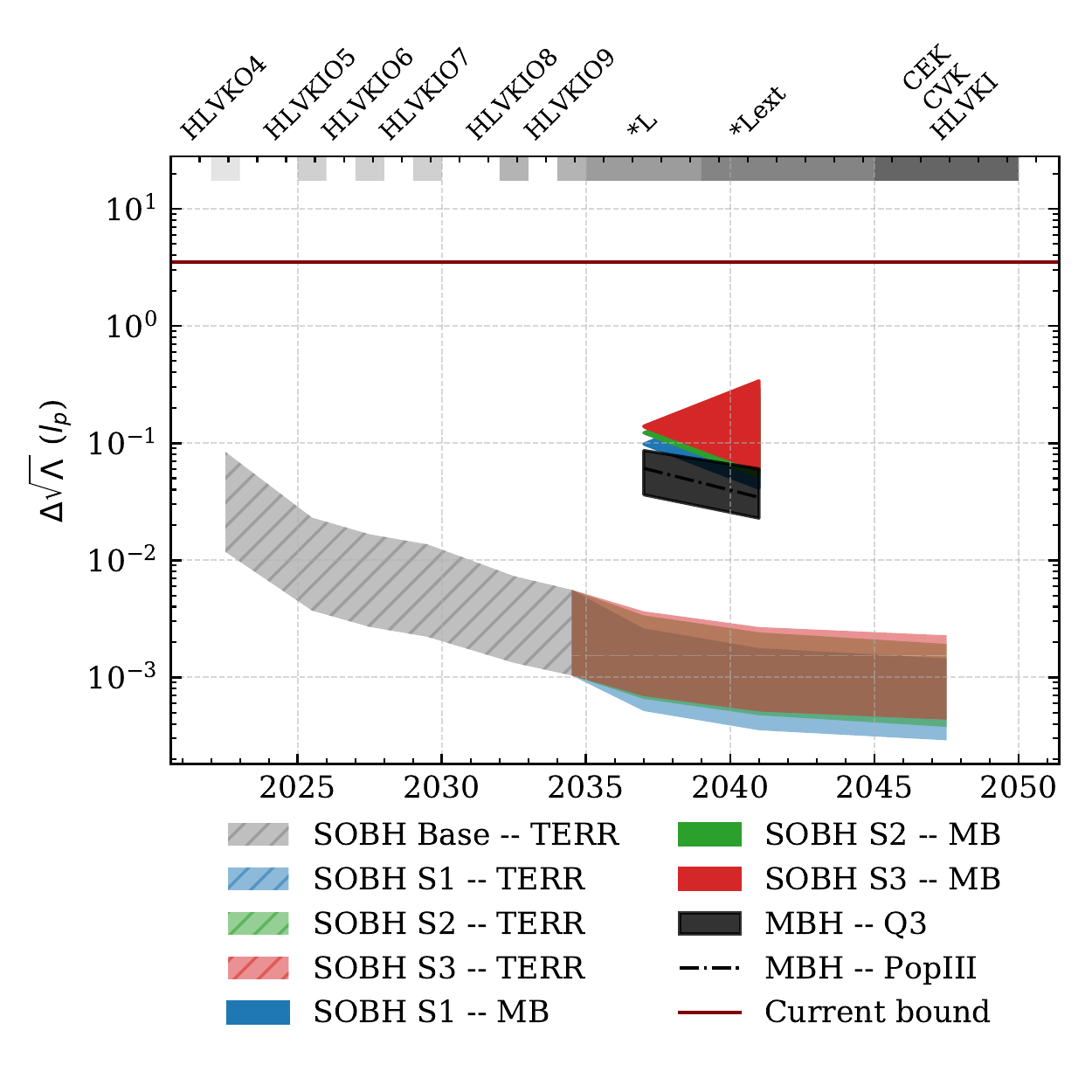}
\caption{
Projected cumulative constraints on $\sqrt{\Lambda}$ for the detector networks and population models examined in this paper. 
Terrestrial-only catalogs, with their populations of millions of sources, seem to dominate any future constraint on this particular deviation, with an improvement by 1--2 orders of magnitude over any other source classification.
This conclusion seems independent of the particular terrestrial scenario we pick, with comparable performance from all three.
}\label{fig:noncom}
\end{figure}

\begin{figure*}
\includegraphics[width=\linewidth]{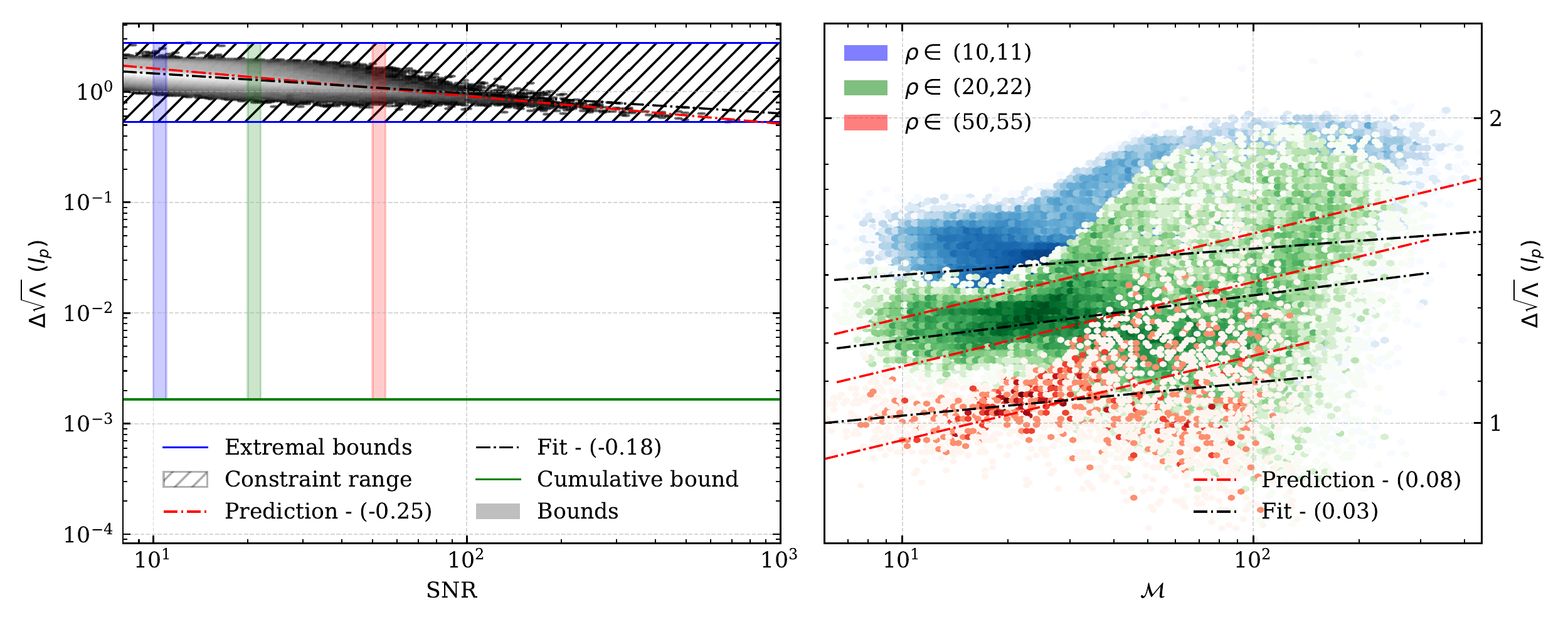}
\caption{
Result of the scaling analysis outlined in Sec.~\ref{sec:res_LV} performed on the data synthesized with the HLVKIO8 network and the SPOPS 0 population.
The plotting style is the same as in Fig.~\ref{fig:dipole_scaling}.
The left panel shows a heat map of the constraint on $\sqrt{\Lambda}$ versus the SNR of the source. 
The right panel shows the density of the constraint versus the chirp mass, with empirical trends shown in black and predicted trends shown in red.
The small range of constraints from the catalog lead to considerable enhancements of the cumulative bound when stacking observations, and the weak scaling with chirp mass and moderate scaling with SNR further benefit SOBH sources over other source classes.
}\label{fig:NC_scaling}
\end{figure*}

\subsubsection{Lorentz Violation -- Noncommutative Gravity}\label{sec:res_LV}

If a commutation relation is enforced between momentum and position, as in quantum mechanics, the leading order effect occurs at 2PN.
Predictions for the constraints on the scale of the noncommutative relation are shown in Fig.~\ref{fig:noncom}.
The Jacobian of the transformation found in Appendix~\ref{sec:theories} is given by
\begin{equation}
\left(\frac{\partial \beta}{\partial \Lambda^2 }\right)^{2} \propto  \eta^{-4/5} ( 2 \eta - 1) \,.
\end{equation}
The Jacobian only introduces source-dependent terms of $\order{1}$, and as such, bounds on $\Lambda^2$ should generally follow the scaling trends found in Sec.~\ref{sec:general_mod}.
Given that this modification comes at 2PN, we would expect the terrestrial-only source catalogs to constrain non-commutative gravity the strongest: the power of large catalogs is enhanced, and the effect of LISA observations of the early inspiral is less relevant for positive PN effects.

The bounds predicted by our models are shown in Fig.~\ref{fig:noncom}.
As expected, the terrestrial networks contribute the most to any future bound on non-commutative gravity. 
Even when just considering the three terrestrial-only scenarios, the differences are minimal.
Furthermore, the other source classes (MBH and multiband) perform almost identically. 
All of these trends further solidify our conclusion that the key to future constraints on this particular modification is large catalogs of observations, as opposed to single, favorable sources.
Future constraints from all source classes should improve by 1--3 orders of magnitude over present constraints.

Continuing our analysis to explore the more subtle trends we are seeing, we can repeat the analysis outlined in Sec.~\ref{sec:general_mod}.
This gives us the following approximation for the variance on $\sqrt{\Lambda}$:
\begin{equation}
\Delta \sqrt{\Lambda} \approx \left( \frac{32768}{1875} \right)^{1/8} \frac{\eta^{1/5} \left(\pi \mathcal{M} f_{\text{low}}\right)^{1/12}}{ \left( 1-2 \eta \right)^{1/4} \rho^{1/4}}\,.
\end{equation}
Although the bound on $\Lambda^2$ scales as expected from Sec.~\ref{sec:general_mod}, approximating our bound on $\sqrt{\Lambda}$ given our constraint on $\Lambda^2$ introduces modifications to the trends we would not have expected from a straightforward extrapolation from constraints on generic modifications. 
Namely, we see that the bound should generically scale with the SNR as $\rho^{-1/4}$, and the constraint should scale with the chirp mass as $\mathcal{M}^{1/12}$.

Pertinent trends related to this approximation are shown in Fig.~\ref{fig:NC_scaling}, where the HLVKIO8 network and the SPOPS 0 model were used to do the analysis.
The left panel shows a heat map in the constraint-SNR plane, with the extremal, single source bounds shown as solid blue lines. 
The cumulative bound for only this network-population combination is shown as the solid green line.
Our predicted trend for the constraint with respect to the SNR is shown in red, while the empirically determined trend is shown in black.
The right panel shows a heat map in the constraint-chirp mass plane, where we have separately analyzed three different slices of sources with specific SNRs, denoted by the colors red, blue, and green.

In the left panel of Fig.~\ref{fig:NC_scaling}, we can see that our approximation for the relation between the constraint and the SNR does fairly well relative to the empirically determined trend.
Furthermore, we see that the range of constraints is considerably tighter than even the generic constraints at 2PN. 
The largest and smallest bound for non-commutative gravity are separated by one order of magnitude, leading to a significant improvement of the cumulative bound over the tightest single-observation bound.
This feature further explains to some degree the discrepancy between LISA sources and terrestrial-only sources in Fig.~\ref{fig:noncom}.

In the right panel of Fig.~\ref{fig:NC_scaling}, we see much wider distributions in the constraint-chirp mass plane, as compared to the previously analyzed modifications.
Our predicted trends are moderately accurate, although with noticeably lower accuracy.
This is consistent with the fact the constraint scales very weakly with chirp mass ($\mathcal{M}^{1/12}$), and other correlations are widening the distribution and complicating the relation.

\begin{figure}
\includegraphics[width=\linewidth]{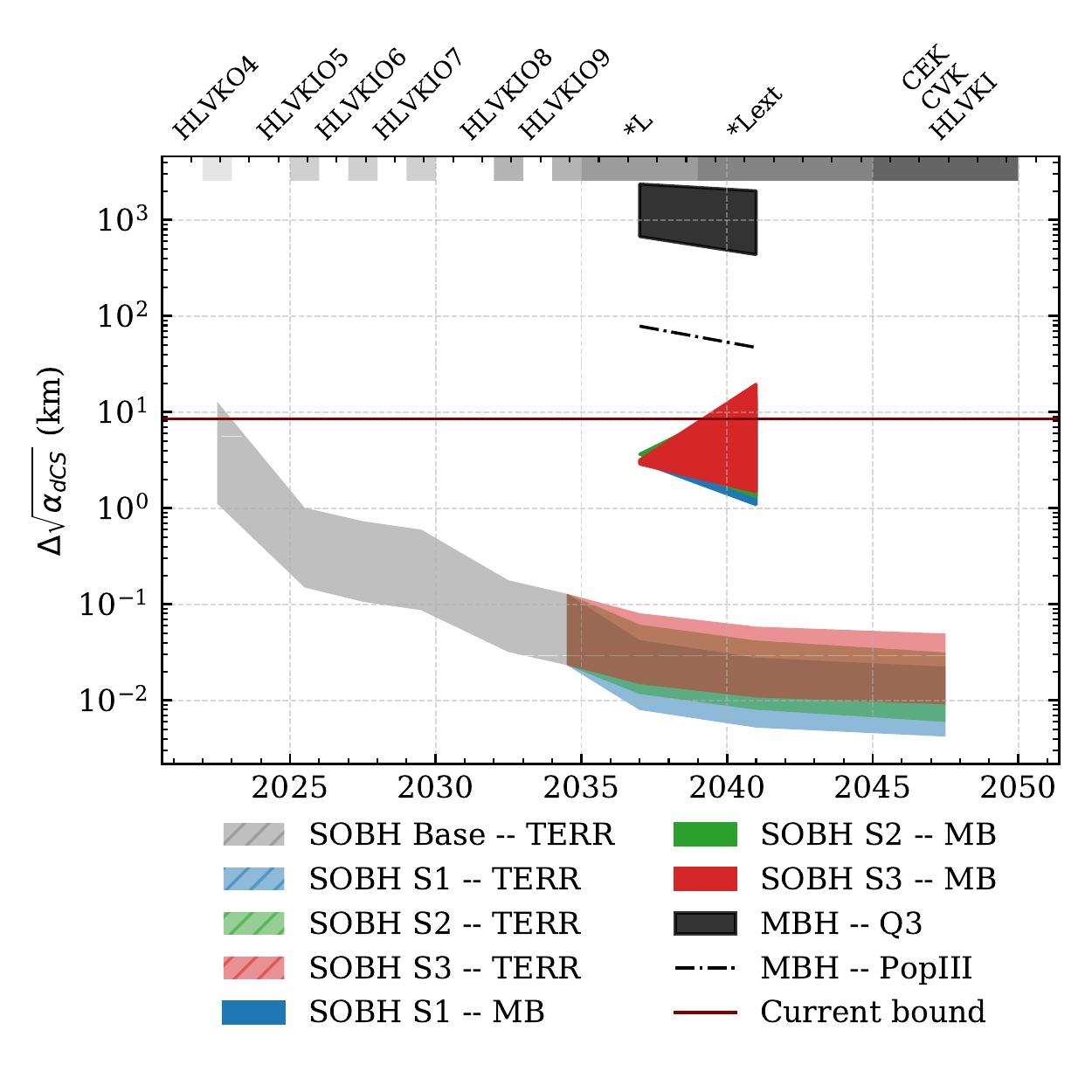}
\caption{
Projected cumulative constraints on $\sqrt{\alpha_{\dCS}}$ for the detector networks and population models examined in this paper. 
Terrestrial-only catalogs, with their populations of millions of sources, dominate any future constraint on this particular deviation, with an improvement of 2-5 orders of magnitude over other source classification.
This conclusion is independent of the terrestrial scenario we pick, with comparable performance from all three.
Multiband sources, with their low chirp masses, seem to perform the next best.
}\label{fig:dcs}
\end{figure}

\begin{figure*}
\includegraphics[width=\linewidth]{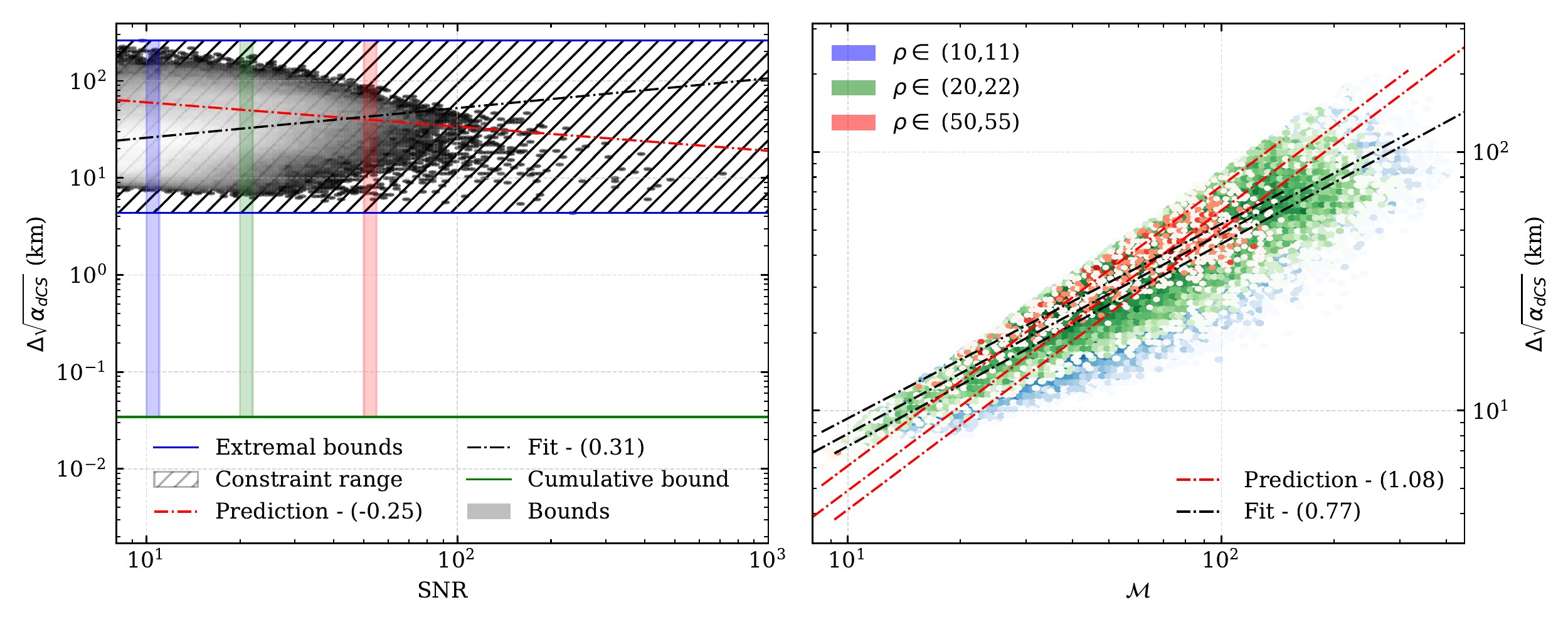}
\caption{
Result of the scaling analysis outlined in Sec.~\ref{sec:res_PV} performed on the data synthesized with the HLVKIO8 network and the SPOPS 0 population.
The plotting style is the same as in Fig.~\ref{fig:dipole_scaling}.
The left panel shows a heat map of the constraint on $ \sqrt{\alpha_{\dCS}}$ versus the SNR of the source. 
The right panel shows the density of the constraint versus the chirp mass, with empirical trends shown in black and predicted trends shown in red.
Our prediction for the SNR scaling is considerably less accurate than for previous theories, presumably from covariances with other source parameters and competing scaling trends with the chirp mass.
The tight range of constraints and large improvement of the cumulative bound over all other single source constraints, seen in the left panel, indicate strong dependence on the total number of sources in the catalog.
}\label{fig:dCS_scaling}
\end{figure*}

\subsubsection{Parity Violation -- Dynamical Chern Simons}\label{sec:res_PV}

One of the fundamental tenets of GR is the parity invariance of the gravitational action.
Dynamical Chern-Simons (dCS) gravity includes a parity-odd, second-order curvature term in the action, known as the Pontryagin density, coupled to a scalar field through a dimensionful parameter $\alpha_{\dCS}$. 
The fact that the Pontryagin density is parity-odd necessarily restricts the scalar field to also be odd in vacuum, making it an axial field.
The leading-order effect in the GW phase sourced by these deviations enters at 2PN order.
In Appendix~\ref{sec:theories} we recall that the following mapping holds: 
\begin{equation}\label{eq:dcs_jac}
\left(\frac{\partial \beta}{\partial \alpha_{\dCS}^2}\right)^{2} \propto \frac{\left[ \hat{m}_1 s_2^{\dCS} - \hat{m}_2 s_1^{\dCS} \right]^4 \eta^{8/5} }{(1+z)^{-8}   \mathcal{M}^8 } \,,
\end{equation}
where $s_i^{\dCS}$ is the BH sensitivity, defined in Eq.~\eqref{eq:dCS_sens}, and $\hat{m}_i = m_i/\mathcal{M} = \eta^{-3/5}(1\pm \sqrt{1-4\eta}) / 2$ for the larger ($+$) and smaller ($-$) mass.
Here, we have only shown the Jacobian to leading order in spin, and we have transformed the mass components to explicitly show the chirp mass dependence.
As the mass ratio and spin factors are bounded to a magnitude of $\order{1}$, the dependence of the Jacobian on $\mathcal{M}^{-8}$ should have the most significant effect on $\Delta\alpha_{\dCS}$ and \emph{strongly} favor low-mass systems, suggesting that SOBHs would be considerably more effective than MBHs. 
Furthermore, as this is a positive PN modification, we would expect to see a sizeable benefit from large catalogs, given the analysis in Sec.~\ref{sec:multiple_source_scaling}, and the impact of LISA observations of the early inspiral should be considerably less important.
All of these factors point to the terrestrial-observation only scenarios outperforming LISA detections of MBH sources and LISA-terrestrial joint detections of multiband sources.

Our predictions for the constraints on the strength of this coupling are shown in Fig.~\ref{fig:dcs}.
Indeed, terrestrial-only detections perform the best at constraining dCS modifications to GW, with bounds up to $\sim 2$ orders of magnitude tighter than multiband sources and $\sim 4\text{-}5$ orders of magnitude better than MBH sources.
As expected, MBH sources detected by LISA are severely inhibited by the particular Jacobian for this specific modification.
Furthermore, we also see little variation between the three terrestrial scenarios, indicating that a significant weight lies with the size of the catalogs, as opposed to the source properties of a select minority of favorable observations.
As the power of constraining this particular modification to GR benefits strongly from large numbers of sources, we can expect to slowly push the current bound down by $\sim 3$ orders of magnitude, with minimal dependence on the actual detector schedule, over the course of the next thirty years.

Further analysis using the techniques in Sec.~\ref{sec:ind_scaling} leads to the following approximate form of the variance:
\begin{align}\nonumber
\Delta \sqrt{\alpha_{\dCS}} &\approx \left(\frac{3584 \sqrt{6}}{5 \pi}\right)^{1/4} \frac{ \left( \pi \mathcal{M} f_{\text{low}} \right)^{1/12} \mathcal{M} }{(1+z)\eta^{1/5} \rho^{1/4}  } \\ \nonumber
\times &\left( |3015 \chi_2^2 \hat{m}_1^2 - 5250 \chi_1 \chi_2 \hat{m}_1 \hat{m}_2 +3015 \chi_1^2 \hat{m}_2^2 \right. \\ 
 &\left. - 14 (\hat{m}_2 s^{\dCS}_1 - \hat{m}_1 s^{\dCS}_2)^2| \right)^{-1/4}\,.
\end{align}
Beyond the additional terms coming from the Jacobian of the parameter transformation, we now see additional deviations from our analysis on generic modifications in Sec.~\ref{sec:general_mod}.
Raising the bound on $\alpha_{\dCS}^2$ to the one-fourth power to obtain our further approximated bound on $\sqrt{\alpha_{\dCS}}$ has introduced new dependence of the constraint on all the source parameters of interest. 
Namely, the dependence on $\rho$ has been amended to scale as $\rho^{-1/4}$, and the dependence on the chirp mass is now $\mathcal{M}^{13/12}$. 

Results related to this analysis are shown in Fig.~\ref{fig:dCS_scaling}, derived from data produced with the HLVKIO8 network and the SPOPS 0 model.
The left panel shows a heat map of the sources in the catalog in the $\Delta \alpha_{\dCS}$--SNR plane, with the extremal bounds shown in blue, and the cumulative bound (for this single catalog) shown in green.
The right panel shows a heat map of the sources in the $\Delta \alpha_{\dCS}$--$\mathcal{M}$ plane for three slices in SNR-range (in red, blue, and green).
The trends we have predicted are shown in red, while the empirically determined trends are shown in black, for both panels.

Starting in the left panel, the range in single-observation constraints on $\sqrt{\alpha_{\dCS}}$ is quite small.
The tight range of the constraints (just 1--2 orders of magnitude between the strongest and weakest constraints) helps to explain the enhanced effectiveness of the terrestrial networks at constraining this modification, as the constraint scales favorably with large numbers of observations.
This is explicitly seen by the sizable improvement of the cumulative constraint over the constraint coming from the strongest single observation.

Furthermore, in the left panel, we see that our prediction for the SNR trend does not accurately reflect what we observe in the synthetic data.
This is in stark contrast with non-commutative gravity, where the modification enters at the same PN order and predicts identical scaling with respect to the SNR. 
Notably, this deviation also occurs in EdGB gravity, detailed below, which has a similarly complicated Jacobian.
The primary differences between the modification introduced by dCS and non-commutative gravity are (i) the scaling of the constraint with respect to the chirp mass, and (ii) covariances between the modified gravity coupling constant and all other sources parameters (such as the spins and mass ratio).

\begin{figure}
\includegraphics[width=\linewidth]{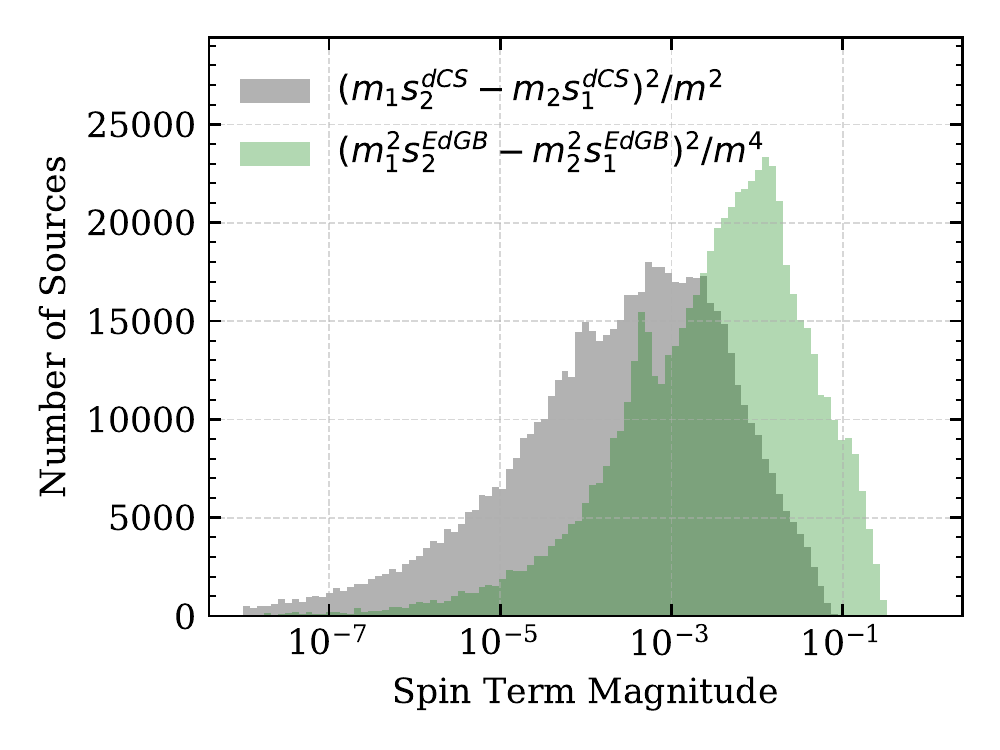}
\caption{
Histogram of spin-related terms contributing to the relevant Fisher element for dCS and EdGB.
The sources were taken from the catalog derived from the HLVKIO8 network and SPOPS 0 population model.
For dCS, this only includes the term to first order in spin.
The wide range of magnitudes that this term can take (5--6 orders of magnitude) helps to explain the breakdown of our ability to predict trends concerning the constraints on these theories.
From Fig.~\ref{fig:dCS_scaling} we see that the SNR and chirp mass only span a range of 1 or 2 orders of magnitude, and as such, the trends we would expect to see for these parameters could be completely washed out by this additional spin-dependent term, which we have neglected in our simple analysis.
}\label{fig:quad_grav_hist}
\end{figure}

\begin{figure}
\includegraphics[width=\linewidth]{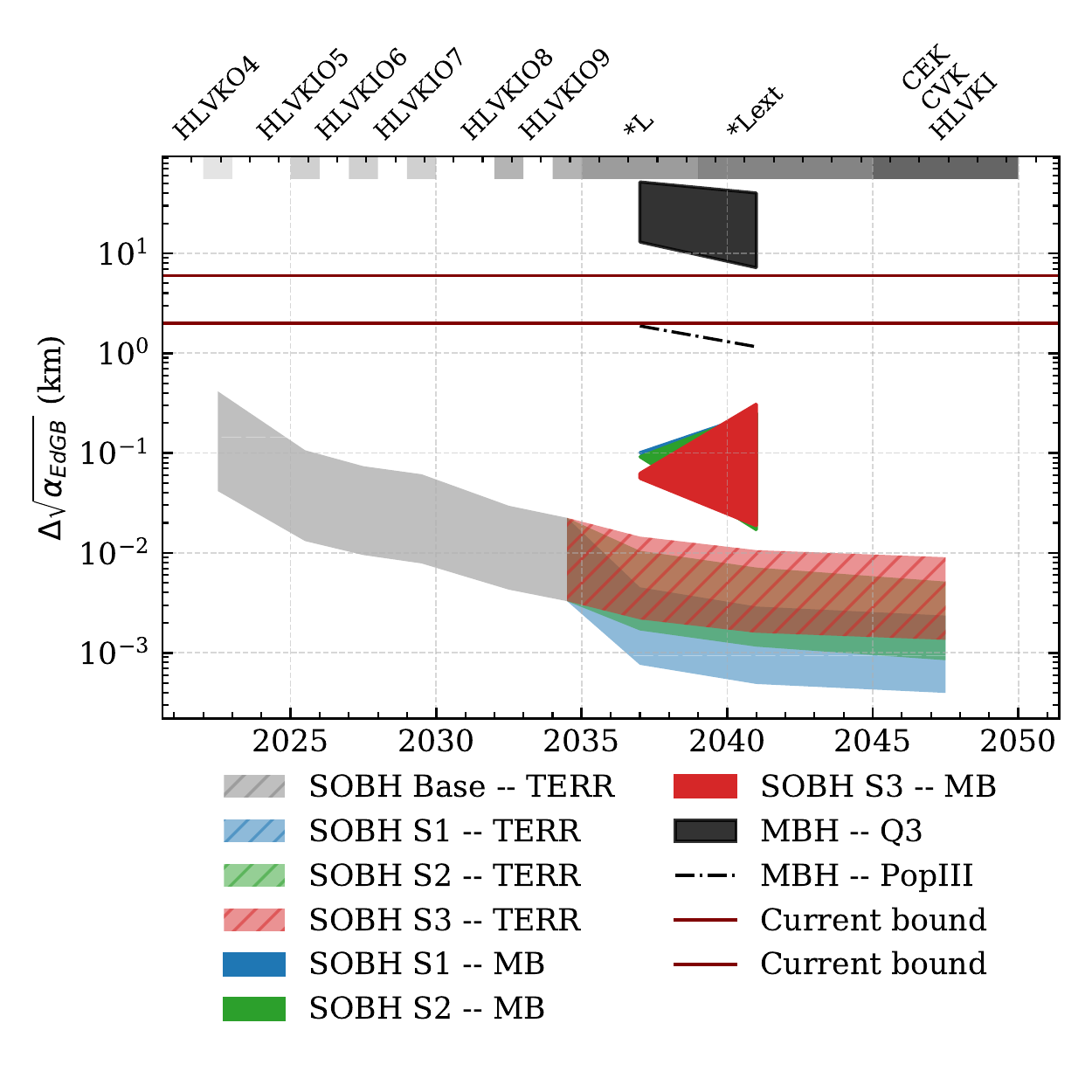}
\caption{
Projected cumulative constraints on $\sqrt{\alpha_{\EdGB}}$ for the detector networks and population models examined in this paper. 
Terrestrial-only catalogs, with their populations of millions of sources, seem to most efficiently constrain EdGB, but multiband sources are not far behind.
The modified scaling of the constraint with SNR and chirp mass work in favor of terrestrial networks, but the fact that EdGB produces a negative PN modification to leading order benefits multiband sources.
MBHs are not effective at constraining EdGB, and will not contribute much to future bounds on this theory.
}\label{fig:edgb}
\end{figure}

\begin{figure*}
\includegraphics[width=\linewidth]{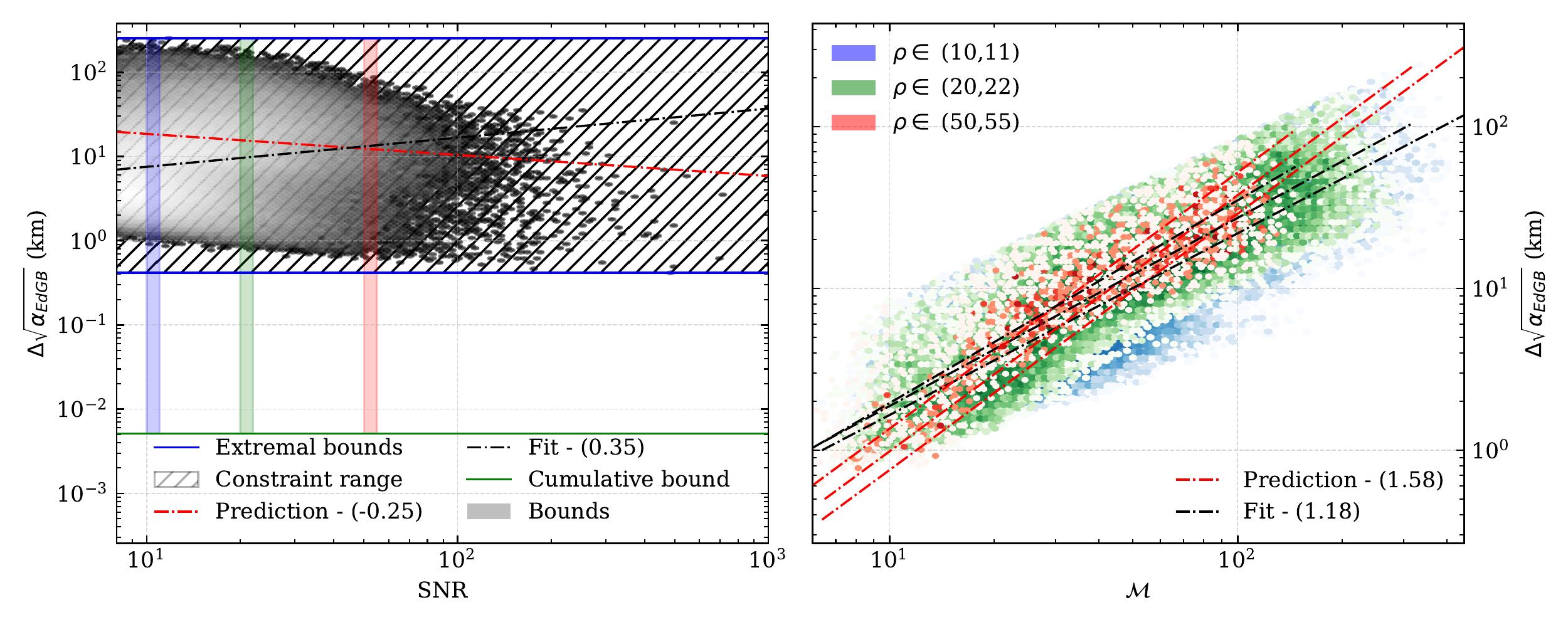}
\caption{
Result of the scaling analysis outlined in Sec.~\ref{sec:res_edgb} performed on the data synthesized with the HLVKIO8 network and the SPOPS 0 population.
The plotting style is the same as in Fig.~\ref{fig:dipole_scaling}.
The left panel shows a heat map of the constraint on $ \sqrt{\alpha_{\EdGB}}$ versus the SNR of the source. 
The right panel shows the density of the constraint versus the chirp mass, with empirical trends shown in black and predicted trends shown in red.
Because of the small range in single-observation constraints (about 1--2 orders of magnitude), the cumulative bound greatly benefits from large numbers of observations, despite this being a negative PN effect that would typically be dominated by a small cadre of favorable sources.
The predicted trend for the constraint-SNR relationship fails, presumably due to covariances introduced through the Jacobian.
The predicted trend for the constraint-$\mathcal{M}$ relationships performs fairly well, as the correlation is enhanced through the Jacobian.
}\label{fig:EdGB_scaling}
\end{figure*}

For difference (i), we can examine the right panel of Fig.~\ref{fig:dCS_scaling}, where we see moderate agreement with our predicted scaling trend for the chirp mass and much tighter correlations for dCS than for non-commutative gravity. 
Not only is the trend more accurately predicted, but the scaling with chirp mass in dCS, as compared with non-commutative gravity, is considerably stronger ($\mathcal{M}^{13/12}$ as opposed to $\mathcal{M}^{1/12}$).
Considering there is a negative correlation between the constraint and the SNR, a positive correlation between the constraint and the chirp mass, and a positive correlation between the SNR and chirp mass, a shift in the different trends as significant as that found in dCS may lead to the observed deterioration in our predictions.

For difference (ii), the mild agreement of the chirp mass scaling in the right panel suggests that covariances between parameters are degrading the accuracy of all of our approximations, not just the SNR. 
To further explore this idea, we can look at the typical range of values that the other source-dependent terms from the Jacobian in Eq.~\eqref{eq:dcs_jac} can take. 
For the final bound from a given source, the magnitude of these additional terms in an absolute sense is important, but in terms of the trends we expect to see, the range of values these terms can take is the quantity of interest.
If certain sources with comparable SNR and chirp mass have Jacobian transformations that span several orders of magnitude because of these additional terms, our simple analytical approximations cannot be expected to accurately match the synthetic data.
A histogram of the spin- and mass ratio-dependent terms for both dCS and EdGB are shown in Fig.~\ref{fig:quad_grav_hist}, where we do indeed see a non-negligible range of values.
Figure~\ref{fig:dCS_scaling} shows that the SNR and chirp mass both span approximately 1--2 orders of magnitude for this particular catalog, while the complicated Jacobian factors that we have neglected in our analysis span approximately 4--5 orders of magnitude.
A range this large can easily erase any structure we would hope to see with our simple approximations, and helps to explain why our simple analytical approximation fails for dCS (and for EdGB, as we will discuss below).

Between these two factors, our ability to predict scaling trends of the constraint on $\sqrt{\alpha_{\dCS}}$ as a function of source parameters has moderate success with regards to the chirp mass, but is definitely degraded in general when compared with the same analysis for general modifications. The dCS example provides direct evidence that conclusions derived from generic constraints may be highly misleading when focusing on a particular modified theory.   

\subsubsection{Quadratic Gravity -- Einstein-dilaton-Gauss-Bonnet}\label{sec:res_edgb}

Similar to dCS, Einstein-dilaton-Gauss-Bonnet (EdGB) gravity is also quadratic in curvature at the level of the action. 
In this case, a scalar field is coupled to the Gauss-Bonnet invariant through a dimensionful coupling constant $\alpha_{\EdGB}$.
In contrast to dCS, the scalar field in EdGB is parity-even in vacuum (because the Gauss-Bonnet invariant is also parity-even), and the leading order correction to the GW phase comes at $-1$PN order, because the dominant modification to the generation of GWs is the introduction of dipolar radiation.
The Jacobian for this particular theory is
\begin{equation}
\left(\frac{\partial \beta}{\partial \alpha_{\EdGB}^2}\right)^{2} \propto  \frac{\left[ \hat{m}_2^2 s_1^{\EdGB} - \hat{m}_1^2 s_2^{\EdGB} \right]^4 \eta^{12/5} }{(1+z)^{-8}  \mathcal{M}^8 } \,,
\end{equation}
where $s_i^{\EdGB}$ is the BH sensitivity defined in Eq.~\eqref{eq:EdGB_sen}, and we again use the mass parameters $\hat{m}_i = m_i/\mathcal{M} = \eta^{-3/5}(1\pm \sqrt{1-4\eta}) / 2$ for the larger ($+$) and smaller ($-$) mass.
Given the new dependencies on source parameters introduced by the Jacobian, we would expect to see SOBH sources receive a sizeable boost due to the chirp mass scaling. 
Furthermore, this is a negative PN effect, which already tends to favor small chirp masses (cf. Sec.~\ref{sec:ind_scaling}).
Both of these considerations imply that multiband and terrestrial networks should outperform LISA MBH sources.

Constraints on $\sqrt{\alpha_{\EdGB}}$ are shown in Fig.~\ref{fig:edgb}.
Indeed, we see SOBH sources of all kinds outperforming MBH sources.
Within the SOBH source classes, terrestrial networks outperform multiband sources by 1--2 orders of magnitude.
While multiband sources benefit from long early inspiral observations from LISA, which encodes much information for a negative PN effect, the large catalogs of sources in the terrestrial-only catalogs are enhanced by the modified dependence on the SNR, discussed below.
As a further consequence of the adjusted SNR dependence, we also see fairly minor variations between the three terrestrial network scenarios.
After approximately thirty years of observations, our models indicate that we could see $\sim 2$--$4$ orders of magnitude improvement on previous constraints on $\sqrt{\alpha_{\EdGB}}$. This conclusion is fairly robust under variations of the terrestrial network.

Analyzing the constraints on $\sqrt{\alpha_{\EdGB}}$ with the machinery of Sec.~\ref{sec:general_mod}, we obtain the following approximation on the variance of the coupling parameter:
\begin{align}\nonumber
\Delta \sqrt{\alpha_{\EdGB}} &\approx \left(\frac{903168}{25\pi^6 }\right)^{1/8} \frac{ \left( \pi \mathcal{M} f_{\text{low}} \right)^{7/12} \mathcal{M}}{(1+z) \eta^{3/10} \rho^{1/4}} \\ 
& \times \left( \hat{m}_2^2 s_1^{\EdGB} - \hat{m}_1^2 s_2^{\EdGB} \right)^{-1/2} \,.
\end{align}
We now see additional modifications to the dependencies on source parameters, beyond the Jacobian shown above.
Just as in the cases of dCS and non-commutative gravity, we must transform from $\alpha_{\EdGB}^2$ to $\sqrt{\alpha_{\EdGB}}$, which forces the constraint to scale with $\rho^{-1/4}$ and $\mathcal{M}^{19/12}$.

Trends related to this approximation are shown in Fig.~\ref{fig:EdGB_scaling}, produced from our simulations based on HLVKIO8 and SPOPS 0.
The left panel shows a heat map of all the sources in the $\Delta \sqrt{\alpha_{\EdGB}}$-SNR plane, with extremal single-source constraints shown in blue, and the cumulative constraint for this catalog shown in green.
The right panel shows a heat map in the $\Delta \sqrt{\alpha_{\EdGB}}$-$\mathcal{M}$ plane, for three different slices of SNR, shown as blue, green, and red.

In the left panel, we again see that our prediction for the SNR scaling is not accurate.
Just as in dCS gravity, this discrepancy lies in covariances complicating the relationships beyond the point where our simple approximations are valid.
For comparison, we can examine what we found for generic dipole radiation constraints in Sec.~\ref{sec:res_dipole}, where we saw a much better agreement with our predictions for the constraint-SNR relationship.
Referring again to the histogram in Fig.~\ref{fig:quad_grav_hist}, we see that the terms related to the BH sensitivity in EdGB span several decades, washing out the trends we would expect to see from the analysis of Sec.~\ref{sec:general_mod}.
As a by product, these complications lead to a tight range in single-observation constraints, spanning 1--2 orders of magnitude. 
This in turn leads to a large enhancement for terrestrial networks: cumulative bounds from tightly grouped populations of constraints benefit from large numbers of sources, which is not typically expected from a modification at $-1$PN.

In the right panel, we see moderate agreement between our prediction for the $\Delta \sqrt{\alpha_{\EdGB}}$--$\mathcal{M}$ relationship, but again, covariances seem to degrade the quality of simple analytical scaling relationships between the constraint and the source parameters. In contrast, for generic dipole radiation we see a much tighter correlation between the constraint and the chirp mass.
The difference between the two trends further confirms our explanation: more complex Jacobians tend to complicate the source parameter-constraint relation we identified in Sec.~\ref{sec:general_mod}.

\begin{figure}
\includegraphics[width=\linewidth]{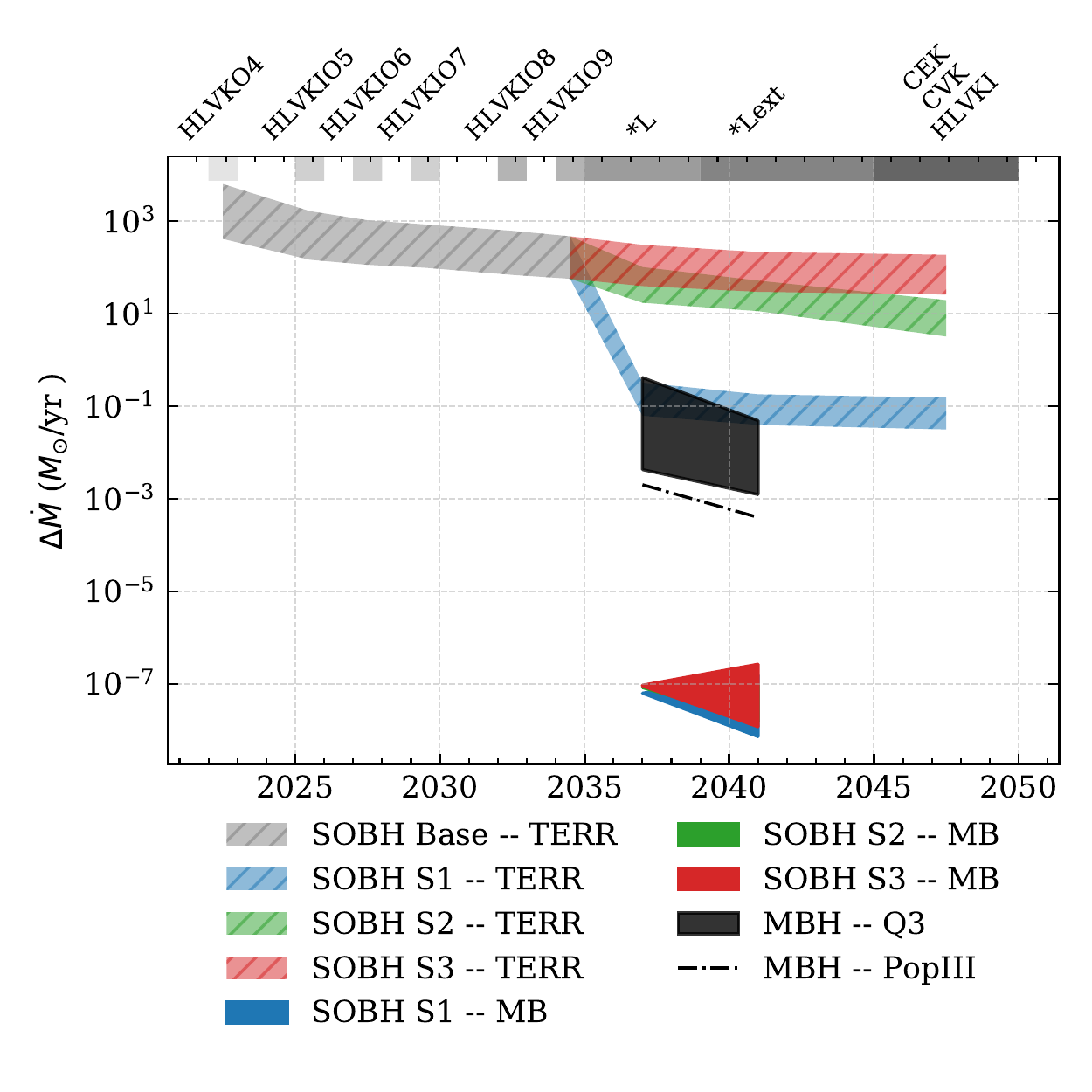}
\caption{
Projected cumulative constraints on the rate of black hole evaporation $\dot{M}$, for the detector networks and population models examined in this paper. 
Our models predict multiband sources to perform the best from the three classes of sources examined in this paper, followed next by MBH observations by LISA. 
Terrestrial-only observations from the most optimistic scenario are competitive with LISA's MBH sources, but the other two scenarios considered in this work trail behind by 2-3 orders of magnitude.
}\label{fig:BHE}
\end{figure}

\begin{figure*}
\includegraphics[width=\linewidth]{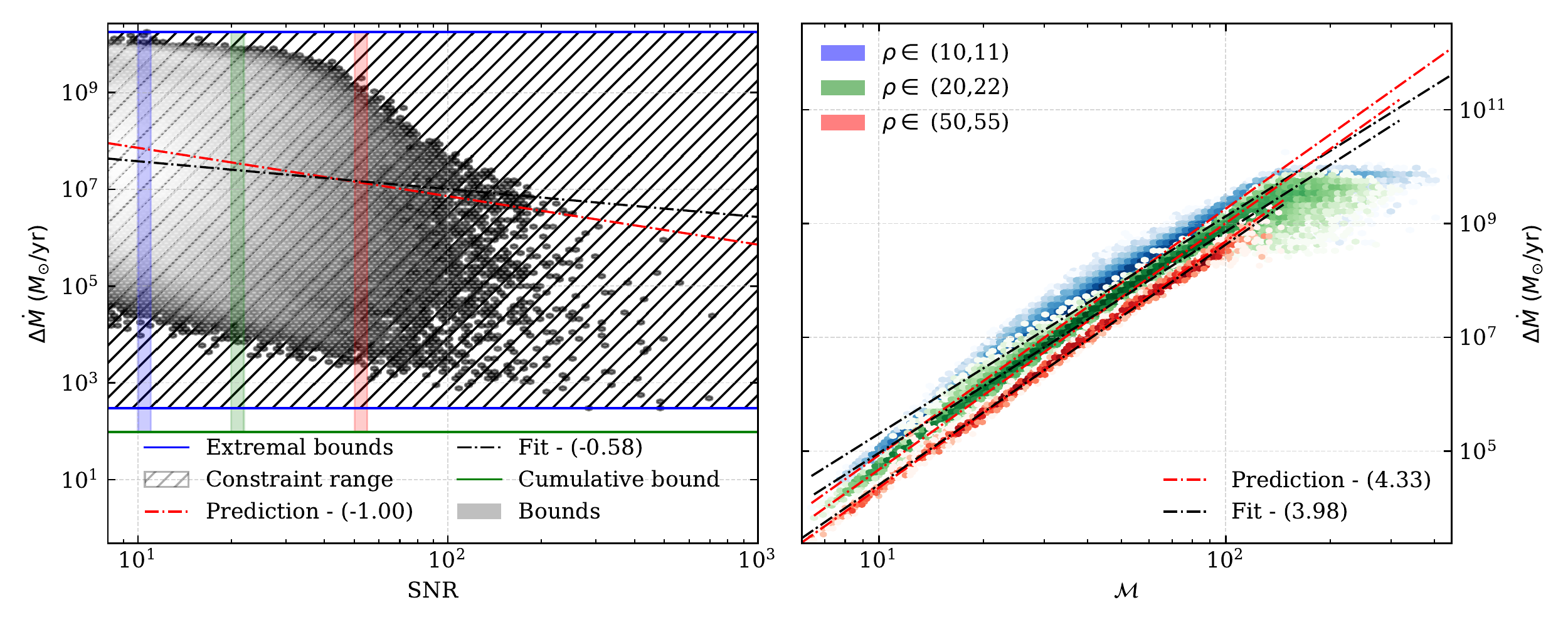}
\caption{
Result of the scaling analysis outlined in Sec.~\ref{sec:res_BHE} performed on the data synthesized with the HLVKIO8 network and the SPOPS 0 population.
The plotting style is the same as in Fig.~\ref{fig:dipole_scaling}.
The left panel shows a heat map of the constraint on $\dot{M}$ versus the SNR of the source. 
The right panel shows the density of the constraint versus the chirp mass, with empirical trends shown in black and predicted trends shown in red.
The wide distribution of constraints in this catalog indicate that the benefit of large catalogs is minimal, and the total bound is dominated by a select few, highly favorable observations.
The distribution of the sources in the $\Delta \dot{M}$-$\mathcal{M}$ plane is to a very good approximation linear, showing a tight correlation between the two quantities.
The $\Delta \dot{M}$-SNR relationship also agrees fairly well with our predictions.
}\label{fig:BHE_scaling}
\end{figure*}

\subsubsection{Black Hole Evaporation}\label{sec:res_BHE}

In the case of BH evaporation, the modification first enters the GW phase at $-4$PN order.
The Jacobian from the ppE parameter to this particular process, as shown in Appendix~\ref{sec:theories}, is given by
\begin{equation}\label{eq:bhe_jac}
\left(\frac{\partial \beta}{\partial \dot{M}}\right)^{2} \propto \left[\frac{3 - 26 \eta + 34\eta^2}{ \eta^{2/5} ( 1 - 2 \eta)}\right]^2 \,.
\end{equation}
As the Jacobian only depends on the system parameters through the symmetric mass ratio (bounded to $(0,0.25]$), no parameters specific to a given system will induce large changes in the attainable bound.
This fact leads us to the conclusion that the driving factors in the constraint magnitude will be the chirp mass (benefitting SOBH sources) and the SNR (benefitting LISA MBH sources and the most sensitive ground-based detector networks).
Furthermore, as this modification also enters at a highly negative PN order, multiband sources can also be expected to perform competitively. 

Constraints on the rate of BH evaporation are show in Fig.~\ref{fig:BHE}.
As expected, multiband sources constrain BH evaporation the tightest, with MBH sources from LISA's catalog trailing by 4-6 orders of magnitude.
The most sensitive terrestrial network scenario examined in this paper is also competitive with the LISA MBH sources, but the other two scenarios we have considered fall behind by 2-3 orders of magnitude.

By using the machinery of Sec.~\ref{sec:general_mod}, we obtain the following approximate form of the bound on $\dot{M}$:
\begin{equation}
\Delta \dot{M} \approx \frac{425984}{5}\sqrt{\frac{6}{5}}\frac{\left(f_{\text{low}} \pi \mathcal{M} \right)^{13/3}\eta^{2/5} }{\rho}\left|\frac{1-2\eta}{3-26\eta + 34 \eta^2}\right|\,.
\end{equation}
The Jacobian does not depend on the total mass and the phase modification scales linearly with the modifying parameter, so we see a scaling relation as expected from Sec.~\ref{sec:general_mod}.

Results related to this approximation are shown in Fig.~\ref{fig:BHE_scaling}.
The left panel depicts a heat map of the sources in the HLVKIO8 network and the SPOPS 0 population model in the $\Delta \dot{M}$--SNR plane.
The solid blue lines correspond to the strongest and weakest constraints coming from single observations, while the green line represents the cumulative bound for the entire catalog.
The right panel shows a heat map in the $\Delta \dot{M}$--$\mathcal{M}$ plane for different slices of SNR (in red, blue, and green).
The empirically determined scaling trends are shown in black, while our predictions for the trends are shown in red.

The left panel of Fig.~\ref{fig:BHE_scaling} shows good agreement between the trends predicted by our simple, analytic calculations and the data from our fully numerical treatment. 
The wide distribution in constraints coming from single sources in the catalog indicates weak scaling with the size of the catalog, giving a relative boost in power to the smaller source populations in the MBH LISA and MB catalogs.
This conclusion is supported by the very modest improvement of the cumulative bound for the catalog over the strongest single-source constraint.
In the right panel, we see good agreement with our predicted chirp mass scaling relation.
The correlation between the chirp mass and the constraint is quite tight for this particular modification, due to the strong scaling and the highly negative PN order (reducing correlations that widen the distribution).

\begin{figure}
\includegraphics[width=\linewidth]{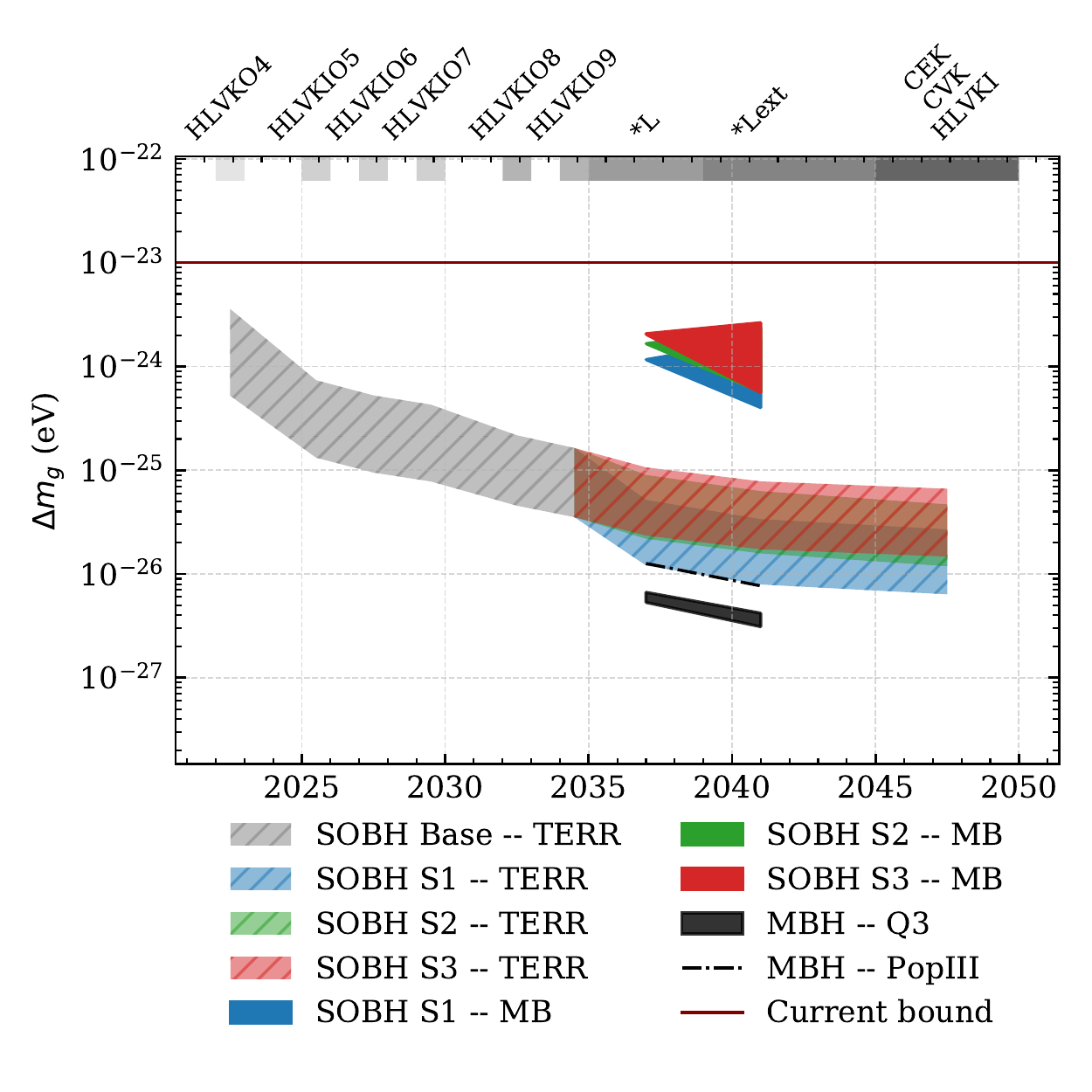}
\caption{
Projected cumulative constraints on the mass of the graviton, $m_g$, for the detector networks and population models examined in this paper. 
Our models show that MBH sources observed by LISA will perform the best at constraining this modification, but only slightly better than the terrestrially-observed only sources.
Multiband sources perform the worst, as they received no benefits from the Jacobian and already perform only moderately well for positive PN order effects.
}\label{fig:mg}
\end{figure}

\begin{figure*}
\includegraphics[width=\linewidth]{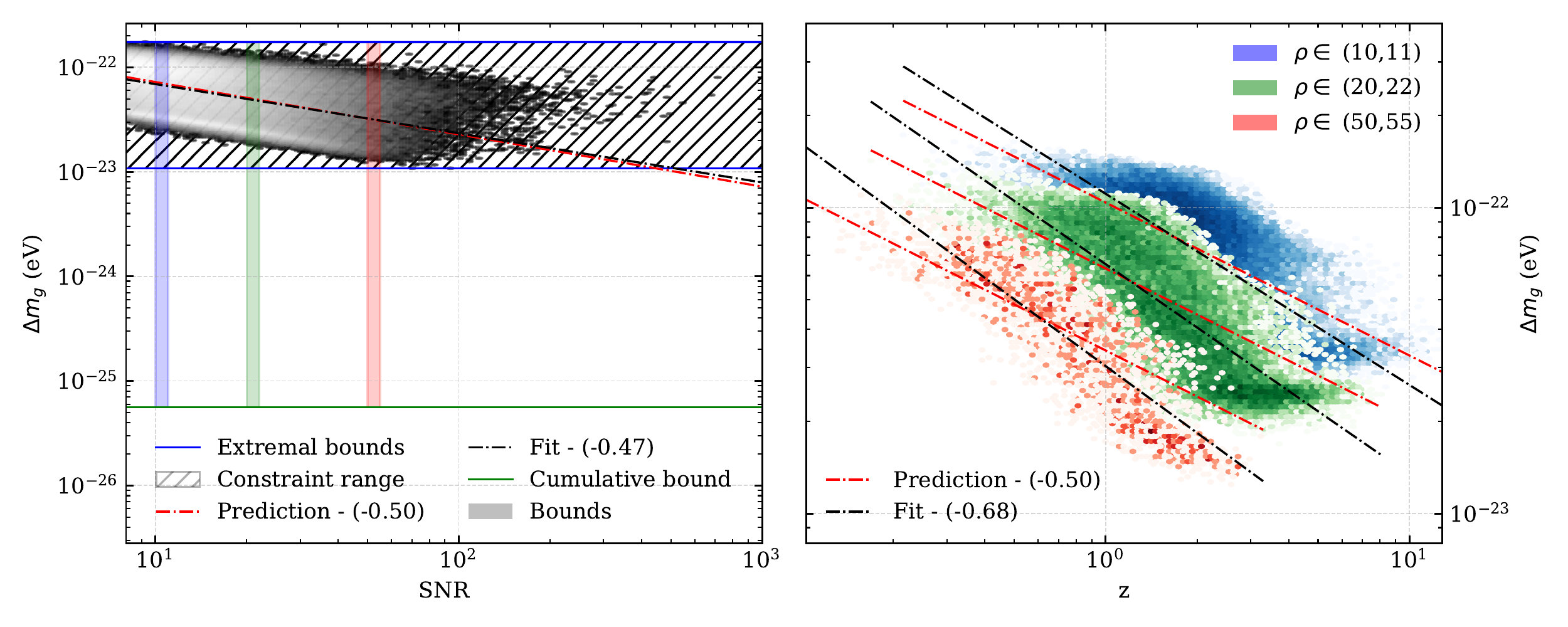}
\caption{
Result of the scaling analysis outlined in Sec.~\ref{sec:res_mg} performed on the data synthesized with the HLVKIO8 network and the SPOPS 0 population.
The plotting style is the same as in Fig.~\ref{fig:dipole_scaling}.
The left panel shows a heat map of the constraint on $m_g$ versus the SNR of the source. 
The right panel shows the density of the constraint versus the redshift $z$, with empirical trends shown in black and predicted trends shown in red.
Because of the narrow range of constraints in the catalog and the large enhancement of the cumulative bound over the strongest single observation, stacking observations is quite efficient for this modification.
The right panel shows that there is indeed a trend in the $\Delta m_g$-$z$ relation (although the distributions are moderately wide)
which would favor sources far from Earth, and would primarily benefit MBH sources.
}\label{fig:mg_scaling}
\end{figure*}

\subsubsection{Modified Dispersion -- Massive Graviton}\label{sec:res_mg}

If the graviton were massive, contrary to what is predicted when considering GR as the classical limit of a quantum theory of gravity, the leading order effect would enter the GW phase at 1PN.
The Jacobian of the transformation from the ppE framework to this particular modification is 
\begin{equation}
\left(\frac{\partial \beta}{\partial m_g^2}\right)^{2} \propto \left( \frac{\mathcal{M} D_0}{1+z}\right)^2  \,,
\end{equation}
where the quantity $D_0$ is a new cosmological distance defined in Appendix~\ref{sec:theories}.
We get modified scaling with the chirp mass, and similarly to the variable-$G$ mapping, this Jacobian causes the constraint to inversely scale with the mass.
As a result, this new mass factor will benefit MBHs over SOBHs.
Furthermore, we now have strong dependence on the distance to the source, $D_0$, where constraints from farther sources will be enhanced as compared to those sources closer to Earth (see e.g.~\cite{Berti:2004bd}).
These facts benefit LISA MBH sources, which therefore should provide the best constraints.

This is confirmed in Fig.~\ref{fig:mg}. 
The MBH sources observed by LISA do indeed perform the best, but only marginally. 
The effectiveness of stacking is seen to still be quite high for this particular modification, as the three terrestrial scenarios all perform comparably.
Furthermore, as this is a positive PN effect, terrestrial networks receive a boost from the generic scaling effects discussed in Sec.~\ref{sec:ind_scaling}. 
Multiband sources perform the worst, as they receive little benefit from early inspiral observation, they typically have low mass, and are located at low redshifts.
Ultimately, we can expect to improve on the current bound on $m_g$ by 2--3 orders of magnitude over the next thirty years, and this conclusion is robust under variations of the terrestrial detector schedule.
This improvement will be insufficient to rule out a massive graviton as a possible explanation of the late-time acceleration of the Universe: in a cosmological context, the graviton would need a mass of the order of the inverse of the Hubble constant, $H_0^{-1}$, which is of the order of $10^{-30}$ eV, much smaller than our predicted final constraints.

To explore these relations deeper, we can apply our approximation from Sec.~\ref{sec:general_mod}, giving us the following approximation for the constraint on $m_g$:
\begin{equation}
\Delta m_g \approx \frac{h}{\pi}\left( \frac{5}{2} \right)^{1/4} \sqrt{ \frac{(1+z)}{D_0} \frac{ \pi f_{\text{low}}}{\rho}}\,.
\end{equation}
This approximation has produced a notably different scaling relation than what has been seen previously.
Namely, the constraint no longer scales with the chirp mass, as the Jacobian factor has cancelled the chirp mass dependence from the generic ppE scaling.
While this final form of the constraint does not explicitly benefit MBH systems, generic constraints scale with the chirp mass as $\mathcal{M}$.
The removal of this chirp mass dependence benefits MBH sources much more than SOBH sources.
Also different from previous constraints, we have strong scaling with the distance to the source.
For low redshifts, the distance parameter $D_0 \approx z H_0$ to lowest order in redshift.
Extending this expansion to the constraint, the leading-order term should scale as $z^{-1/2}$ for low-redshift sources.

The results related to this approximation are shown in Fig.~\ref{fig:mg_scaling}.
The left panel shows a heat map of the sources in the catalog created from the HLVKIO8 network and SPOPS 0 population model in the $\Delta m_g$-SNR plane, with the solid blue lines denoting the extremal, single observation constraints. 
The solid green line represents the cumulative bound from this particular catalog.
We see good agreement between our predicted scaling for the SNR, after accounting for the Jacobian above. 
There is a narrow range for the constraints, only spanning one order of magnitude between all sources.
This leads to sizeable benefits for large catalogs, also evident from the overlap between the different terrestrial network scenarios.

The right panel shows a heat map of the sources in the $\Delta m_g$-redshift plane.
We do indeed see a trend in this particular relationship, although the distributions are moderately wide.
Our predictions for the scaling relation agrees fairly well with the synthetic data.

\subsection{Effect of Precession on the Constraints}\label{sec:precession}

The differences between the two SOBH population models go beyond the size of the catalogs, which has been our focus so far.
An aspect differentiating the SPOPS 0 and SPOPS 265 catalogs, that could have a large impact on our analysis, is the typical magnitude of the in-plane component of the binary's spins, which is the cause of relativistic precession.
The question we now address is whether the stronger constraints coming from the SPOPS 0 catalog over the SPOPS 265 catalog are entirely due to the larger catalog sizes, or if the difference in source parameter distributions also impacts the cumulative bounds attainable through GWs.

\begin{figure}
\includegraphics[width=\linewidth]{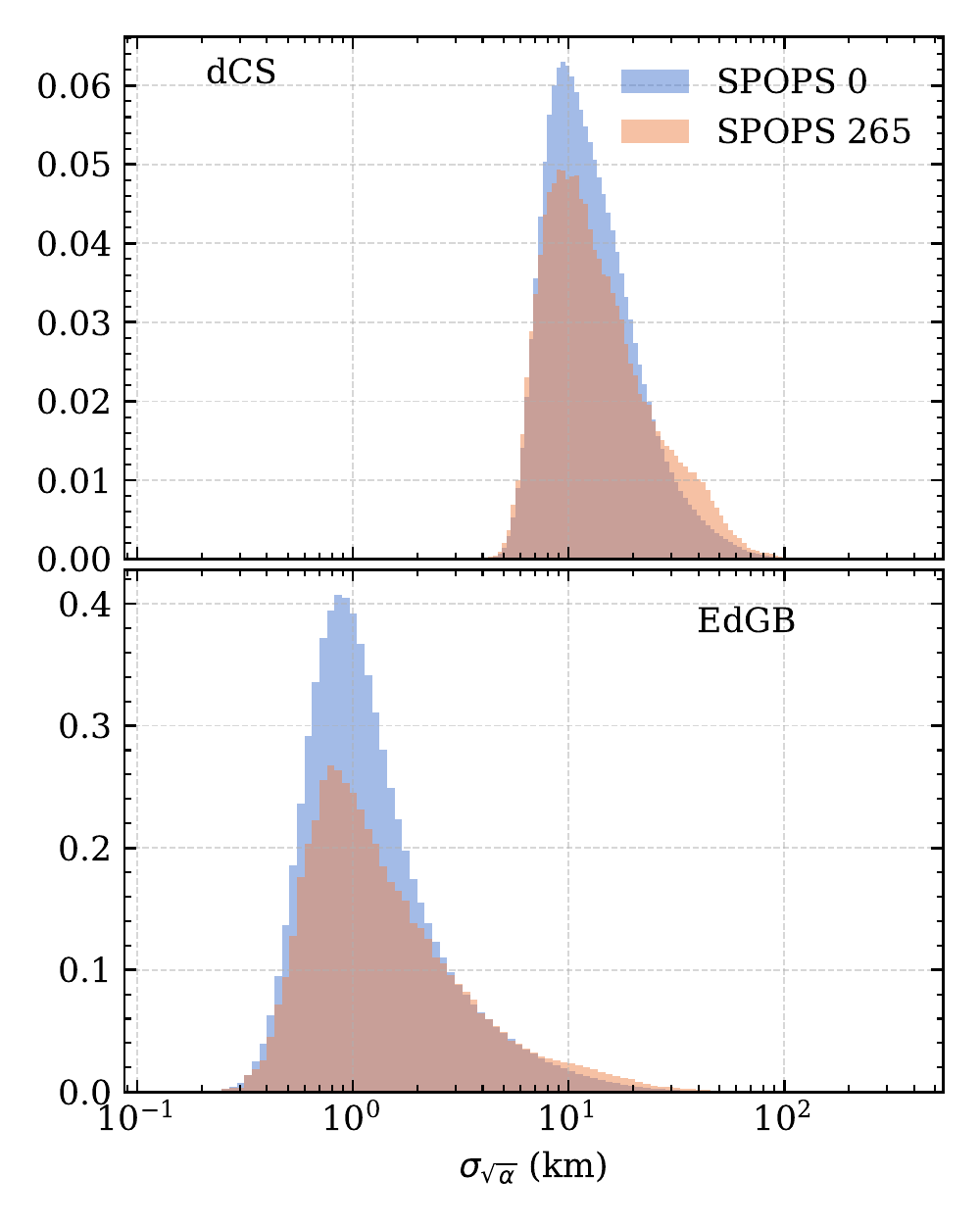}
\caption{
Distributions of single-source constraints on the GR-modifying parameters $\sqrt{\alpha}_{\dCS}$ (top) and  $\sqrt{\alpha}_{\EdGB}$ (bottom) from the two population models SPOPS 0 (blue) and SPOPS 265 (orange) as detected by the CEK network.
The histograms are normalized to provide a comparison of the shapes of the distributions, as opposed to the raw numbers of sources.
We see that the distributions only diverge slightly, towards the larger-constraint side of the spectrum.
This suggests that the larger precessional effects seen in the SPOPS 265 catalog do not significantly modify the typical constraints attainable by individual sources, or that any effect we may have seen was washed out by the differences in the distributions of other source parameters, such as the total mass and mass ratio.
This lack of difference could also be an artifact of our waveform model (\software{IMRPhenomPv2}), which is not the most up-to-date waveform available, or of the Fisher approximation, which could be improved upon by a full MCMC analysis.
}\label{fig:spops_comp}
\end{figure}

Previous work has shown that the inclusion of precessional effects can break degeneracies in various source parameters when considering a full MCMC analysis, allowing for significantly tighter constraints on various source properties~\cite{Chatziioannou:2014bma}.
To determine if this effect can be seen in our data, in Fig.~\ref{fig:spops_comp} we show histograms of the individual source constraints on dCS and EdGB, using the two different catalogs (SPOPS 0 and SPOPS 265) and the CEK network.
These two theories in particular were chosen because conventional thinking would suggest that they would be the most sensitive to precessional effects, due to the dependence of the ppE parameter on spins. 

The figure shows little deviation between the two population models for these theories.
The distribution changes slightly on the larger-constraint side of the histogram, but the difference is negligible when considering cumulative constraints.
Furthermore, these small deviations in the distributions of constraints cannot be solely attributable to precessional effects, as the parameter distributions shown in Fig.~\ref{fig:source_properties} are all modified as well.

\begin{figure}
\includegraphics[width=\linewidth]{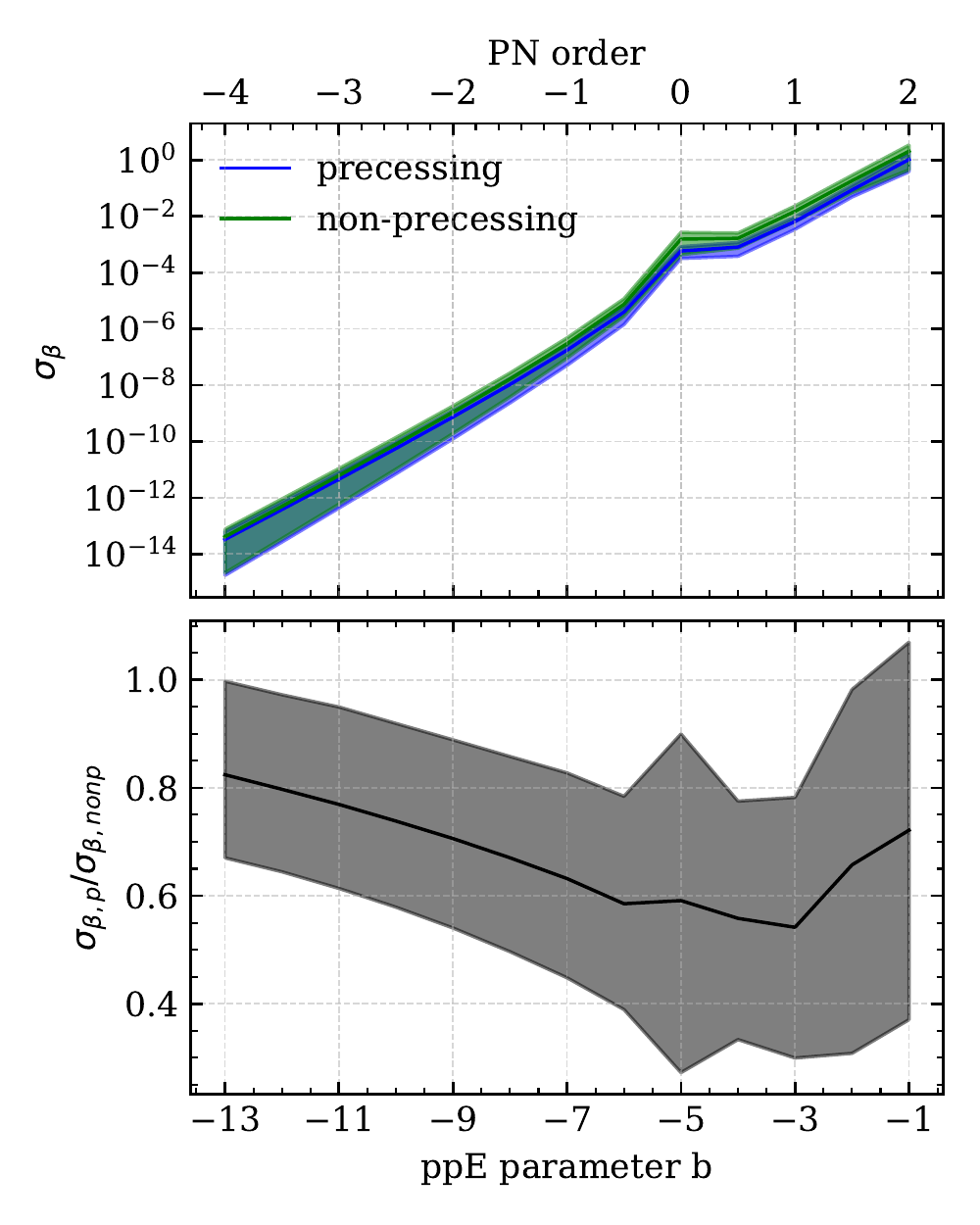}
\caption{
To create the data involved in this figure, we have created a set grid in parameter space with total mass ranging from 5\,$M_{\odot}$ to 20\,$M_{\odot}$, mass ratio in the range 0.05 to 1, and aligned-spin components for each binary ranging from $-0.8$ to $0.8$.
The rest of the parameters were populated with random numbers in the usual ranges, and the luminosity distance was set such that the typical SNRs ranged from $\sim 20$--$150$.
We computed Fisher matrices for each set of parameters, with the in-plane component of the spin set to 0, and then we recomputed them setting the in-plane component of the spin $\chi_p = \sqrt{1 - \chi_1^2}$, so that the binary is approximately ``maximally'' precessing.
The top panel shows the distribution of the bounds for the two binary subsets -- precessing (blue) and non-precessing (green) -- as a function of PN order.
The solid line denotes the average of the synthetic catalog, while the shaded region denotes the $1\sigma$ interval.
The lower panel shows the ratio $\sigma_{\beta,\,\rm p}/\sigma_{\beta,\,\rm nonp}$.
Each ratio is calculated for a single parameter set, and the mean of these ratios is shown as a solid black line, with the $1\sigma$ spread shown by the shading.
Even in this more extreme comparison, the improvement in constraint as the result of larger precession effects only amounts to a factor of $\sim2$.
However, more drastic difference may be possible if we performed a full MCMC analysis, or if we used different waveform models.
}\label{fig:precession_investigation}
\end{figure}

To explore the impact of precession on generic modifications in a more controlled environment, we did a direct comparison between systems with zero precession and ``maximal'' precession (in a sense to be defined shortly), but which are otherwise identical.
The results of this analysis are shown in Fig.~\ref{fig:precession_investigation}.
The methodology we implemented to produce Fig.~\ref{fig:precession_investigation} began with a set grid in the total mass, ranging from 5\,$M_{\odot}$ to 20\,$M_{\odot}$, mass ratio in the range [0.05,\,1], and aligned-spin components for each BH ranging from $-0.8$ to $0.8$.
With this grid of intrinsic source parameters, we populated the other extrinsic parameters using randomly generated numbers in the conventional ranges.
The range on the luminosity distance was chosen such that the SNRs would range from $\sim~$20 to 150.
Once a set of full parameter vectors had been created, we calculated one set of Fishers for a fixed detector network with the in-plane component of the spin, $\chi_p$, set to 0.
Then, without changing any other parameters, the in-plane spin component was increased to $\chi_p = \sqrt{1 - \chi_1^2}$, which is approximately the maximal spin one can achieve while still maintaining a total spin magnitude less than 1.
The top panel shows the mean constraint for both configurations as a solid line, with the $1\sigma$ interval of the distribution of constraints shown as the shaded region.
In the bottom panel we compare the constraints from each configuration (precessing and non-precessing) for each individual source. 
The mean of this ratio is then plotted as a solid line, and the $1\sigma$ region is shown as the shaded region.

The conclusion from Fig.~\ref{fig:precession_investigation} is that precession seems to have a moderate influence, but one that could be easily washed out by other physical effects. 
In the most favorable scenario where the binary is maximally precessing, our analysis suggests an improvement of at most a factor of $\sim2$.
Given previous work (see e.g.~\cite{Chatziioannou:2014bma}), one may expect more significant improvements when considering even mild precession.
While we do predict improvements from the use of precessing templates, our more restrained conclusions could be the result of two facets of our analysis.
Our use of a more rudimentary statistical model, the Fisher matrix, does not capture all the more nuanced artifacts in the posterior surface, like a full MCMC analysis would. 
Furthermore, we here use the \software{IMRPhenomPv2} waveform, which is in some ways more limited in modeling precession with respect to the waveforms used in Ref.~\cite{Chatziioannou:2014bma}.
Future studies of precession could focus on these two areas in particular.

\section{Conclusions}\label{sec:conclusions}

In this work, we have constructed forecasts of what constraints can be placed on a variety of modifications to GR, both generic and theory-specific, using astrophysical population models and the most current projections for detector development over the next thirty years.
Our analysis spans several topics of interest to the GW community concerned with tests of GR. 

We investigate what fundamental physics can be done with a variety of source populations (heavy-seed MBHs, light-seed MBHs, terrestrially observed SOBHs, and multiband SOBHs) and plans for detector development.
All of these aspects are connected to what fundamental science is achievable. Ours is the first robust study of this breadth and scope that is capable of quantifying the effects of detector development choices and astrophysical uncertainties.

We identify trends and scaling relationships of constraints for individual GW observations, studying how they evolve with PN order and how they depend on the target source class (MBHs, terrestrially observed SOBHs, and multiband SOBHs).
We also quantify the effect of combining constraints from a full, synthetic catalog, appropriately informed by robust population models.
We find that the effectiveness of stacking observations is a PN-dependent conclusion. The techniques developed here have important implications for the future of GW-based tests of GR, especially in the era of 3g detectors.
The two components of our analysis (individual scaling and studies of the stacking of multiple observations) combine to create a full picture of some of the most important aspects involved in testing GR with GWs.
We hope that this information will be valuable in driving design choices for future detector development. 

We map our generic constraints to theory-specific constraints, where we analyze specific parameters in viable, interesting theories.
Repeating some of the scaling analysis done in previous sections leads, in some cases, to a reversal of the conclusions drawn for generic modifications.
This reinforces the need to incorporate theory-specific waveforms in future analyses, when available. 

This work opens up several new avenues of research.
We focused on BBH systems, neglecting future contributions from neutron star-neutron star and neutron star-BH binaries.
These binaries have much longer inspiral signals relative to typical BH mergers observed by the LVC, and they could provide crucial information concerning early inspiral, negative PN effects.
Beyond the signal length, neutron stars are sometimes treated on unequal footing in the context of specific theories, such as scalar-tensor gravity, EdGB and dCS.
This could provide other insights into specific theories that do not affect BBH mergers.

Because of the scale of the catalogs involved we used simple Fisher matrix forecasts, running $\sim 10^8$ Fisher matrix calculations. 
A more thorough analysis using MCMC, or other more robust data analysis techniques, could provide more information about some of the trends we have identified.
An MCMC population study on the scale of this work is currently intractable, but even an analysis of a subset of sources could be enlightening.

Our work has focused on estimating only a single PN modification at a time, but any modified theory of gravity will correct the waveform at all orders in a PN expansion. Recent work studied how constraints are affected when one attempts to simultaneously constrain ppE deformations that enter at multiple orders~\cite{Gupta:2020lxa}. Here we have chosen to limit ourselves to a single parameter at a time for the following reasons. While allowing for multiple parameters to vary in a completely independent way at several PN orders is a more robust and general framework, this treatment is probably overly pessimistic. Past work~\cite{Sampson:2013lpa,Arun:2006yw} showed that, indeed, varying multiple generic parameters simultaneously drastically lowers our ability to constrain them. However, in the context of a given, physically motivated theory there should be some relation between the different ppE modifications. Any PN expansion should converge in the appropriate domains, ensuring a hierarchy on the size of the modifications. Moreover, the modification at each PN order should at least depend on the coupling parameters of the theory, ensuring that no two PN orders are totally independent from each other. These criteria suggest that the overall bound on a given modification, in the context of a physically motivated theory, should not be significantly weakened by the inclusion of higher-order corrections (except in the most unfortunate of fine-tuning scenarios). Therefore our conclusions should be robust under the inclusion of higher-order PN corrections to the waveform. 

Our investigation of the effects of precession on modified GR constraints could be improved in at least three ways.
While we did include a full inspiral/merger/ringdown model of precession by implementing  \software{IMRPhenomPv2}~\cite{Hannam:2013oca,Khan:2015jqa,Husa:2015iqa}, more recent and complex waveform models (such as \software{IMRPhenomPv3}~\cite{Khan:2018fmp}, \software{IMRPhenomXPHM}~\cite{Pratten:2020ceb} or \software{SEOBNRv4PHM}~\cite{Ossokine:2020kjp}) could encode more information in the signal, helping to break degeneracies.
A more robust statistical analysis, such as a full MCMC, could explore the posterior space more thoroughly, shedding light on the effects of precession.
Last but not least, the astrophysical SOBH models considered here only allow for isolated field formation under restrictive assumptions. Dynamical formation generally predicts a larger fraction of precessing systems~\cite{Rodriguez:2016vmx}, and it is important to consider other pathways for producing BBHs with large misaligned spins even within the isolated formation channel~\cite{Gerosa:2018wbw,Steinle:2020xej}.

\acknowledgments

We thank Vishal Baibhav, Davide Gerosa, Gabriela Gonz\'alez, Bangalore Sathyaprakash, Sashwat Tanay and Kaze Wong for many useful discussions on various aspects of this work.
N.Y. acknowledges support from NSF Grants No. PHY-1759615, PHY-1949838 and NASA ATP Grant No. 17-ATP17-0225.
E.B. is supported by NSF Grants No. PHY-1912550 and AST-2006538, NASA ATP Grants No. 17-ATP17-0225 and 19-ATP19-0051, and NSF-XSEDE Grant No. PHY-090003. This work has received funding from the European Union’s Horizon 2020 research and innovation programme under the Marie Skłodowska-Curie grant agreement No. 690904. The authors would like to acknowledge networking support by the GWverse COST Action CA16104, ``Black holes, gravitational waves and fundamental physics.''
This work made use of the Illinois Campus Cluster, a computing resource that is operated by the Illinois Campus Cluster Program (ICCP) in conjunction with the National Center for Supercomputing Applications (NCSA) and which is supported by funds from the University of Illinois at Urbana-Champaign. It also used computational resources at the Maryland Advanced Research Computing Center (MARCC). 
The following software libraries were used at various stages in the analysis for this work, in addition to the packages explicitly mentioned above: \software{GSL}~\cite{gough2009gnu}, \software{numpy}, \software{scipy}, \software{filltex}~\cite{2017JOSS....2..222G}.

\appendix

\section{Bayesian Theory and Fisher Analysis Details}
\label{app:Fisher}

Signal analysis in GW science is usually based on Bayes' theorem:
\begin{equation}
p(\vec{\theta},d) = \frac{p(d,\vec{\theta}) p(\vec{\theta})}{p(d)}\,,
\end{equation}
where $p(\vec{\theta},d)$ is the posterior probability of the vector of parameters $\vec{\theta}$ given some data set $d$. 
The quantity $p(\vec{\theta})$ is the prior information about the source parameters, reflecting any initial beliefs held before the data was taken.
The evidence, $p(d)$, is the normalization of the posterior, which also generally holds valuable information about the signal, but will not be the focus of this work.
The quantity $p(d,\vec{\theta})$ is the likelihood of the data, and describes the probability that one would see a data set $d$ given some some set of parameters $\vec{\theta}$. 
For GW data analysis, this is given by
\begin{equation}\label{eq:likelihood}
p(d,\vec{\theta}) \propto \exp \left[-\frac{1}{2}\sum_i^{N_{\rm detector}} (d_i-h_i|d_i-h_i) \right] \,,
\end{equation}
for each data series $d_i$ and detector response template $h_i$ from the $i$-th detector, where the noise-weighted inner product is given by 
\begin{equation}\label{eq:inner_product2}
(d-h|d-h) = 4 \Re \left[ \int \frac{ (d-h) (d-h)^{\ast}}{S_n (f)} df \right]\,.
\end{equation}

To estimate the posterior using real data from LIGO, one would use a Markov Chain Monte Carlo ~\cite{Abbott:2016blz,LIGOScientific:2019hgc} to explore the parameter space of the signal.
This would yield a set of independent samples from the posterior that quantifies not only the most likely values for the vector $\vec{\theta}$, but also includes information about our confidence in those estimates.
This approach is the most reliable and accurate, but it is too computationally expensive for our purposes.
Even the most optimized algorithms would take considerable computational resources to analyze the number of sources examined in this paper.
We therefore turn to a commonly used approximation of the posterior to estimate the confidence intervals on $\vec{\theta}$ that is much more computationally tractable: the Fisher information matrix.

We calculate the Fisher matrices for each detector and combine them to construct a total Fisher matrix for each source according to Eq.~\eqref{eq:total_cov}.
To properly reflect the ability of a terrestrial network to localize sources in the sky, we incorporate a time delay between detectors that is $\alpha$- and $\delta$-dependent.
That is, for each detector besides the reference detector, we append the following factor to the phase:
\begin{equation}
t_{c,i} \rightarrow t_{c,\text{ref}} + \delta t_{c,i}(\alpha,\delta)\,,
\end{equation}
where $\delta t_{c,i}$ is defined as 
\begin{equation}
\delta t_{c,i}(\alpha,\delta) = \frac{\mathbf{x}_{\text{ref}} \cdot \mathbf{\hat{x}}_{\text{source}}(\alpha, \delta)  - \mathbf{x}_{i} \cdot \mathbf{\hat{x}}_{\text{source}}(\alpha,\delta)}{c}\,.
\end{equation}
The detector positions $\mathbf{x}_{\text{ref}}$ and $\mathbf{x}_{i}$ are in Earth-centered coordinates, the unit vector $\mathbf{\hat{x}}_{\text{source}}$ points to the source in the sky in the same coordinates,
and we have reintroduced the speed of light $c$ for clarity. 
The positions of the detectors in these Earth-centered coordinates were taken from \software{LALSuite}~\cite{lalsuite}.
This procedure is neglected when considering LISA, as sky localization comes from the orbital motion of the satellites and long signal durations for space-based detectors.

An additional concern in the context of utilizing Fisher metrices with consistent parameters is the description of the binary's orientation. 
There are three coordinate systems that naturally arise in the description of terrestrial and space detectors. 
The natural coordinate system to use for LISA is the ecliptic coordinate system, specifically the parameters $\theta_j$ and $\phi_j$, as these are the quantities that show up in LISA's response function. 
For terrestrial detectors, the polarization angle $\bar{\psi}$ and the inclination angle $\iota$ naturally arise in the response function, where the polarization angle is naturally defined in the equatorial coordinate system.
Finally, the source properties themselves are stipulated in the source frame, aligned with the orbital angular momentum $\mathbf{L}$, and subsequently used to calculate the waveform.
Any choice is valid as long as it is consistently enforced, so we chose to use the equatorial coordinates, and we accounted for the coordinate transformation in the calculation of the derivative of the response function. 
An equally simple solution would be to compute the Fisher matrices in their respective, natural coordinates, then use the Jacobian matrix to transform them as follows:
\begin{equation}
\Gamma_{i'j'} = \frac{\partial x^i}{\partial x^{i'}} \Gamma_{ij} \frac{\partial x^j}{\partial x^{j'}}\,,
\end{equation}
which is exactly how we transform our bounds on generic modifications to theory-specific modifications.

The actual transformation relies on the construction of an explicit rotation matrix between the different frames of reference. 
Transforming between ecliptic and equatorial coordinates is a trivial rotation by a constant angle, so we will instead just describe the transformation between the source frame and the equatorial system.

The first frame in question is the equatorial frame, which is the frame that defines the parameters $\theta_{\rm L}$, $\phi_{\rm L}$, $\alpha$, and $\delta$. 
From these quantities, one can construct two vectors: the direction of propagation $\mathbf{\hat{N}}$ (which points from the solar system to the source), and the direction of the orbital angular momentum $\mathbf{\hat{L}}$ at some reference frequency. 
These two vectors also define the inclination angle of the orbital angular momentum
\begin{equation}\label{eq:frames_iota}
\cos \iota = - \mathbf{\hat{L}}\cdot \mathbf{\hat{N}}\,,
\end{equation}
which will be needed in the next frame.

The second frame is the source frame, in which the waveform is naturally constructed. 
This frame is defined by a coordinate system with $\mathbf{\hat{L}} = \mathbf{\hat{z}}$, while
the other two Cartesian axes are chosen such that the direction of propagation $-\mathbf{\hat{N}}$ (where $\mathbf{\hat{N}}$ points from the solar system to the source) lies in the $x$-$z$ plane when the reference phase $\phi_{\text{ref}}=0$.
The vector $\mathbf{\hat{N}}$ is then rotated azimuthally by an angle $\phi_{\text{ref}}$ for nonzero reference phases.
The angle between $\mathbf{\hat{L}}$ and $\mathbf{\hat{N}}$ in the source frame is just the inclination defined in Eq.~\eqref{eq:frames_iota}, which fully specifies this vector in the second frame.
Using these two vectors, we can construct a third, orthogonal vector as the cross product of these two, which we will call $\mathbf{\hat{K}} = \mathbf{\hat{L}}\times \mathbf{\hat{N}}$.

With three vectors in each frame, we can construct an explicit rotation matrix to transform any quantities from one frame to the other by the set of equations
\begin{align}
\mathbf{\hat{L}}_{eq} &= \mathbf{R}\cdot \mathbf{\hat{L}}_{SF}\,, \nonumber \\ 
\mathbf{\hat{N}}_{eq} &= \mathbf{R}\cdot \mathbf{\hat{N}}_{SF}\,, \nonumber \\ 
\mathbf{\hat{K}}_{eq} &= \mathbf{R}\cdot \mathbf{\hat{K}}_{SF}\,,
                        \label{eq:rotation_matrix}
\end{align}
where $\mathbf{R}$ is the unspecified rotation matrix and the subscripts ``eq'' and ``SF'' correspond to equatorial coordinates and source-frame coordinates, respectively.
The system~\eqref{eq:rotation_matrix} can be inverted analytically, resulting in analytical expressions for the rotation matrix $\mathbf{R}$.

This rotation matrix allows us to transform any quantity between the two frames. This can be used to calculate the ecliptic angles of the total angular momentum $\mathbf{\hat{J}}$, which is needed for the LISA response function.
The vector $\mathbf{\hat{J}}$ is easily constructed in the source frame, as the spins are defined in this frame and the orbital angular momentum already defines the coordinate system.
The vector is simply rotated into the equatorial frame, and subsequently into the ecliptic frame, to compute the LISA response function.

We also need to specify the polarization angle for the terrestrial network. 
We simply use the relation~\cite{Chatziioannou:2017tdw}
\begin{equation}\label{eq:psi}
\tan \bar{\psi} = \frac{\mathbf{\hat{J}}\cdot \mathbf{\hat{z}} - (\mathbf{\hat{J}}\cdot \mathbf{\hat{N}} ) (\mathbf{\hat{z}} \cdot \mathbf{\hat{N}})}{ \mathbf{\hat{N}} \cdot ( \mathbf{\hat{J}} \times \mathbf{\hat{z}}) } \,,
\end{equation}
where $\mathbf{\hat{z}}$ is the unit vector of the equatorial coordinate system aligned with the axis of rotation of the Earth, defining a globally consistent polarization angle.
These transformations allow us to use the vector of parameters outlined above, where all the quantities are consistently defined.

Once a combined Fisher for each source is calculated, the inversion of each Fisher results in the individual covariance matrices, which effectively acts as marginalization. 
We extract the variance of the ppE parameter $\beta$  by taking the diagonal element $\sigma_{\beta \beta}$, which gives us a marginalized posterior on $\beta$ for a single source. 
Finally, to combine the sources, we multiply the marginalized posteriors together (because each source is completely independent), which for a series of Gaussians becomes
\begin{align}
p(\beta| \vec{\theta}) &\propto \prod_i^N \exp \left(- \frac{1}{2} \frac{\beta^2}{\sigma_{\beta,i}^2} \right) \\ \nonumber
&\propto \exp \left(- \frac{1}{2} \beta^2 \sum_i^N \frac{1}{\sigma_{\beta,i}^2} \right)\,.
\end{align}
Therefore, our resulting bound on $\beta$ is simply given by 
\begin{equation}
\label{eq:PPEcombined}
\sigma_{\beta}^2 = \left( \sum_i^N \frac{1}{\sigma_{\beta,i}^2} \right)^{-1}\,.
\end{equation}

\section{Mapping to Specific Theories}\label{sec:theories}

The main goal of this appendix is to map parameterized deviations, that do not necessarily have a physical interpretation, to specific parameters appearing in beyond-GR theories.  

\subsubsection{Dipole Radiation}\label{sec:dipole}

In GR, the generation of GWs is sourced from the second time derivative of the mass quadrupole moment, resulting in quadrupolar radiation. 
This connection to the quadrupole moment is tied to the conservation of the stress energy tensor, rooted in the Bianchi identities (a purely geometrical constraint). 
If additional fields were added to the gravitational sector that were not subject to such energy conditions, one would generically expect dipolar radiation, providing an additional avenue of energy loss for the system.
An additional channel for outgoing power would drive the binary to inspiral faster than what would be predicted by GR, and this faster inspiral would produce a measurable effect on the waveform.

To determine this effect on the waveform, we can write the time derivative of the gravitational binding energy of the system as~\cite{Chamberlain:2017fjl}
\begin{equation}\label{eq:dip_energy}
\dot{E} = \dot{E}_{\GR} + \delta \dot{E}\,, 
\end{equation}
where $\dot{E}_{\GR}$ is the GW power output in GR, and $\delta \dot{E}$ is our generic deviation.
In terms of these parameters, our modification to the waveform  becomes (in the language of ppE parameters)~\cite{Chamberlain:2017fjl}
\begin{equation}\label{eq:dipole_beta}
\beta_{\DIP} = \frac{-3}{224} \eta^{2/5} \delta \dot{E}\,, 
\end{equation}
where $\eta=m_1 m_2 / ( m_1 + m_2)^2 $ is the symmetric mass ratio of the binary system.

Of course, $\delta \dot{E}$ is written generically in Eq.~\eqref{eq:dip_energy}.
Once a specific theory has been selected, this term will be a function of the source parameters and of any fundamental constants of the theory in question.
For example, in Einstein-dilaton Gauss-Bonnet gravity (EdGB)~\cite{Alexander:2009tp} the waveform modification can be calculated to be~\cite{Nair:2019iur}
\begin{align}\label{eq:EdGB_beta}
\beta_{\EdGB} &= -\frac{5}{7168} \frac{\zeta_{\EdGB}}{\eta^{18/5} } \frac{\left(m_1^2 s_2^{\EdGB} - m_2^2 s_1^{\EdGB} \right)^2}{m^4}\,, \\
s_i^{\EdGB} &= \frac{2 \left[ \left(1-\chi_i^2\right)^{1/2} - 1+\chi_i^2 \right]}{\chi_i^2}\,,\label{eq:EdGB_sen}
\end{align}
where $\zeta_{\EdGB}$ is related to the coupling parameter of the theory $\alpha_{\EdGB}$ by $\zeta_{\EdGB} = 16 \pi \alpha_{\EdGB}^2 (1+z)^4/ m^4$, and $m=m_1+m_2$ is the total redshifted mass of the system. The quantities $s_i^{\EdGB}$ given in Eq.~\eqref{eq:EdGB_sen} are the sensitivities of the BHs, and $\chi_i$ are the dimensionless, (anti-)aligned spin components of the $i$th BH. 

Because of the approximations used to derive Eq.~\eqref{eq:EdGB_beta}, this particular formula is only valid when $\sqrt{\alpha_{\EdGB}} \leq m_s/2$, where $m_s$ is the smallest length scale of the system (see e.g.~\cite{Julie:2019sab}). 
For this work, the smallest length scale will be the mass of the smaller BH, $m_2$.

\subsubsection{Black Hole Evaporation}\label{sec:BHE}

High-energy theories that might be candidates for quantum theories of gravity often involve the embedding of our four-dimensional spacetime in a higher-dimensional space, where the extra dimensions are often compactified. 
For example, Arkani-Hamed, Dimopoulos, and Dvali proposed a model which had implications for the hierarchy problem between the electroweak and Planck scale~\cite{ArkaniHamed:1998rs,ArkaniHamed:1998nn}. 
Another set of models proposed by Randall and Sundrum (RS-I/II)~\cite{Randall:1999ee,Randall:1999vf} postulate a braneworld model where the four-dimensional brane we occupy resides in a five-dimensional anti-de Sitter bulk spacetime.
In RS-II, BHs were initially predicted to evaporate much faster as compared with analogous situations in four dimensions, with an evaporation rate given by~\cite{Emparan:2002jp,Berti:2018cxi}
\begin{equation}\label{eq:evaporation_rate}
\frac{dm}{dt} = -2.8\times 10^{-7} \left( \frac{1 M_{\odot} (1+z)}{m}\right)^2 \left( \frac{l}{10\mu m}\right)^{2} \frac{M_\odot}{\text{yr}}\,,
\end{equation}
where $l$ is the lengthscale of the extra dimension and $m$ is the detected mass. However, more recent work has shown that black holes in RS-II are actually stable and evaporation does not occur~\cite{Figueras:2011gd,Abdolrahimi:2012qi}. 

Regardless of the physical origin of the evaporation, it is still interesting to consider its effect on the gravitational waveform. Let us imagine that either the volume or the area of a BH changes with time due to some quantum or classical extension of GR. The volume and the area are common geometric quantities associated with a BH, so it is plausible that if BH solutions become time-dependent, then it is these quantities that acquire the time dependence. Assume then that $dV/dt = c_{V} \ell^{2}$ or $dA/dt = c_{A} \ell$, where $c_{V,A}$ are dimensionless constants and $\ell$ is a new length that controls the scale at which time dependence kicks in. If so, using that $V = (32 \pi/3) m^{3}$ and $A = 16 \pi m^{2}$ for a Schwarzschild BH, we then have that $dm/dt = [c_{V}/(32 \pi)] (\ell/m)^{2}$ or $dm/dt = [c_{A}/(32 \pi)] (\ell/m)$. On general grounds, then, one would expect $dm/dt \sim (\ell/m)^{q}$ with $q=1$ or $q=2$, depending on whether the time dependence acts on the area or the volume of the BH, and a constraint on $dm/dt$ would then imply a constraint on the evaporation scale $\ell$.

Regardless of the process that leads to evaporation, the waveform modification has the form~\cite{Yagi:2011yu}
\begin{equation}\label{eq:BHE_beta}
\beta_{\BHE} = \frac{25}{851968}\dot{M}\left(\frac{ 3- 26 \eta + 34 \eta^2}{\eta^{2/5}\left(1-2\eta\right)}\right)\,,
\end{equation}
where $\dot{M} = d M / dt = d m_1 / dt + d m_2 / dt$ is the anomalous evaporation rate.

\subsubsection{Local Position Invariance Violation}\label{sec:LPP}

In the case where Newton's gravitational constant is promoted to a time-dependent quantity, conspicuous additional accelerations could be experienced by binaries inspiralling together. 
This phenomenon could come about, for example, because the gravitational constant is tied to a background scalar field which evolves on cosmological timescales.
This effect can be observed as alterations to the binding energy of the binary, and it has a mapping to the ppE framework~\cite{Yunes:2009bv}:
\begin{equation}\label{eq:Gdot_beta}
\beta_{\Gd} = \frac{-25}{65526} \frac{ \dot{G}  \mathcal{M}}{(1+z) }\,,
\end{equation}
where $\dot{G}=d G / dt$ is the time derivative of the gravitational constant and $\mathcal{M}$ is the redshifted chirp mass.

\subsubsection{Parity Violation}\label{subsusbsec:PV}

Many attempts to unify quantum mechanics and gravity involve terms quadratic in curvature at the level of the action in the low-energy limit, as well as additional fields coupled to these higher-order terms. 
The strength of this coupling is determined by the coupling parameter of the theory, and therefore determines the magnitude of the effect on the waveform.
EdGB (discussed above) is an example of this type of modification where the modifying parameter comes at a negative PN order because of dipolar radiation. 
EdGB, however, preserves parity because the term added to the action is parity-even, introducing a scalar field that is also parity-even.
A quadratic theory that does not preserve parity is dynamical Chern-Simons (dCS) gravity~\cite{Alexander:2009tp}, which incorporates an additional quadratic curvature term into the action that is parity-odd.
In order to keep the action invariant under parity transformations, this odd-parity term must be coupled to an odd-parity scalar field,
leading to a variety of implications in different gravitational interactions~\cite{Alexander:2009tp}.

This modification affects the waveform as follows~\cite{Nair:2019iur}:
\begin{align}\label{eq:dCS_beta}\nonumber
\beta_{\dCS} &= -\frac{5}{8192} \frac{\zeta_{\text{dcs}}}{\eta^{14/5}} \frac{\left(m_1 s_2^{\text{dCS}} - m_2 s_1^{\text{dCS}} \right)^2}{m^2}\\ & 
+ \frac{15075}{114688} \frac{\zeta_{\text{dCS}}}{\eta^{14/5}} \frac{\left(m_2^2 \chi_1^2 - \frac{350}{201} m_1 m_2 \chi_1 \chi_2 + m_1^2 \chi_2^2\right)}{m^2} \,, \\
s_i^{\dCS} &\equiv \frac{ 2 + 2\chi_i^4 - 2\left( 1 - \chi_i^2\right)^{1/2} - \chi_i^2 \left[ 3 - 2 \left( 1 - \chi_i^2\right)^{1/2}\right]}{2\chi_i^3}\,,\label{eq:dCS_sens}
\end{align}
where $\zeta_{\dCS}$ is related to the coupling parameter by $\zeta_{\dCS} = 16\pi \alpha_{\dCS}^2 (1+z)^4 / m^4$. The quantity $s_i^{\dCS}$ given in Eq.~\eqref{eq:dCS_sens} is the sensitivity of the $i$th BH in dCS. 

As the result of the approximations involved in the derivation of the flux, Eq.~\eqref{eq:dCS_beta} is only valid if $\sqrt{\alpha_{\dCS}} \leq m_s/2$, where $m_s$ is the smallest length scale of the system, just as in EdGB.
Here we are interested in BBHs, and $m_s$ is the mass of the smaller BH.

\subsubsection{Lorentz Violation}\label{sec:LV}

Noncommutative gravity promotes the coordinates in GR to operators with a nontrivial commutation relation defined by
$\left[ \hat{x}^\mu, \hat{x}^\nu\right] = i \theta^{\mu\nu}$,
where $\theta^{\mu\nu}$ is a real, constant antisymmetric tensor~\cite{Tahura:2018zuq,Kobakhidze:2016cqh}. 
This tensor plays a role analogous to the role of Planck's constant in quantum mechanics, and
it defines a length scale at which there is a fundamental uncertainty between physical parameters.

Defining the quantity $\Lambda^2 = \theta^{0i}\theta_{0i}/(l_p t_p)^2$, where $l_p$ and $t_p$ are the Planck length and time, respectively, one can derive the modification to the waveform as~\cite{Tahura:2018zuq,Kobakhidze:2016cqh}
\begin{equation}\label{eq:NC_beta}
\beta_{\NC} = -\frac{75}{256} \eta^{-4/5} \left( 2 \eta - 1 \right) \Lambda^2\,.
\end{equation}
In this parameterization, $\sqrt{\Lambda}$ defines the energy scale of noncommutativity, relative to the Planck scale.

\begin{figure}
\includegraphics[width=\linewidth]{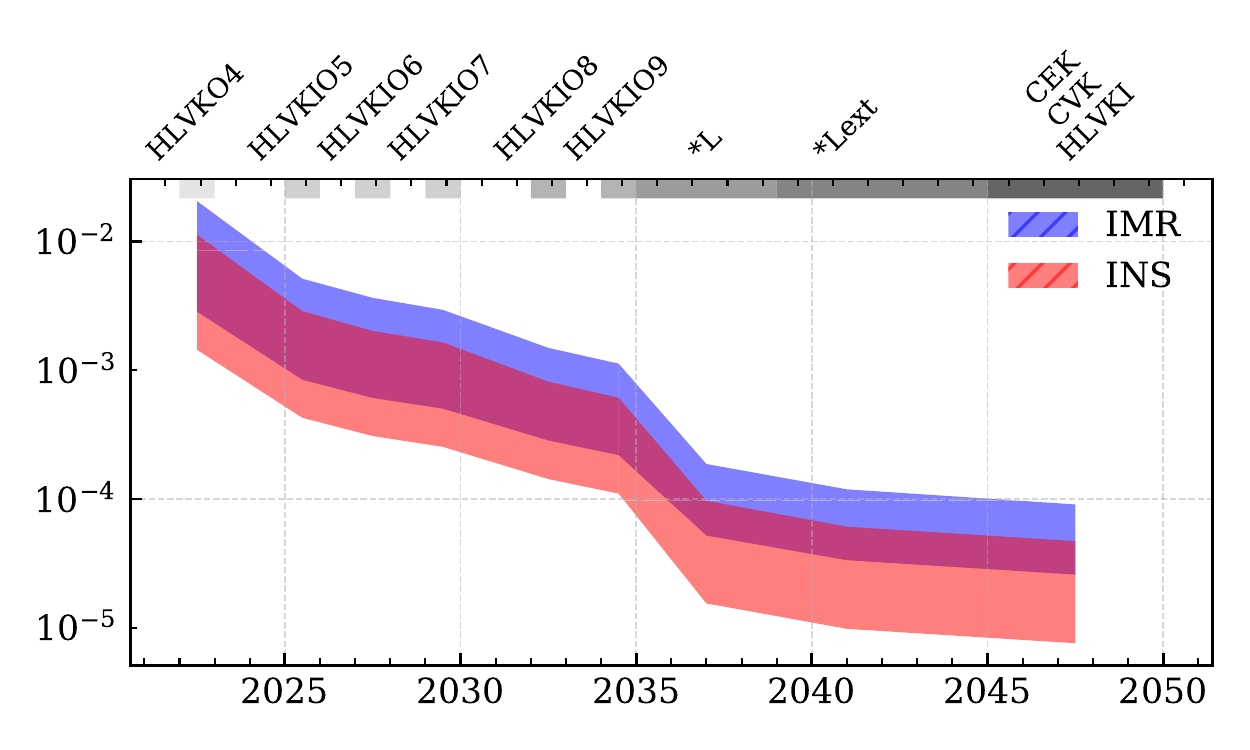}
\caption{Comparison between the constraints on $\beta$ at 1.5PN predicted by using the generation modification (INS), as opposed to the propagation modification (IMR).
	We used the same catalogs and networks (SOBH Base and SOBH S1) in both cases.
	The difference is negligible when considering the order-of-magnitude constraints of interest in this work.
	}\label{fig:IMR_INS}
\end{figure}

\subsubsection{Modified Dispersion}\label{sec:MGR}

Another assumption made by GR is that gravitons are massless.
If this is not assumed, the leading-order correction to the measured GW signal would come about through the propagation of GW~\cite{Will:1997bb,Yunes:2016jcc}.
The graviton would be ascribed a massive-particle dispersion relation 
$E^2 = p^2 + m_g^2$,
where $E$ is the graviton energy, $p$ is the graviton momentum, and $m_g$ is the graviton mass.
With a nonlinear relation between energy and momentum, one would expect that the group velocity would become frequency-dependent. 
This introduces an additional term in the GW phase~\cite{Will:1997bb}:
\begin{align}\label{eq:MG_beta}
\beta_{\MG} &= \pi^2 \frac{D_0}{1+z}\frac{\mathcal{M}_z}{\lambda_{\text{\MG}}^2}\,, \\
D_0 &\equiv \left(1+z\right)\int^{z}_{0} \frac{1}{H(z')}\frac{dz'}{\left(1+z'\right)^2}\,,
\end{align}
where $D_0$ is a new cosmological distance similar to the luminosity distance, and $\lambda_g$ is the Compton wavelength of the graviton, related to the mass by $\lambda_g=h/m_g$. To evaluate the Hubble parameter $H(z)$ we use the cosmological parameters inferred from the Planck Collaboration~\cite{Ade:2015xua} and software from the \software{Astropy} python package~\cite{2013A&A...558A..33A,2018AJ....156..123A}. 

\section{Inspiral/merger/ringdown vs. inspiral waveforms}\label{sec:imr_vs_ins}

Concerning the deviations away from GR that we have injected into the waveforms, we examine two families of modifications: those that affect GW propagation and those that modify GW generation. 
The difference between these two mechanisms arises from our lack of knowledge about the dynamics of BBHs close to merger in modified theories of gravity. 
To reflect this ignorance, we include the modification due to generation effects in the inspiral portion of the waveform only. 
Propagation effects are under no such shroud as the mechanism responsible acts in the low-curvature regions between galaxies and should equally affect the waveform across the entire frequency range. 
We therefore include modifications due to propagation effects in the entire waveform. 
As we are only ever looking at one effect at a time, these two families of effects are never examined concurrently. 
To incorporate these modifications, we utilize the ppE methodology~\cite{Yunes:2009ke,Cornish:2011ys,Sampson:2013lpa,Chatziioannou:2012rf}.
In the case of precessing systems, the modifications are treated slightly differently.
For generational effects, we append a phase modification to the waveform in the coprecessing frame, where the physics of GW generation are approximately the same as those for a nonprecessing binary.
The waveform is then ``twisted-up'' in the usual fashion for \software{IMRPhenomPv2} waveforms, but with the modified coprecessing waveform.
For propagation effects, we append the modification to the waveform at all frequencies, after the waveform has been transformed to the inertial frame. In equations:
\begin{align}
\tilde{h}_{\text{coprec},\text{gen}} &= 
\begin{cases} 
\tilde{h}_{\text{coprec},\text{GR}} e^{i \beta \left(\mathcal{M} \pi f \right)^{-b/3}} & f<0.018m \\ 
\tilde{h}_{\text{coprec},\text{GR}} &  0.018m<f \\ 
\end{cases}\\
\tilde{h}_{\text{inertial},\text{prop}} &= 
\tilde{h}_{\text{inertial},\text{GR}} e^{i \beta \left(\mathcal{M} \pi f \right)^{-b/3}}\,.
\end{align}

A comparison between the two methods is shown in Fig.~\ref{fig:IMR_INS}, which illustrates that the difference is small.
Because of this, we used the full inspiral-merger-ringdown modification in all of this paper.

\bibliography{refs}

\end{document}